# Few layer 2D pnictogens catalyze the alkylation of soft nucleophiles with esters.

Vicent Lloret,[a,†] Miguel Ángel Rivero–Crespo,[b,†] José Alejandro Vidal–Moya,[b] Stefan Wild,[a] Antonio Doménech–Carbó,[c] Bettina S.J. Heller,[d] Sunghwan Shin,[d] Hans-Peter Steinrück,[d] Florian Maier,[d] Frank Hauke,[a] Maria Varela,[e] Andreas Hirsch,[a] Antonio Leyva–Pérez[b,]* and Gonzalo Abellán.[a,f]*

[a] Department of Chemistry and Pharmacy and Joint Institute of Advanced Materials and Processes (ZMP) Friedrich–Alexander–Universität Erlangen–Nürnberg (FAU) Henkestrasse 42, 91054 Erlangen and Dr.–Mack Strasse 81, 90762 Fürth, Germany. Phone: +49 91165078–65031; Fax: +49 91165078–65015.

[b] Instituto de Tecnología Química. Universidad Politècnica de València–Consejo Superior de Investigaciones Científicas. Avda. de los Naranjos s/n, 46022, Valencia, Spain. Phone: +34963877800; Fax: +349638 77809.

[c] Departament de Química Analítica. Universitat de València. Dr. Moliner, 50, 46100 Burjassot (València) Spain.

[d] Chair of Physical Chemistry II, Friedrich–Alexander–Universität Erlangen–Nürnberg (FAU) Egerlandstr. 3, 91058 Erlangen, Germany.

[e] Universidad Complutense de Madrid, Instituto Pluridisciplinar, Instituto de Magnetismo Aplicado & Departamento de Física de Materiales, Madrid 28040, Spain.

[f] Instituto de Ciencia Molecular (ICMol), Universidad de Valencia, Catedrático José Beltrán 2, 46980, Paterna, Valencia, (Spain).

[†] These authors contributed equally to this work.

Corresponding authors: anleyva@itq.upv.es, gonzalo.abellan@fau.de.

**Abstract.**

Group 15 elements in zero oxidation state (P, As, Sb and Bi), also called pnictogens, are rarely used in catalysis due to the difficulties associated in preparing well–structured and stable materials. Here, we report on the synthesis of highly exfoliated, few layer 2D phosphorene and antimonene in zero oxidation state, suspended in an ionic liquid, with the native atoms ready to interact with external reagents while avoiding aerobic or aqueous decomposition pathways, and on their use as efficient catalysts for the alkylation of nucleophiles with esters. The few layer pnictogen material circumvents the extremely harsh reaction conditions associated to previous superacid–catalyzed alkylations, by enabling an alternative mechanism on surface, protected from the water and air by the ionic liquid. These 2D catalysts allow the alkylation of a variety of acid–sensitive organic molecules and giving synthetic relevancy to the use of simple esters as alkylating agents.

## Introduction.

Two–dimensional (2D) materials have attracted great attention in the last years due to their outstanding physical properties and their potential applications in optoelectronics, sensors, energy storage and catalysis.[1] In contrast to the most studied material graphene, the layered allotropes of group 15 elements (P, As, Sb and Bi, also called pnictogens) have been fairly less developed. 2D pnictogens exhibit a marked puckered structure[2–4] with dative electron lone pairs located on the surface atoms, which results in semiconducting character and good electronic mobility,[4,5] and also in the ability to easily adsorb and stabilize, particularly well, unsaturated organic molecules through van der Waals interactions.[6,7] Thus, 2D pnictogens might, in principle, act as catalysts in synthetic organic transformations involving unsaturated molecules, in a completely different way as graphene does.[8,9] This concept, however, requires a new methodology to synthesize large amounts of exfoliated material, thus exposing most of the catalysts atoms to the outer space for maximizing interaction with substrate molecules.

Alkylation reactions are fundamental in biochemistry and organic synthesis. Nature makes use of alkyl phosphates, sulphonates and esters as alkylating agents, under metal–free physiological conditions.[10,11] In contrast, synthetic methods generally employ energetically higher alkyl halides and alcohols as alkylating agents under very strong basic or acidic conditions (*i.e.* Williamson synthesis),[12] and the synthetic alkylation protocols reported with poly–oxygenated compounds need expensive and toxic metal catalysts, such as the palladium–catalyzed Tsuji–Trost allylation reaction,[13] the Hantzsch ester–assisted hafnium–catalyzed alkylation of quinones[14] and the gold–catalyzed alkylation with alkynylbenzoic acids,[15] among some others.[16,17] Thus, the discovery of a simple, metal–free, biomimetic alkylation reaction with readily–available poly–oxygenated molecules[18] remains a challenge in organic synthesis and catalysis, furthermore attractive if selective and functional–group tolerant.[19,20]

Here, we show the synthesis of two different exfoliated, few layer 2D pnictogens, phosphorene (few layer–black phosphorous, FL–BP) and antimonene (FL–Sb), and their use as catalysts in the alkylation of alcohols, thiols and indoles with simple esters, in good yields and with excellent selectivity. To our knowledge, this is the first organic reaction catalysed by pristine 2D-pnictogens reported so far. Mechanistic studies unveil that the catalytic FL pnictogen selectively adsorbs the nucleophile and ester on surface, with the help of the electronic stabilization generated by the few layers underneath. FL–Sb exhibits a better performance than FL–BP, in accordance with its higher polarizability, enabling acid–sensitive aromatic derivatives to be selectively alkylated with simple esters.

## Results and Discussion.

**Synthesis and characterization of FL–BP and FL–Sb in bmim–$BF_4$.** Figures 1a and 2a show the structure of FL–BP and FL–Sb nanosheets, respectively, produced by liquid phase exfoliation (LPE).[21,22] This technique is often carried out in amide solvents such as *N*–cyclohexyl–2–pyrrolidone (CHP) or *N*–methyl–2–pyrrolidone (NMP). Here, the ionic liquid (IL) bmim–$BF_4$ is used on the basis of its excellent oxidation protection behavior for FL–BP.[23] Sonication of ground BP or Sb crystals dispersed in 1–butyl–3–methylimidazolium tetrafluoroborate (bmim–$BF_4$) was performed in an argon–filled glovebox (<0.1 ppm of $H_2O$ and $O_2$) to yield brownish, open–air stable suspensions of unoxidized FL–BP or Sb nanosheets, after removing the unexfoliated particles by a two–step centrifugation process, 14.000 g during 1 min, and then at 2.000 and 100 g for 60 min for FL–BP and FL–Sb, respectively. The samples were stored under ambient conditions over weeks with no signature of degradation. In order to provide statistical information of the thicknesses and lateral dimensions of the as–prepared nanosheets, topographic AFM characterization and spectroscopic micro-Raman mapping of >150 nanosheets, spin–coated onto $SiO_2$/Si wafers, was performed, and the results showed that the BP particles have median values of *ca*. 150 nm in lateral dimensions and average thicknesses of 13 nm, with thinner particles down to a few nanometers being predominant (see Figures S1–7). Particles smaller than 2 nm were excluded from statistics since capillary and adhesion effects of the IL account for average motifs of *ca*. 60 nm in lateral dimensions and *ca*. 1.8 nm in thickness (see Fig. S8–9).[24] It is worth noting the general difficulties associated to AFM measurements in the presence of ILs, due to the high viscosity, adhesion forces and formation of IL aggregates. The corresponding scanning Raman microscopy (SRM) spectra (>14000 single point spectra), with an excitation wavelength of 532 nm, unambiguously showed the characteristic modes of BP, labeled $A_g^1$, $B_{2g}$, and $A_g^2$, with no signature of oxidation attending to the $A_g^1/A_g^2 > 0.6$ intensity ratio statistics, independent of the orientation (see Figure 1 and Figures S1–7 in Supporting Information for additional Raman and AFM characterization).[23,25] Aberration corrected scanning transmission electron microscopy (STEM) combined with electron energy–loss spectroscopy (EELS) was used to investigate the local structure and chemistry of the flakes. An atomic–resolution high angle annular dark field (HAADF) STEM image of the FL–BP sample acquired down the [110] crystallographic direction, with the electron beam perpendicular to the platelet plane, is shown in Figure 1g (both raw and Fourier filtered data, Figure S10 displays a low magnification image of the flake). The samples exhibit a very high degree of crystallinity, showing the characteristic puckered structure over regions of hundreds of nanometers. The lattice shows high uniformity with the presence of very few defects or dislocations. Analysis by electron energy–loss spectroscopy (EELS) shows that the IL locates and nanometrically covers the edges of the 2D material, as assessed by the P $L_{2,3}$–edges, the C *K*–edge, the N *K*–edge and the O *K*–edge chemical maps in Figure 1h, with onsets near 132, 284, 401 and 532 eV, respectively. X–ray diffraction (XRD) of a FL–BP sample, measured after washings with tetrahydrofurane (THF) under nitrogen atmosphere, ultracentrifugation and evaporation of the solvent, shows a spectrum consistent with BP, with the typical 020, 040 and 060 planes and without any sign of degradation nor oxidation, and when the sample was exposed to the ambient, the peak intensities rapidly decreased (Figure S10). These results infer the high surface area of the FL–BP synthesized here.

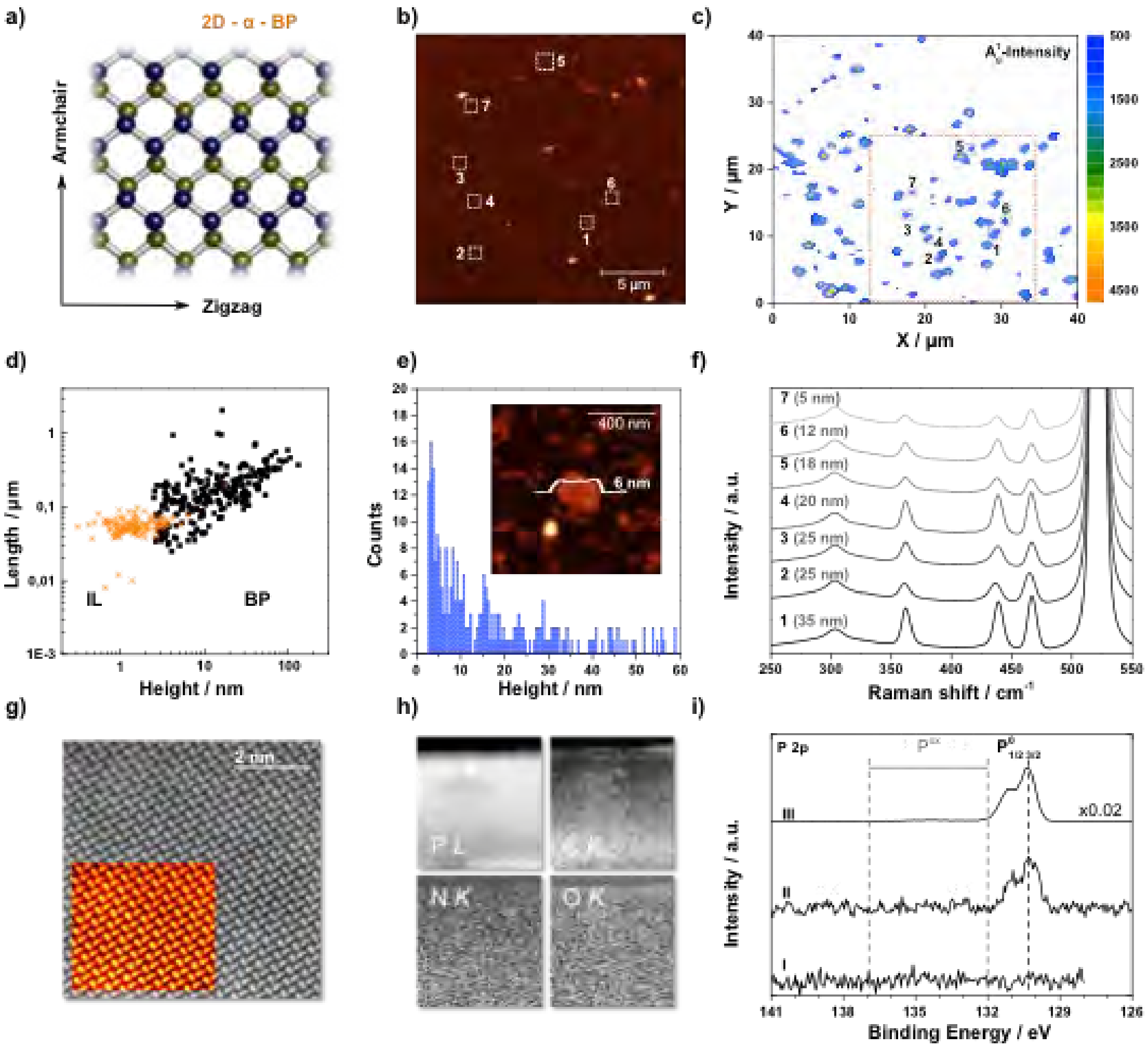

**Figure 1**. **Characterization of exfoliated black phosphorus.** (a) Top view of the orthorhombic structure ($C_{mce}$ space group) of BP. Upper plane P atoms marked in blue, lower plane in yellow. (b) Representative AFM topography image (inset, scale bar 5 µm) of the exfoliated sample spin–coated onto $SiO_2$/Si substrates. (c) The corresponding Raman $A^1_g$ ($\lambda_{exc}$=532 nm) mapping of the same BP flakes (>14000 single point spectra over a surface area of 40 µm$^2$ using a step size of 0.5 µm). The numbers denote the areas in which the Raman spectra shown in (f) were recorded. (d) Plot of the nanosheet length as a function of the flake height (obtained from AFM) considering a total amount of 252 replicates as well as the data corresponding to the IL blank (sample size=116) Supporting Information S8–9). (e) Histogram of the apparent thickness of the exfoliated FL–BP obtained from AFM (sample size=252). The inset shows an AFM image of a nanosheet along with its corresponding height profile of *ca*. 6 nm (inset, scale bar 400 nm) . (f) Raman spectra of the nanosheets indicated in b and c (the numbers labeling the spectra in f correspond to the nanosheets marked by numbers in b and c). (g) Atomic resolution HAADF image acquired down the [110] axis, from the edge of a free–standing portion of a flake. The inset exhibits the raw image and a Fourier filtered (FFT) version, in false color. The scale bar represents

2 nm. (h) Compositional maps derived from electron energy–loss spectroscopy (EELS) measurements acquired on the free–standing portion of the BP flake. The P $L_{2,3}$, C *K*, N *K* and O *K* maps for this area are shown, the scale bar represents 20 nm. Data acquired at 80 kV. (i) XPS P 2*p* region of the neat bmim-$BF_4$ IL (I), the highly-concentrated FL-BP suspension (II) showing only P in oxidation state zero at P $2p_{3/2}$ = 130.2 eV (region for oxidized P species is indicated), and after removal of most of the IL by heating in UHV (III); spectra are offset and re-scaled for sake of clarity. Source data are provided as a Source Data file.

To confirm the zero oxidation state of P and rule out partial reduction of oxidized P species in EELS by the electron beam, X-ray photoelectron spectroscopy (XPS) studies have been carried out under ultra-high vacuum (UHV) conditions on highly concentrated IL FL-BP suspensions ($FL\text{-}BP_{sus}$) coating a clean Au foil as support. The overview spectrum of $FL\text{-}BP_{sus}$ shows the expected IL core levels (Figure S11) and, additionally, typical Si/O/C signals of trace bulk contamination after contact with glassware grease, due to surface enrichment effects as it is often the case in XP spectra for IL systems.[26] At around 130 eV binding energy, a small signal of the spin-orbit split P $2p_{1/2,3/2}$ signal is detected that is absent for the neat IL (Figure 1I, spectrum II). The binding energy position of the P $2p_{3/2}$ level at 130.2 eV can be unambiguously assigned to BP in oxidation state zero and the absence of signals between 132 eV and 137 eV rules out significant oxidized P species being present.[27] By heating the $FL\text{-}BP_{sus}$ sample above 150 °C for 1 h in UHV, most of the IL was gone by thermal desorption (and partial decomposition), which led to an increase of the BP signal intensity as dominating remaining species by a factor around 50 (Figure 1i, spectrum III); again, no oxidized P species could be detected. In order to check if the heated $FL\text{-}BP_{sus}$ sample with most of the protecting IL removed was now prone to oxidation, the sample was exposed to UHV and, then, to air under ambient conditions for about one day. XPS clearly revealed a broad oxide P component around 134 eV (Figure S12, spectrum IV) as has already been observed for *in situ* oxidation studies of BP.[27]

In the case of FL–Sb, the shorter out–of–plane atom–to–atom distances, which are indicative of stronger interlayer interactions, usually hampers mechanical exfoliation. However, the LPE approach here used was able to give median values of 310 nm in lateral dimensions and *ca*. 32 nm in thickness (extracted from >150 flakes), as it can be observed in Figures 2 and S13–18, with a minimum observed apparent thickness of 4 nm.[22] The SRM mappings revealed the characteristic main phonon peaks, the $A^1_g$ mode at 149.8 $cm^{-1}$ and $E_g$ mode at 110 $cm^{-1}$, even for the thinnest particles with no signature of oxidation (peaks related to the formation of $Sb_2O_3$ or $Sb_2O_5$). A phonon softening effect (blueshift) was observed when the sample thickness decreases from the bulk to *ca*. 10 nm, in good agreement with theoretical predictions and recent reports (Figures 2f and S19–20).[4,22,28,29] The $E_g/A^1_g$ intensity ratio (measured using 532 nm excitation wavelength) increases from 0.37 to 0.79 with thickness decreasing from 80 nm to 10 nm (Figure S19).[29]

FL–Sb electron microscopy images denote irregularly shaped sub–micrometric flakes, with lateral sizes in the range of hundreds of nm, as assessed by a low–magnification high angle annular dark field (HAADF) image of a flake and the atomic–resolution image of the crystal structure (Figures 2g and S21), both obtained at an acceleration voltage of 80 kV to prevent beam–induced damage. This structure agrees with that of β–antimony along the [2 1 0] direction. Again, the samples are highly crystalline, and no major defects were observed. EELS maps exhibit a *C*–rich coating consisting of an amorphous layer a few nanometers thick, as well as the presence of O, N and C mostly located around the edges (Figure 2h). The seemingly preferential location of the IL molecules along the edges is in good agreement with the expected higher polarity of the unsaturated atoms of the 2D material.[6,23,30]

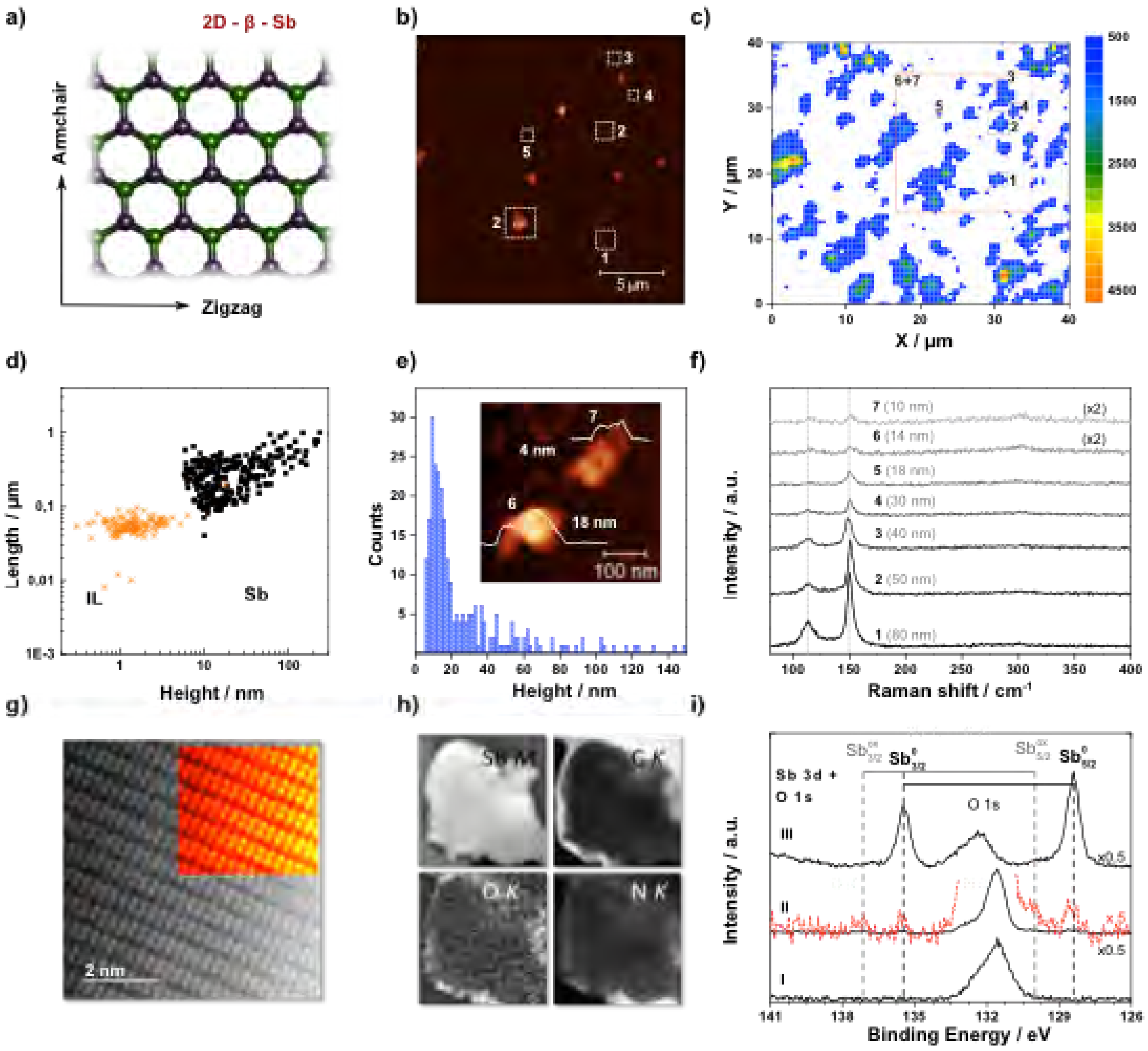

**Figure 2**. **Characterization of exfoliated antimonene.** FL–Sb analysis. (a) Top view of the rhombohedral structure (R3m space group) of Sb. Upper plane atoms marked in green, lower plane in purple. (b) Representative AFM topography image (inset, scale bar 5 µm) of the exfoliated sample spin–coated onto $SiO_2$/Si substrates. (c) The corresponding Raman $A^1_g$ ($\lambda_{exc}$=532 nm) mapping of the same Sb flakes (>14000 single point spectra over a surface area of 40 µm$^2$ using a step size of 0.5 µm). The numbers denote the areas in which the Raman spectra shown in (f) were recorded. (d) Plot of the nanosheet length as a function of the flake height (obtained from AFM), including the data corresponding to the IL blank (Figures S8–9) considering a total amount of 271 and 116 replicates respectively. (e) Histogram of the apparent thickness of the exfoliated FL–Sb obtained from AFM (sample size=271). The inset shows an AFM image of a nanosheet along with its corresponding height profiles of *ca*. 4 and 18 nm, respectively (inset, scale bar 100 nm). (f) Raman spectra of the nanosheets indicated in b and c (the numbers labeling the spectra in f correspond to the nanosheets marked by numbers in b and c). (g) Atomic resolution HAADF image acquired on the edge of a free–standing portion of a flake, near the edge, along with a Fourier filtered (FFT) image in the inset, acquired down the [210] orientation. The scale bar is 2 nm. (h) Compositional maps, derived from EEL spectrum images of the flake (from the area highlighted with a green rectangle in Figure S21). Sb $M_{4,5}$, C *K*, O *K* and N *K* maps corresponding to this area are shown. Data acquired at 80 kV. (i) XPS Sb 3*d* and O 1*s* region of the neat bmim-$BF_4$ IL (I) showing oxygen signals from an IL related surface contamination layer, of the highly-concentrated FL-Sb suspension (II) showing small signals of non-oxidized (Sb $3d_{5/2}$ at 528.2 eV) and minor contributions from oxidized (530.3 eV) antimony next to the oxygen contamination, and after removal of most of the IL by heating in UHV (III); spectra are offset and re-scaled for sake of clarity. Source data are provided as a Source Data file.

As done for the FL-BP system, highly concentrated IL FL-Sb suspensions (FL-$Sb_{sus}$) were investigated using XPS. Next to the broad O 1s signal at 533 eV from the surface enriched IL contamination (see also overview spectrum shown in Figure S22), the Sb 3*d* region between 525 and 545 eV of the FL-$Sb_{sus}$ sample (Figure 2i, spectrum II) reveals very weak Sb $3d_{3/2,5/2}$ signals from antimony in oxidation state zero could be detected at 528.2 eV for the $3d_{5/2}$ level, along with minor contributions from Sb in higher oxidation state at around 530.3 eV (Figure 2i, magnified red spectrum II).[31,32] Removing most of the excess IL by heating in UHV clearly showed Sb signals originating mostly from bulk antimony zero (spectrum III). Exposing the heated FL-$Sb_{sus}$ sample without the protecting IL medium for several hours to air led to a drastic decrease in Sb(0) and concomitant increase of the oxidized Sb species (Figure S23, spectrum IV); these findings thus strongly supports the role of bmim-$BF_4$ stabilising the P and Sb pnictogens in IL solution against oxidation.

**Catalytic alkylation with esters.** Substitution reactions are fundamental transformations in organic chemistry that, due to their bimolecular nature and in order to be performed selectively, require the use of a catalyst able to activate the substrates orthogonally. Highly polarized Lewis bases are, in principle, suitable species to carry out a bimolecular and orthogonal catalytic activation, and FL–BP and FL–Sb may act in this way due to the intrinsic electron richness of the bulk atoms (base) combined with the expected stabilization of the in-situ generated cationic charge.

As a reaction proof, the *tert*–butylation of alcohols, a longstanding challenge in organic synthesis,[33] was studied. Current methodologies at laboratory[34] or industrial scale[35] for this reaction are still based on Friedel–Crafts type chemistry, with isobutylene or *tert*–butyl alcohol as alkylating agents under very strong reaction conditions. These harsh protocols are substrate–limiting and particularly unselective in the presence of aromatic rings,[36] despite synthetically elaborated, energetically richer and expensive *tert*–butylated reagents have been prepared on purpose to mitigate these drawbacks.[37]

Table 1 shows the results for the reaction between benzyl alcohol **1** and *tert*–butyl acetate **2**, in the presence of different catalysts. FL–BP and FL–Sb exclusively gives *tert*–butyl ether **3** after 20 h at 75 ºC, in reasonable yields and with >99% selectivity (the rest is unreacted material, entries 2 and 3). In contrast, other bifunctional layered materials (entries 4–7) and semiconductors with high surface area (entries 8–10), and also inorganic and organic bases of different strength (entries 11–18), do not show any significant catalytic activity under these reaction conditions.

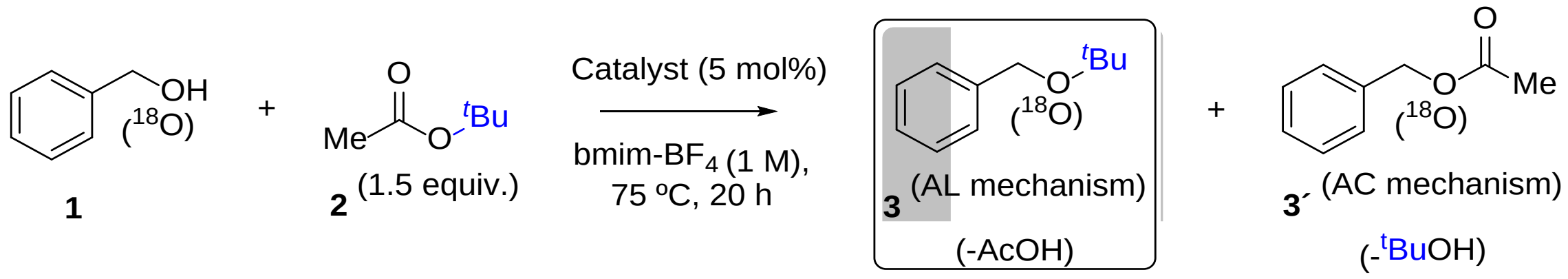

**Table 1** Results for the catalytic alkylation of **1** with **2**. Conversion of **1** is the sum of **3** and **3´** yields, the rest is unreacted material.

| Entry | Catalyst | $TOF_0$ ($h^{-1}$) | Yield of 3 (%) | Yield of 3´ (%) | Selectivity AL/AC (%) |
|---|---|---|---|---|---|
| 1 | none | – | – | – | – |
| 2 | **Phosphorene (FL-BP)** | **20.1** | **46** | <1 | >99 |
| 3 | **Antimonene (FL-Sb)** | **73.2** | **57** | <1 | >99 |
| 4 | Graphene | – | 4 | <1 | – |
| 5 | Boron nitride | – | 3 | <1 | – |
| 6 | Hydrotalcite | – | <1 | <1 | – |
| 7 | Sepiolite | – | 6 | <1 | – |
| 8 | Nano–$TiO_2$ | – | 2 | <1 | – |
| 9 | Nano–$CeO_2$ | – | <1 | <1 | – |
| 10 | Nano–$Fe_2O_3$ | – | <1 | <1 | – |
| 11 | $K_3PO_4$ | – | <1 | <1 | – |
| 12 | $Na_2CO_3$ | – | <1 | <1 | – |
| 13 | Pyridine | – | <1 | <1 | – |
| 14 | $Et_3N$ | – | <1 | <1 | – |
| 15 | KOAc | – | <1 | <1 | – |
| 16 | $KO^tBu$ | – | <1 | <1 | – |
| 17 | DABCO | – | <1 | <1 | – |
| 18 | NanoMgO | – | <1 | <1 | – |
| 19 | HOAc | – | <1 | <1 | – |
| 20 | $H_2SO_4$ | 23.7 | 24 | 46 | 35 |
| 21 | $HNTf_2$ | 7.6 | 8 | 14 | 25 |
| 22 | HOTf | 154.2 | 48 | 52 | 48 |
| 23 | $MnOAc_2$ | - | 0 | - | - |
| 24[a] | $FeCl_2$ | 2.2 | 33 | <1 | >99 |
| 25[a] | $CoCl_2$ | 7.4 | 44 | <1 | >99 |
| 26 | CuCl | - | 2 | - | - |
| 27 | $AgMeSO_4$ | - | 2 | <1 | - |
| 28 | $Pd(OAc)_2$ | - | 6 | - | - |
| 29[a] | $PtCl_2$ | - | 25 | <1 | >99 |
| 30 | $Au(OH)_3$ | - | 0 | - | - |
| 31[a] | $Bi(OTf)_3$ | 6.0 | 37 | 10 | 79 |
| 32[a] | $CeCl_3$ | 7.0 | 52 | <1 | >99 |
| 33 | Sulphated zirconia | - | 11 | <1 | >99 |
| 34 | $SiO_2$-$Al_2O_3$ (13%) | - | 0 | - | - |

[a] No catalytic activity if 2,6–di*tert*–butylpyridine (30 mol%) is added during reaction.

Isotopic experiments with $^{18}O$–**1** confirm that the oxygen atom of the alcohol stays intact in product **3**, which suggests that an ester C–O alkyl cleavage (AL mechanism) operates, as reported with superacids ($H_0$<0) such as $HSO_3F$–$SbF_5$–$SO_2$.[33] Notice that, for any weaker acid, the AL mechanism is rapidly undertook by the more common acyl cleavage (AC) mechanism, to give the trans–esterification reaction. Indeed, acetic acid (AcOH) is too weak to catalyze the reaction between **1** and **2** (entry 19), sulfuric and triflimidic acid show moderate catalytic activity and give ester **3´** as the major product (entries 20 and 21), and only the superacid triflic acid (HOTf) shows catalytic activity but with still poor selectivity towards **3** (entry 22). The increase in selectivity for **3** with acid strength is in good agreement with the need of forming the carbocation intermediate of the AL mechanism, to trigger the unimolecular $A_{AL}1$ reaction and give ether **3**. Strong Lewis acids were also tried (entries 23–32), and while some of them showed some activity (entries 24–25, 29 and 31–32), their activity corresponds exclusively to the Brönsted acidity of the in–situ hydrolyzed anions, as confirmed by the lack of activity when the proton quencher 2,6–di*tert*–butylpyridine is used. Notice that the associated harsh reaction conditions to superacids are incompatible with most functional organic groups, thus any other nucleophile beyond water has not regularly been employed for the superacid–catalyzed alkylation with esters, as far as we know. The initial turnover frequency ($TOF_0$) for FL–BP and FL–Sb (20 and 73 $h^{-1}$, respectively) are in the range of the strong acids (between 8 and 155 $h^{-1}$), which reflects the good intrinsic activity per atom of the pnictogen 2D materials. Strong Brönsted solid acids such as sulphated zirconia and silica–alumina gave only a marginal catalytic activity.

Figure 3 shows the scope for the FL–Sb catalyzed reaction, with different nucleophiles and esters. The results show that a variety of *tert*–butyl esters give ether **3** in >70% yield, up to 2–gram scale, including unsaturated esters and carbonates, and that benzylic (products **4** and **10**) and alkyl alcohols (products **5** and **11**), thiols (products **6**–**7**), indoles, either in the carbon (products **8a**–**9a**) or nitrogen atom (products **8b**–**9b**), and phenols (product **11**) can be alkylated with acetates having either *tert*–butyl, cinnamyl, benzyl and prenyl moieties (in blue). Acid–sensitive functionalities are tolerated under the reaction conditions, such as ether, prenyl, trifluoroacetate and lactone groups (in red). The uniqueness of this synthetic approach is illustrated, for instance, for allylic benzylic alcohols, since product **10** has not been synthesized so far and none of the nine methods reported previously for the generic structure, according to a literature searching, provides a so simple, direct and efficient method as 2D-pnictogens do (Table S1).

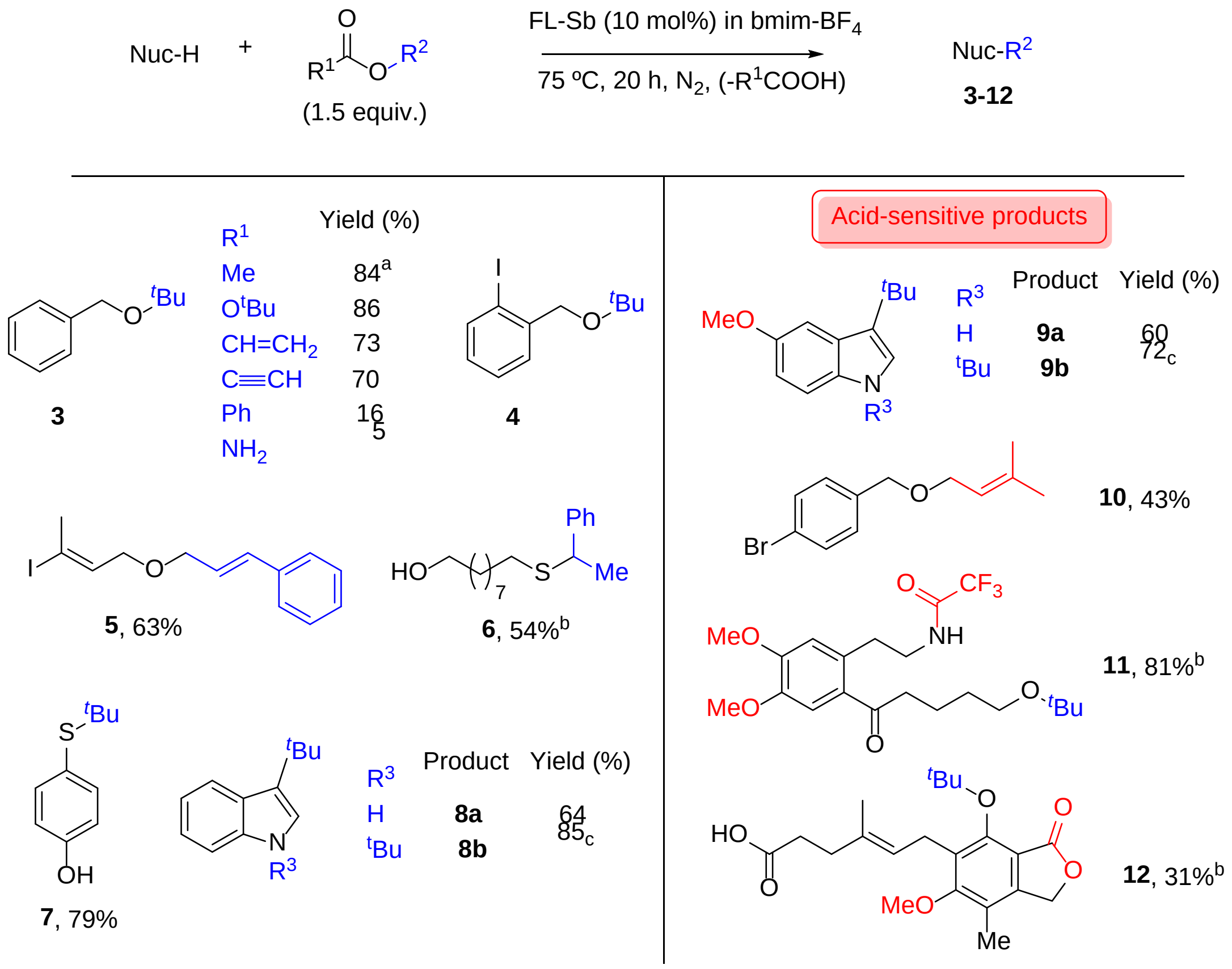

**Figure 3 Substrate scope for the FL–Sb catalyzed alkylation with esters.** $R^1$ is Me (acetate) unless otherwise indicated. In blue, the introduced alkyl moiety; in red, particularly acid–sensitive functional groups. [a] 2–gram scale. [b] 25 mol% catalyst, 48 h. [c] 3 equivalents of ester.

Recovery tests (Figure S24) show that FL–BP and FL–Sb can be reused at least three times under ambient conditions before deactivation occurs. STEM analysis of the reused samples shows a progressive amorphization of the edges (see Figure S25 for further details), which is significantly alleviated by reusing the material under $N_2$ atmosphere, thus prolonging catalyst lifetime.[23,38] In order to further assess the stability of the material *in operando* conditions, $^{31}$P magic angle solid nuclear magnetic resonance (MAS NMR) spectra of the FL–BP catalysts were recorded under reaction conditions, and the results (Figure S26, top) show that the FL–BP catalyst keeps the original signal at 18 ppm during reaction,[39] without any trace of phosphoric acid (0 ppm), phosphonium compounds, or other potential oxidized and hydrolized species. The measurement was also performed in statics (Figure S26, bottom), and the deconvoluted spectrum fits to two components, one corresponding to P in zero oxidation state and the other corresponding to P–O, which accounts for 8% out of the total. Electrochemistry measurements (Figure S27) strongly supports the absence of neat oxidation P or Sb species during reaction,[40] since no significant changes in the 2D pnictogen signals were found, and if any, they correlate with particle aggregation/fractioning processes rather than the formation of phosphonium or further oxidized P and Sb forms. Following this rationale, and in order to increase the catalyst lifetime, sonication of the ionic liquid mixture was carried out before each reuse, and in this way, only a minor loss of final yield was observed after 6 reuses, thus doubling the catalytic performance without sonication.

Figure 4A shows the rate equation of the reaction between **1** and **2** derived from kinetic experiments (Figure S28), with either FL–BP or FL–Sb catalyst. The results give $v_0$=$K_{app1}$[P or Sb][**1**][**2**], that indicates that the three species, i.e. the alcohol, the ester and the pnictogen catalyst, are involved in the rate–determining step of the reaction. This equation rate differs significantly from a classical $A_{AL}1$ mechanism,[33,41,42] where the nucleophile does not participate in the rate–limiting step since only carbocation formation controls the overall reaction rate. Accordingly, kinetic experiments with HOTf, under the reaction conditions employed here, show that **1** does not participate during the rate determining step, with an equation rate $v_0$=$K_{app2}$[$H^+$][**2**]. This result confirms that the classical $A_{AL}1$ mechanism operates with HOTf in BMIM–$BF_4$ (Figure S29), and that it is different from that with FL–BP or FL–Sb. Analysis of the reaction gas phase by gas chromatography coupled to mass spectrometry (GC–MS) shows that nearly 0.5 equivalents of **2** are transformed to isobutylene gas during the HOTf–catalyzed reaction, regardless if **1** is present or not, while only traces of isobutylene gas are detected with the FL–BP catalyst. Control experiments with isobutylene or styrene as reagents, instead of the corresponding esters, discard alkenes as alkylating agents under the present reaction conditions. Notice that isobutylene is a typical by–product of long–lived *tert*–butyl cations, and the preferential formation with HOTf and the different equation rates illustrate the striking differences between the FL–pnictogen and the superacid catalyst mechanisms.

To further check the formation of carbocations or not during 2D pnictogen catalysis, *R*–1–phenylethanol acetate **13** was used as the alkylating agent for methanol. The results in Figure 4B show that a racemic mixture of the alkylated product 1–phenylethyl methyl ether **14a**, together with isomerically pure 1–phenylethanol **14b** and starting **13**, were found as main products with FL–BP catalyst. This result unambiguously demonstrates that the alkyl moiety is transformed on the 2D pnictogen surface into a carbocation, at some point before transferring, since it is able to racemize prior to nucleophile addition. Of course, this also occurs with HOTf catalyst; however, the different equation rates and isobutylene yield found for the 2D pnictogen and HOTf catalysts suggest different carbocation managing during the alkylation reaction. Released acetic acid or traces of water do not act as nucleophiles towards the carbocation in both cases.

Figure 4C shows competitive tests between benzyl alcohol **1** and decyl alcohol **15** with **2**. The relative initial rates and final yields differ dramatically for FL–BP and HOTf; while FL–BP shows one order of magnitude (10.0) higher formation rate for the aromatic than for the alkyl alcohol, HOTf shows a relative rate of 1.7. These values give 6 times higher selectivity for aromatic substrates with FL–BP. Another clear difference is found in the corresponding Hammett plots: while electron withdrawing groups (EWG) on the benzyl alcohol increase the reaction rate with the FL–BP catalyst, electron donor groups (EDG) increase the rate with HOTf catalyst, the latter being the expected behavior for a free nucleophile (Figure S30). Kinetic experiments with isotopically labeled $D_2$–**1** ($PhCD_2OH$) give a significant secondary kinetic isotopic effect of 1.4(5) for the FL–BP catalyst. This last result, together with the Hammett plot, suggests an electron donation of the exfoliated material to the aromatic ring and then, in a lesser extent, to the alcohol by induction effects, in accordance to the known ability of 2D–BP[6] and 2D–Sb[7] to transfer electron charge to planar aromatic molecules.

Calorimetry measurements with **1** and FL–BP (Figure S31) show that the aromatic alcohol strongly adsorbs to the BP surface, even at just 30 ºC. Taft plots (Figure S32) shows a clear positive slope for the 2D materials, which indicates that bulkiness on the aromatic ring decreases reactivity. Decoupling steric and inductive effects by least–squares Taft regression analysis[43] (Tables S2–S4) confirms the need of co–planarity of the aromatic alcohol with the 2D catalyst, since the $\delta$ values obtained by the Taft regression analysis are the same than the Hammett plot, within the experimental error (-0.3±0.5 for HOTf, 0.89±0.08 FL-BP and 0.5±0.3 for Sb-BP). 2D nanosheets with different thickness were prepared by different ultra–centrifugation steps, with nearly one order of magnitude accessible bulk and edge atoms (Fig. S33–S38),[21] and kinetic results for the alkylation of **1** with **2** showed that the $TOF_0$ increases for the centrifuged samples. Since unsaturated P atoms in vertexes and edges are likely oxidized under reaction conditions and do not participate during catalysis, these results indicate that the catalysis is directly related to the number of atoms on the bulk (Figure S39).[21,44] When nitrobenzene was used as an inhibitor reagent, the alkylation rate of **1** with **2** was progressively quenched up to 20 mol% of nitrobenzene (respect to FL–BP, Figure S40), which indicates that the high electron–deficient aromatic ring of nitrobenzene is strongly adsorbing to the bulk atoms of the 2D material. This inhibition value fits well to the number of P atoms present on the whole bulk surface. These results point again to bulk P or Sb atoms as responsible for the catalysis.

P and Sb are very suitable atoms to stabilize carboxoniums, furthermore if the carbocation–oxonium equilibrium on surface is shifted towards the latter and assures a short living of the former. Indeed, low amounts of isobutylene were found during the alkylation reaction with FL–BP and FL–Sb, which supports a short living carbocation. A Hammett plot with different *para*–substituted *tert*–butyl benzoate esters shows that a positive charge is formed in the ester during reaction, thus carboxoniums can be presumed as intermediates (Figure S41).

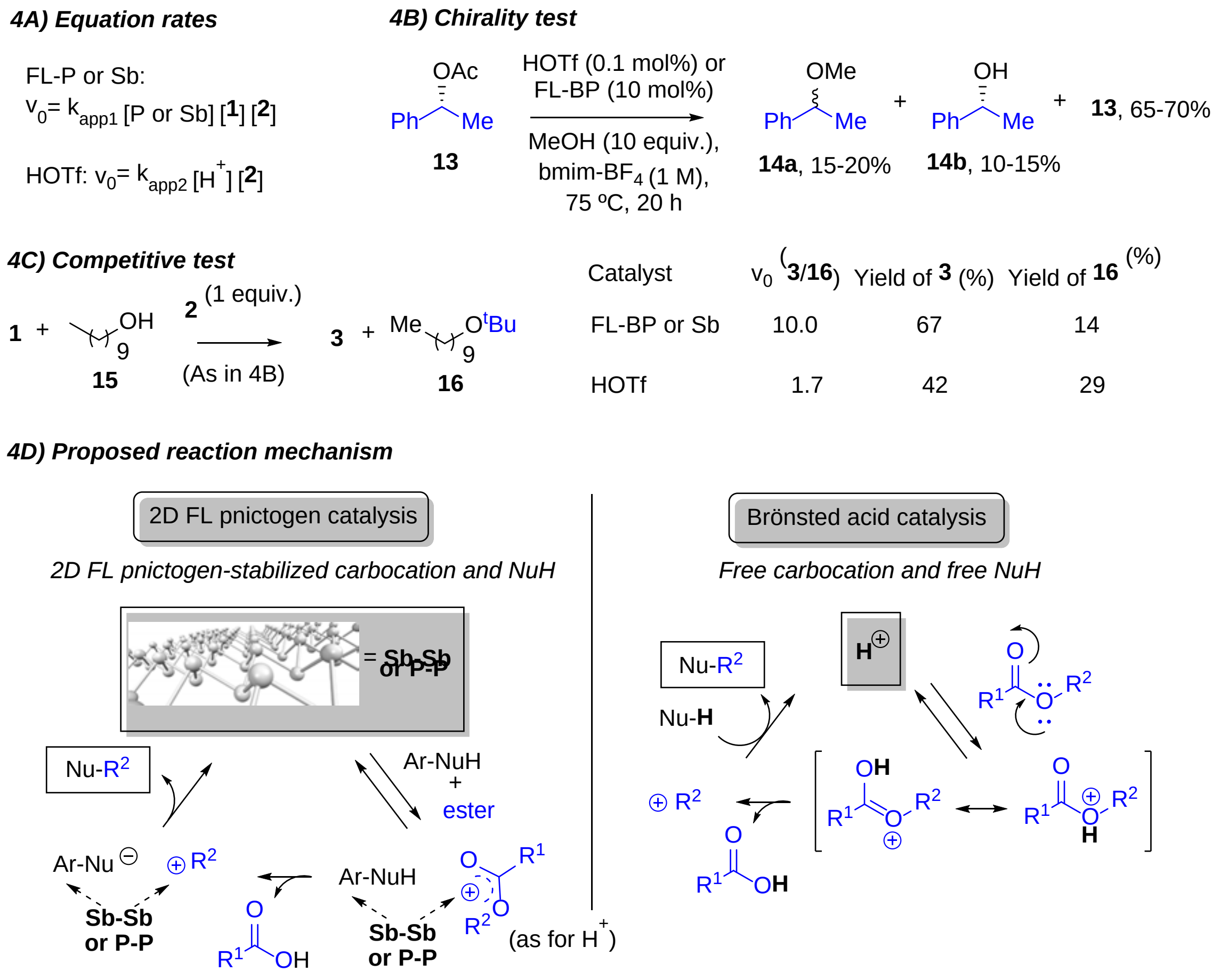

**Figure 4 Mechanistic studies.** Experimental evidences and proposed mechanisms for FL–BP, FL–Sb and HOTf catalyzed–alkylation of nucleophiles with esters in bmim–$BF_4$.

With all these data in hand, a very plausible mechanism for the 2D FL pnictogen–catalyzed alkylation of **1** with **2** is shown in Figure 4D along the classical AL mechanism with HOTf for comparison, are shown in Figure 4D. For the 2D FL pnictogen, the electron–rich 2D surface transfers electron density to the adsorbed aromatic nucleophile, which generates a deficiency of charge in the material stabilized by the few layers underneath[6,7] that activates the ester group and stabilizes a carboxonium intermediate, rapidly trapped by the nucleophile on surface and, thus, regenerating the electro–neutrality of the catalytic material. The higher polarizability of Sb *vs.* P explains the higher catalytic activity of the former. Notice that the activation of the ester occurs only after the bulk atoms act as a Lewis base, the natural behavior of 2D-BP and 2D-Sb, which enables a unique mechanism for the nanometric FL–BP and FL–Sb.

The transition state energetic values for FL–BP, calculated by an Eyring plot with the reaction between **1** and **2**, are $\Delta H^{\#}$= 49.1(4) KJ·mol$^{-1}$ and $\Delta S^{\#}$= 0.0(2) KJ·(mol·K)$^{-1}$, and the unchanged entropy is consistent with both reactants binding to bulk P atoms, to couple rapidly after carbocation formation. In clear contrast, the mechanism for the superacid HOTf shows the formation of a free carbocation in solution, then trapped by the nucleophile. The mechanisms proposed in Figure 4D satisfactorily explain the experimental differences observed for the 2D-pnictogen and HOTf catalysts, and in particular, the equation rate, higher reactivity of EW aromatic nucleophiles and carbocation racemization without alkene formation for the 2D-pnictogen catalyst. Other potential mechanisms for the alkylation reaction such as the formation of phosphonium intermediates, or neat redox processes on the pnictogen surface fostered by their narrow homo–lumo gap, seems not to be pointed out by the experimental evidences shown above.

The particular nature of the 2D-pnictogen materials described here, embedded in water– and oxygen–protecting ionic liquids, and with a high basal area, allows organic catalysis, and not necessarily other expected applications for BP. For instance, water splitting under typical experimental conditions did not give any $H_2$ formation. The suitability of FL–BP and FL–Sb for organic catalysis is further supported by the preliminary positive results found for the catalytic Bailys-Hillman reaction (Figure S42), a representative carbon-carbon bond-forming reaction.

2D-pnictogens widens and complements graphene catalysis (carbocatalysis), since when graphene is doped with precisely pinpointed heteroatoms, typically nitrogen, its catalytic activity equals simple metal catalysts (Table S5), which is not more what 2D-pnictogens make naturally here, without any additional modification. This rationale drives to think that 2D-pnictogens are potential advanced versions of doped graphene and extremely promising catalysts in organic reactions.

**Discussion.**

2D nanosheets of BP and Sb, exfoliated in the ionic liquid bmim-$BF_4$, catalyze the alkylation of soft nucleophiles with alkyl esters, in good yields and selectivity, particularly for aromatic substrates. The 2D few layer materials circumvent superacid–mediated alkylations by enabling a surface mechanism with concomitant activation of the aromatic nucleophile and ester on surface, which allows acid–sensitive molecules to be alkylated. As far as we know, this is the first example of catalytic application of pristine FL-BP and FL-Sb in organic synthesis, beyond recent examples on photonic excitation or with supported metal nanoparticles,[45–49] thus expanding the list of potential applications for these promising materials.

**Methods.**

Materials and exfoliation process: Throughout all experiments, BP and Sb with purity higher than 99.999% (Smart Elements) were used.

Afterwards, the FL–BP flakes were transferred onto Si/$SiO_2$ substrates (300 nm oxide layer). The exfoliation was performed in argon filled LABmasterpro sp glove box (MBraun) equipped with a gas purifier and solvent vapour removal unit (oxygen and water content lower than 0.1 ppm).

Solvent purification: Anhydrous, 99.9% purity 1–butyl–3–methylimidazolium tetrafluoroborate (bmim–$BF_4$) was purchased from Sigma–Aldrich. The bmim–$BF_4$ was pump freezed, and the $O_2$ was removed by vacuum. This procedure was iteratively repeated a minimum of 4 cycles to remove traces of oxygen.

*Exfoliation of layered pnictogens.*

LPE in bmim–$BF_4$ under inert conditions: BP was exfoliated under inert conditions by sonication in an argon–filled glovebox ($O_2$<0.1 p.p.m.; $H_2O$<0.1 p.p.m.) using a Bandelin Sonoplus 3100, 25% amplitude, 12 h, pulse 2 s on, 2 s off. Sb was exfoliated under the same inert conditions by sonication using 40% amplitude, 16 h, pulse 2 s on, 2 s off. The starting concentrations were 1.25 mg·mL$^{-1}$ and 2.5 mg·mL$^{-1}$ for BP and Sb, respectively. The resultant dispersions were decanted and transferred into vials. All solvent transfer was carried out in the glovebox.

*Centrifugation:* Centrifugation was carried out in a MPW–350R centrifuge using 2 mL Eppendorfs. A two–step process was followed; first supernatant dispersion was centrifuged 14.000 g during 1 min, and afterwards the resulting supernatant was submitted to a second step at 2.000 g and 100 g for 60 min for FL–BP and FL–Sb, respectively. The final concentrations were determined by ICP-OES being *ca*. 0.125 $mg \cdot mL^{-1}$ and 0.075 $mg \cdot mL^{-1}$ for FL-BP and FL-Sb, respectively. Longer centrifugations periods were performed to prepare samples of different thickness.

*Photoluminiscence.* PL of the different centrifuged samples was acquired on a Horiba Scientific Fluorolog-3 system equipped with 450 W Xe halogen lamp, double monochromator in excitation (grating 600 lines/mm blazed at 500 nm) and emission (grating 100 lines/mm blazed at 780 nm) and a nitrogen cooled InGaS diode array detector (Symphony iHR 320). Spectra were obtained at 5 °C measured (spectral region of 550–1300 nm) with a 550 nm cutoff filter in emission. Excitation and emission band widths were typically 10 nm and integration times 2 s.

*Surface preparation:* $SiO_2$ surfaces were sonicated for 15 min. in acetone and 15 min. in 2–propanol and then dried under an argon flow.

Immediately after the removal from the inert atmosphere, images of FL–BP flakes were recorded under an optical microscope (Zeiss Axio Imager M1m), using different objectives enabling their re–localization in Raman and AFM measurements. Raman spectra were acquired on a LabRam HR Evolution confocal Raman microscope (Horiba) equipped with an automated XYZ table using 0.80 NA objectives. All measurements were conducted using an excitation wavelength of 532 nm, with an acquisition time of 2 s and a grating of 1800 grooves/mm. To minimize the photo–induced laser oxidation of the samples, the laser intensity was kept at 5 % (0.88 mW). The step sizes in the Raman mappings were in the 0.2–0.5 μm range depending on the experiments. Data processing was performed using Lab Spec 5 as evaluation software. When extracting mean intensities of individual BP Raman modes, it is important to keep each spectral range constant, *e.g.* from 355–370 $cm^{-1}$ and from 460–475 $cm^{-1}$ because otherwise the resulting value of the $A^1{}_g/A^2{}_g$–ratio can be slightly influenced. The same applies to Sb $E_g/A^1{}_g$ ratio analyses.

*Atomic force microscopy (AFM):* AFM was carried out using a Bruker Dimension Icon microscope in tapping–mode. The samples were prepared by spin coating a solution of a given sample at 5.000 rpm. Bruker Scanasyst-Air silicon tips on nitride levers with a spring constant of 0.4 $N \cdot m^{-1}$ were used to obtain images resolved by 512x512 or 1024x1024 pixels.

*Scanning transmission electron microscopy (STEM):* STEM observations were carried out in a JEOL ARM200cF operated at 80 kV and equipped with a spherical aberration corrector and a Gatan Quantum electron energy–loss spectrometer (EELS), at the ICTS–ELECMI Centro Nacional de Microscopía Electronica at UCM (Spain). Compositional maps were produced using a multiple linear least squares fit of the data to reference EEL spectra.

*X-ray photoelectron spectroscopy (XPS):* XPS measurements were carried out in an ultra-high vacuum (UHV, base pressure < $1{\cdot}10^{-10}$ mbar) system dedicated for angle-resolved XPS at UHV-compatible liquid samples such as ILs[50] Spectra were collected in normal emission (information depth in ILs: 7–9 nm, depending on electron kinetic energy) using monochromated Al Kα radiation (1486.6 eV) and a pass energy of the hemispherical electron analyser of 35 eV (overall energy resolution: 0.4 eV). Binding energies are referenced to C 1*s* for aliphatic carbon of the IL (285.0 ± 0.1 eV) and Au $4f_{7/2}$ (84.0 ± 0.1 eV) of clean gold. XPS-detection limit typically is around 1 at-% depending on relative XPS cross-sections and signal-to-background situation for the trace atom;[51] The FL-Sb and FL-BP solutions employed for our catalysis studies with an overall P- and Sb-content around 0.1 mg/mL (that is, below 0.01 at-% for P and Sb) were thus below the XPS detection limit as has been tested: only bmim-$BF_4$ signals could be detected along with a Si/O/C containing trace contamination that commonly shows up in IL-XPS investigations due to surface enrichment effects, and is typically due to contact with glassware grease.[26] Hence, highly-concentrated suspensions (2D-inks) of FL-BP and FL-Sb were prepared by filtering the dispersions through a 0.2 µm reinforced cellulose membrane filter (Sartorius) in the glove box to remove most of the IL. The filtration was stopped just before the initial amount passed the filter, which allowed collecting the 2D-ink consisting of highly concentrated FL-BP/Sb material from the filter surface with a Teflon spatula. The 2D-inks were transferred to air and spread onto clean gold foils that were mounted on XPS sample holders. After exposure for several hours to air, the sample holders were introduced into the UHV system. Spectra were taken for the pristine samples and after heating in UHV in order to remove most of bmim-$BF_4$ by thermal evaporation (and partial decomposition as proven by XPS) to maximise P and Sb signal intensities. The heated samples were then exposed to air for about day and measured again.

*General reaction procedure.* The corresponding nucleophile (0.1 mmol) and alkylating ester (0.15 mmol) were added under ambient conditions to a solution/dispersion of the corresponding catalyst (0.005 mmol for Brönsted acids and organic and inorganic bases, 0.02 for P, 0.01 mmol for Sb, and 1 mg for solids) in bmim–$BF_4$ (100 mg), placed in a 2 ml vial equipped with a magnetic stir bar. The vial was sealed and the resulting mixture was magnetically stirred at the required temperature for 4–20 h. Then, the reaction mixture was cooled and extracted with diethyl ether (1.5 ml). The extracts were analysed by GC and GC–MS after adding dodecane (22.4 μl, 0.2 mmol) as an external standard, and the products were isolated by preparative thin–layer chromatography (TLC).

For kinetics, each point was taken from an individual reaction. For preparation purposes, scale is proportionally increased up to grams of starting material. For reuses, 2 mmol of starting materials are used and volatiles are removed from the extracted reaction mixture, under vacuum at room temperature for 15 min, prior to addition of fresh reactants. For longer catalyst lifetime, sonication after each reaction cycle of the ionic liquid mixture under inert atmosphere was carried out.

*Calorimetry.* 750 mg of ionic liquid-FL-BP (750 mg) was evacuated under vacuum for 30 min in a glass line, and a glass ampule was made in-situ, when still under vacuum. The ampule is submerged in a 1M solution of benzyl alcohol in diethyl ether, inside the calorimetry apparatus, and then broken. Exchanged energy was measured at 30 ºC during 2–4 h.

*Electrochemistry.* Voltammetric experiments were performed using a CH 660c equipment on FL–BP and FL–Sb diluted in the IL, after successive additions of BrOAc and *t*–BuOAc until 1 M concentration. Glassy carbon electrode (geometrical area 0.071 $cm^2$) was used as a working electrode, completing the three–electrode arrangement with a Pt mesh counter electrode and a Pt wire pseudo-reference electrode. A second series of experiments were performed in 0.10 M potassium phosphate buffer at pH 7.0 after forming a fine deposit of FL–BP and FL–Sb on glassy carbon electrode via transfer of 20 μL of each one of the above IL solutions, after 30 min of reaction, plus of 20 μL ethanol, evaporation at air and drying with a smooth paper tissue.

*Photocatalysis.* The photocatalytic water splitting is performed with a solar simulator light source (Newport®, Oriel Instruments, model 69921) equipped with a Xe lamp (1000

W) coupled with an AM1.5 filter that provides simulated concentrated sunlight in UV-visible range. 50 mg of FL–BP or FL–Sb were dispersed in 20 mL pure $H_2O$ and then $N_2$ was purged into the reactor (quartz cell, 50 mL volume). Gas samples were periodically taken and analysed in a micro-GC, using Ar as a standard.

**Acknowledgements.**

We thank the European Research Council (ERC Starting Grant 804110 to G.A., and ERC Advanced Grant 742145 B–PhosphoChem to A.H.) for financial support. The research leading to these results was partially funded by the European Union Seventh Framework Programme under grant agreement No. 604391 Graphene Flagship. G.A. has received financial support through the Postdoctoral Junior Leader Fellowship Programme from "la Caixa" Banking Foundation. G.A. thanks support by the Deutsche Forschungsgemeinschaft (DFG; FLAG-ERA AB694/2-1), the Generalitat Valenciana (SEJI/2018/034 grant) and the FAU (Emerging Talents Initiative grant #WS16-17_Nat_04). Financial support by MINECO through the Excellence Unit María de Maeztu (MDM-2015-0538), Severo Ochoa (SEV-2016-0683) and RETOS (CTQ2014-55178-R) program is acknowledged. M. A. R.–C. thanks MINECO for the concession of a FPU fellowship. We also thank the DFG (DFG–SFB 953 "Synthetic Carbon Allotropes", Project A1), the Interdisciplinary Center for Molecular Materials (ICMM), and the Graduate School Molecular Science (GSMS) for financial support. Research at UCM sponsored by Spanish MINECO/FEDER grant MAT2015–066888–C3–3–R and ERC–PoC–2016 grant POLAR–EM. H.-P.S. thanks the European Research Council (ERC) under the European Union's Horizon 2020 research and innovation program for financial support, in the context of an Advanced Investigator Grant granted to him (Grant Agreement No. 693398-ILID). B.S.J.H. and S.S. acknowledge financial support by the DFG within the Cluster of Excellence "Engineering of Advanced Materials" (project EXC 315, Bridge Funding). F.M. acknowledges R. Ransom for very helpful discussions.

**Supporting Information Available.**

Additional experimental data, schemes and figures, synthesis of starting materials, and characterization of reactants and products, including NMR copies, can be found free of charge via the Internet at….

## Data Availability.

The authors declare that all other data supporting the findings of this study are available within the paper and its supplementary information files.

The source data underlying Figs 1b, 2b–f, Figs 1, 2h,I, and Supplementary Figs S1-S9, Figs S11-S17, Figs S19,20, Figs S22-S24, Figs S28-S30, Figs S32, Figs S34-S40 are provided as a Source Data file. The Source Data file can be found in: Materials Cloud Archive [https://doi.org/10.24435/materialscloud:xxx/v1]

### Author Contribution

G.A. and A.L.–P. conceived the research, designed the experiments, analysed the data, supervised the project and wrote the manuscript. V.L. and S.W. synthesized the samples. V.L., S.W. and G.A. performed AFM and Raman characterization; M.R.–C. and A.L.–P. performed the XRD experiments and the catalytic and kinetic studies; J.V.–M. contributed with the $^{31}P$ MAS–NMR measurements; A.D.–C. contributed with the electrochemical studies. M.V. performed STEM and EELS. F.H. and A.H. supervised the project. B.S.J.H, S.S., H.-P.S, and F.M. performed XPS characterisation. All the authors discussed the results and contributed to writing the manuscript. The authors declare no competing interests.

## References.

# Supplementary Information

**Few layer 2D pnictogens catalyze the alkylation of soft nucleophiles with esters.**

**Lloret et al.**

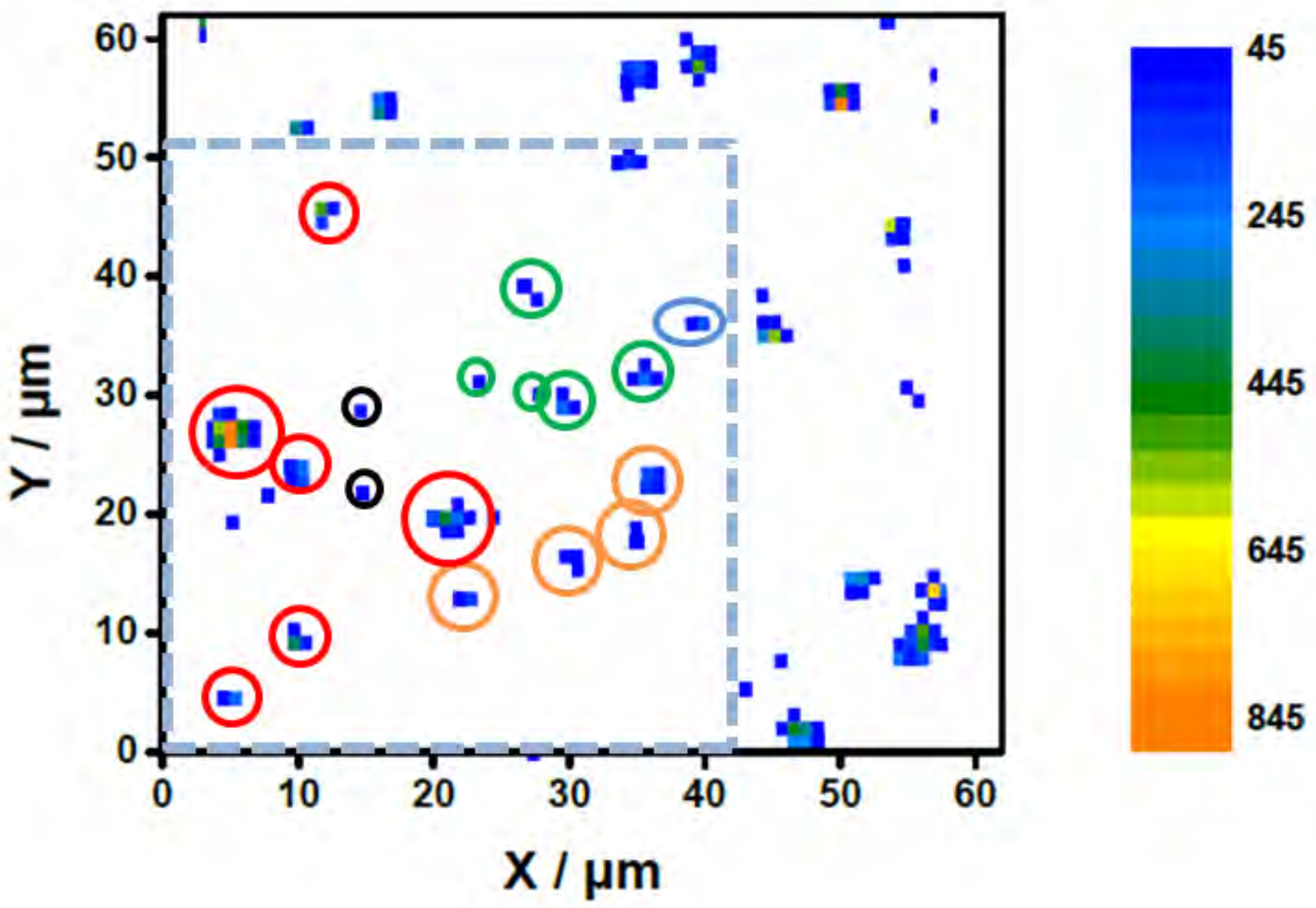

**Supplementary Figure 1.** Raman map of the $A^2_g$ band of BP which correlates to AFM image shown in Supplementary Figure 2. Source data are provided as a Source Data file.

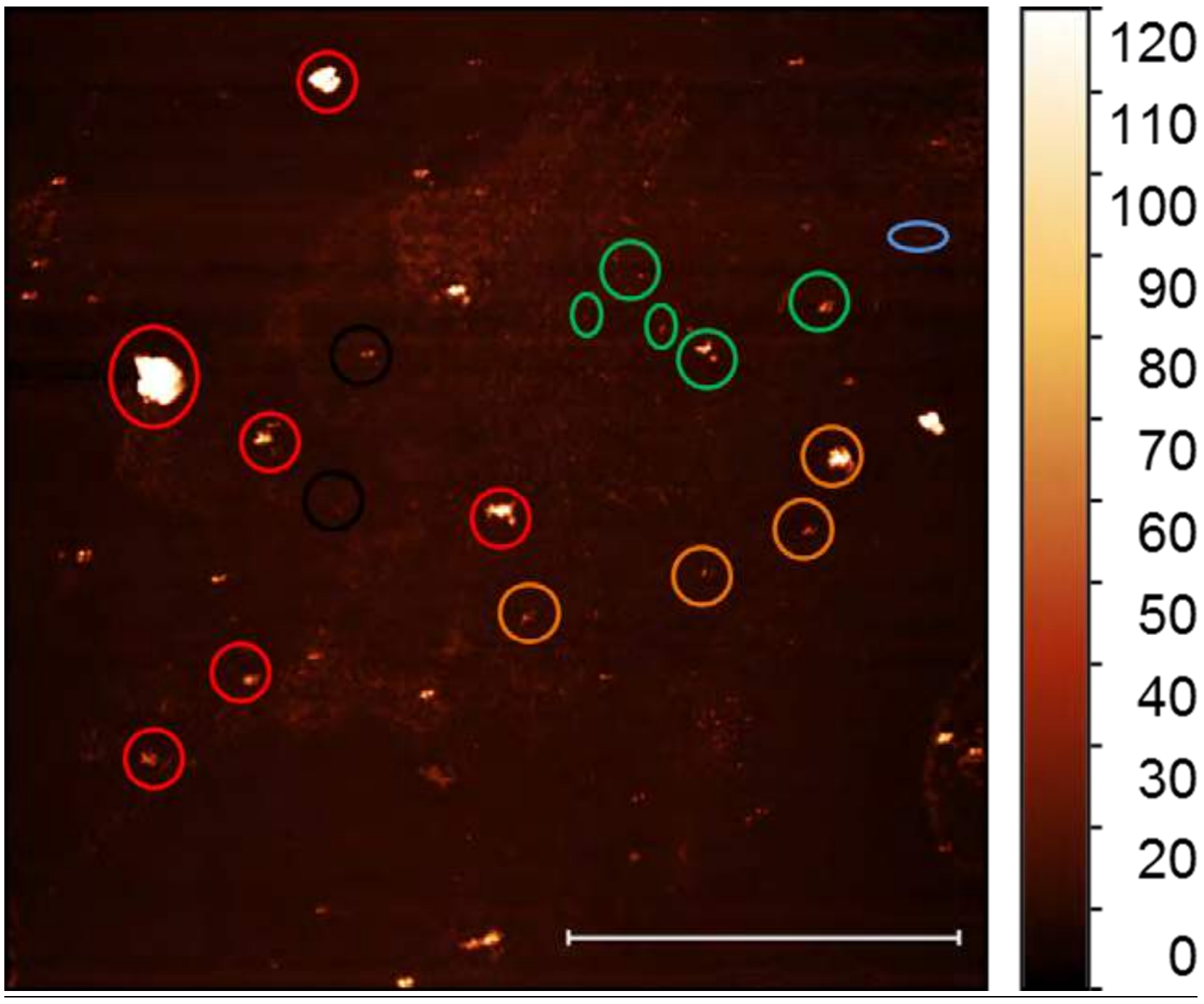

**Supplementary Figure 2.** AFM image corresponding to the Raman mapping shown in Supplementary Figure 1 (scale bar represents 20 μm). Source data are provided as a Source Data file.

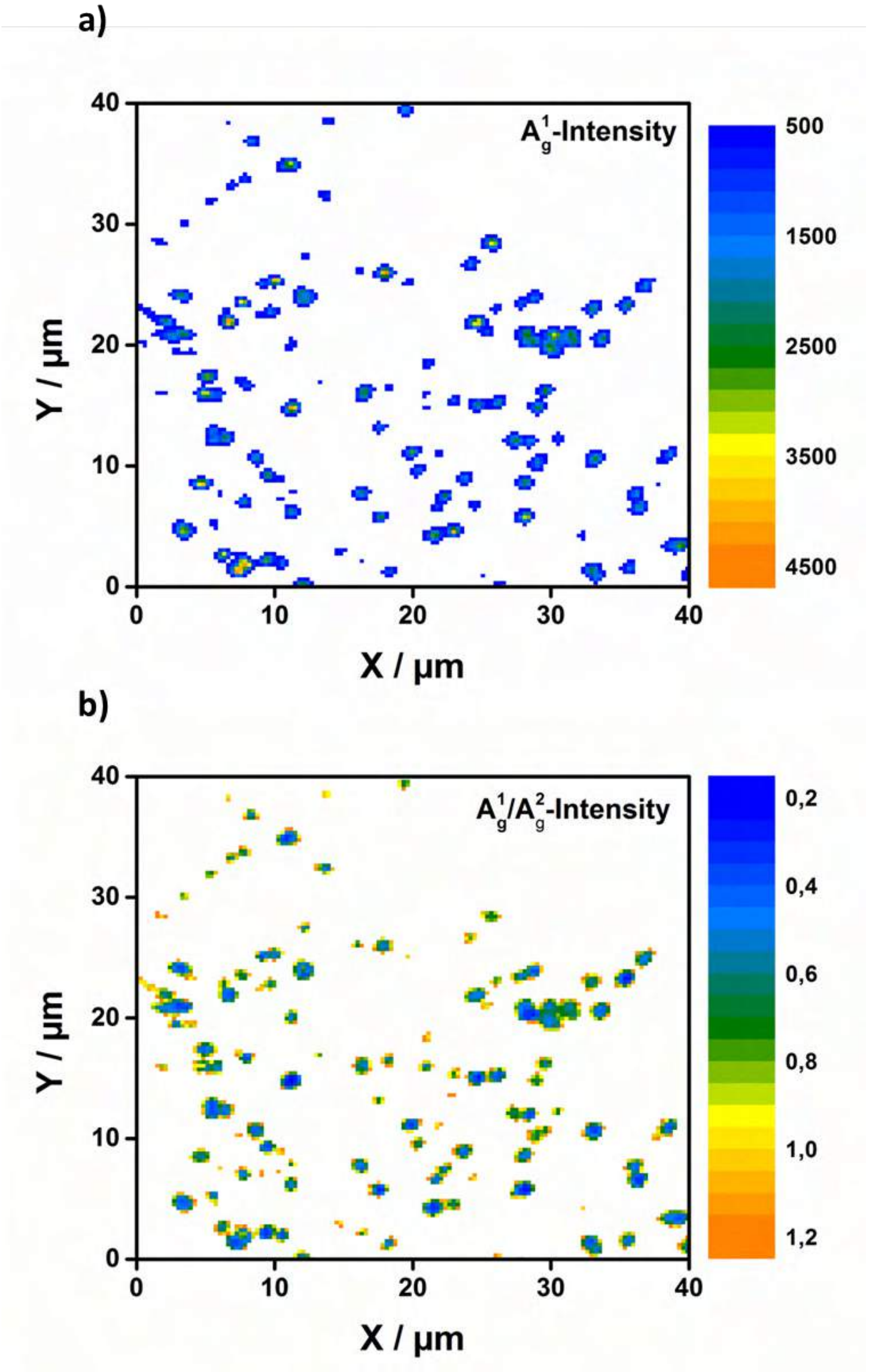

**Supplementary Figure 3.** a) Raman mapping of the $A^1_g$ band of BP and b) Raman mapping of $A^1_g/A^2_g$ ratio. Source data are provided as a Source Data file.

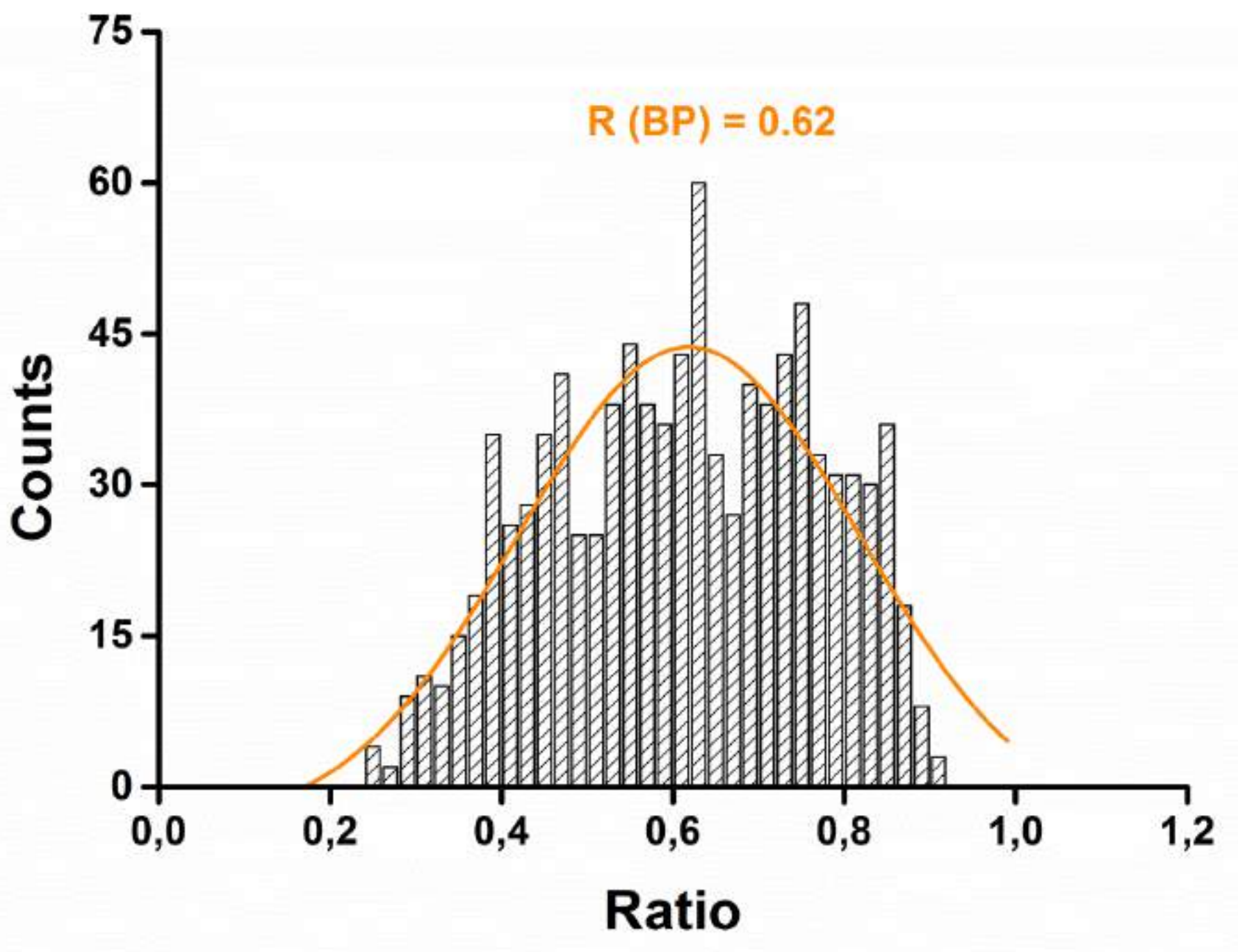

**Supplementary Figure 4**. a) Histogram showing the $A^1_g/A^2_g$ ratio of BP. The orientation of the flakes is random and therefore the ratio varies between values of 0.3 and 0.9. Source data are provided as a Source Data file.

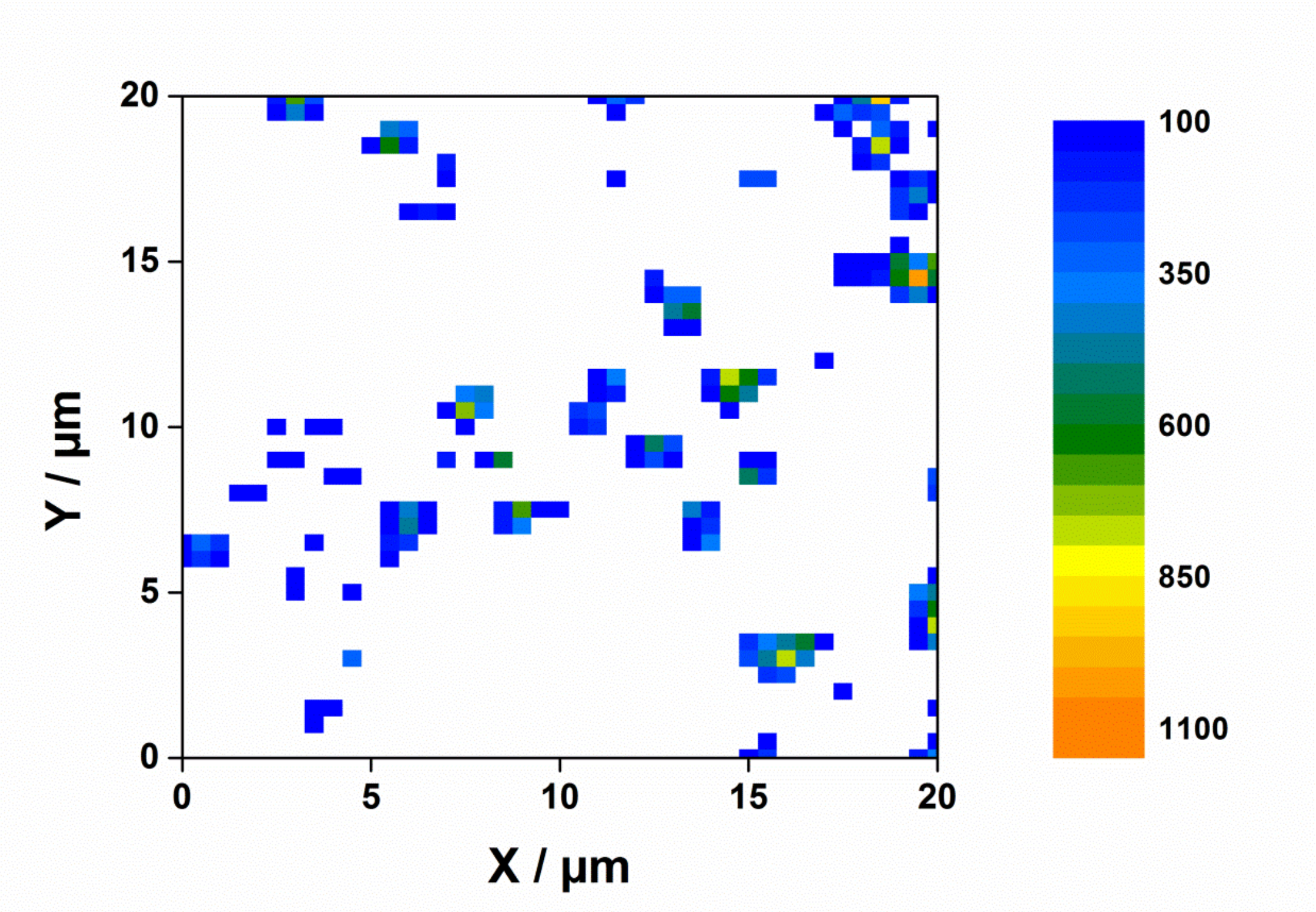

**Supplementary Figure 5.** Raman map of the $A^2_g$ band of BP which correlates to AFM image shown in Supplementary Figure 6. Source data are provided as a Source Data file.

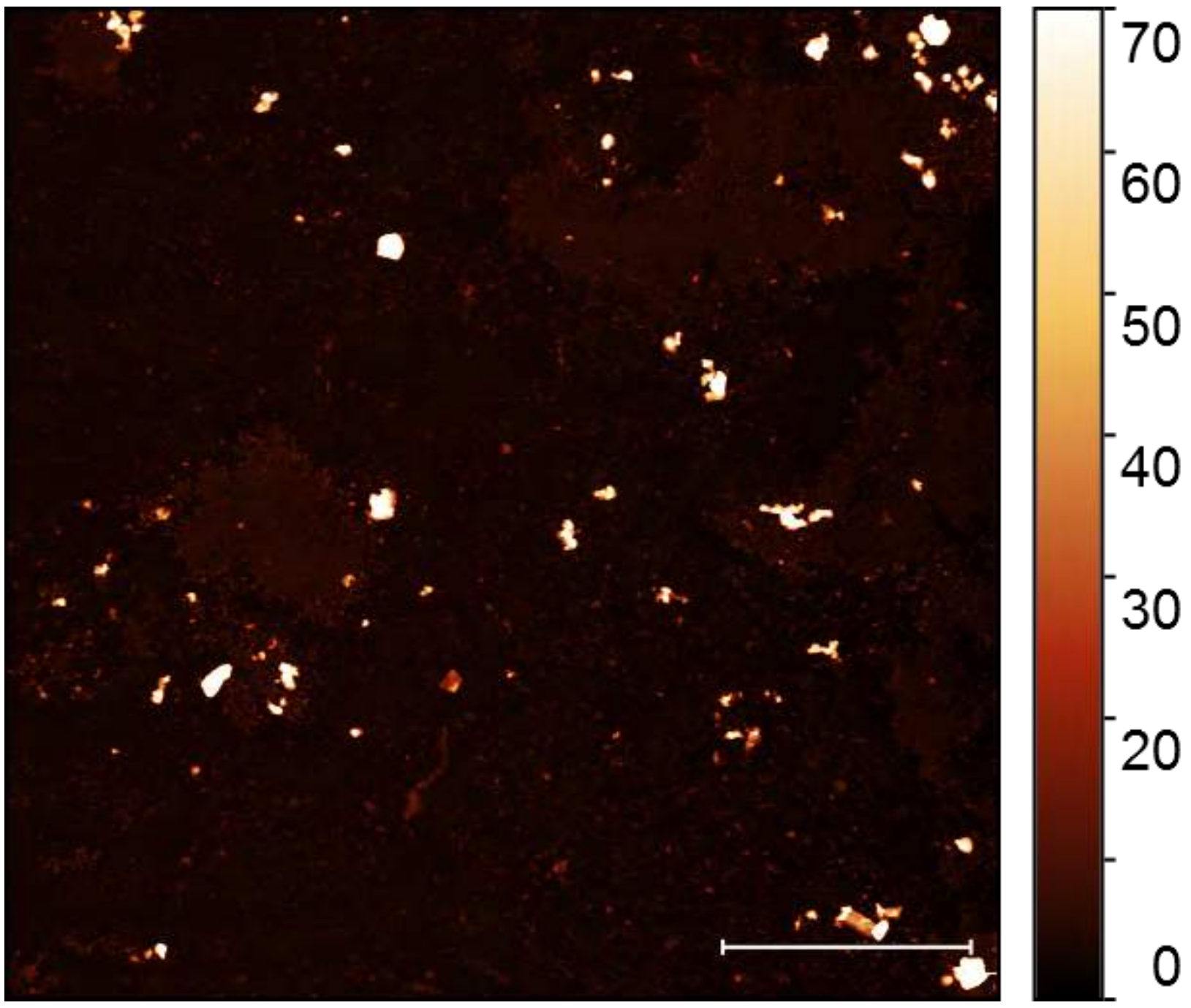

**Supplementary Figure 6.** AFM image corresponding to the Raman mapping shown in Supplementary Figure 5 (scale bar represents 5 µm). Source data are provided as a Source Data file.

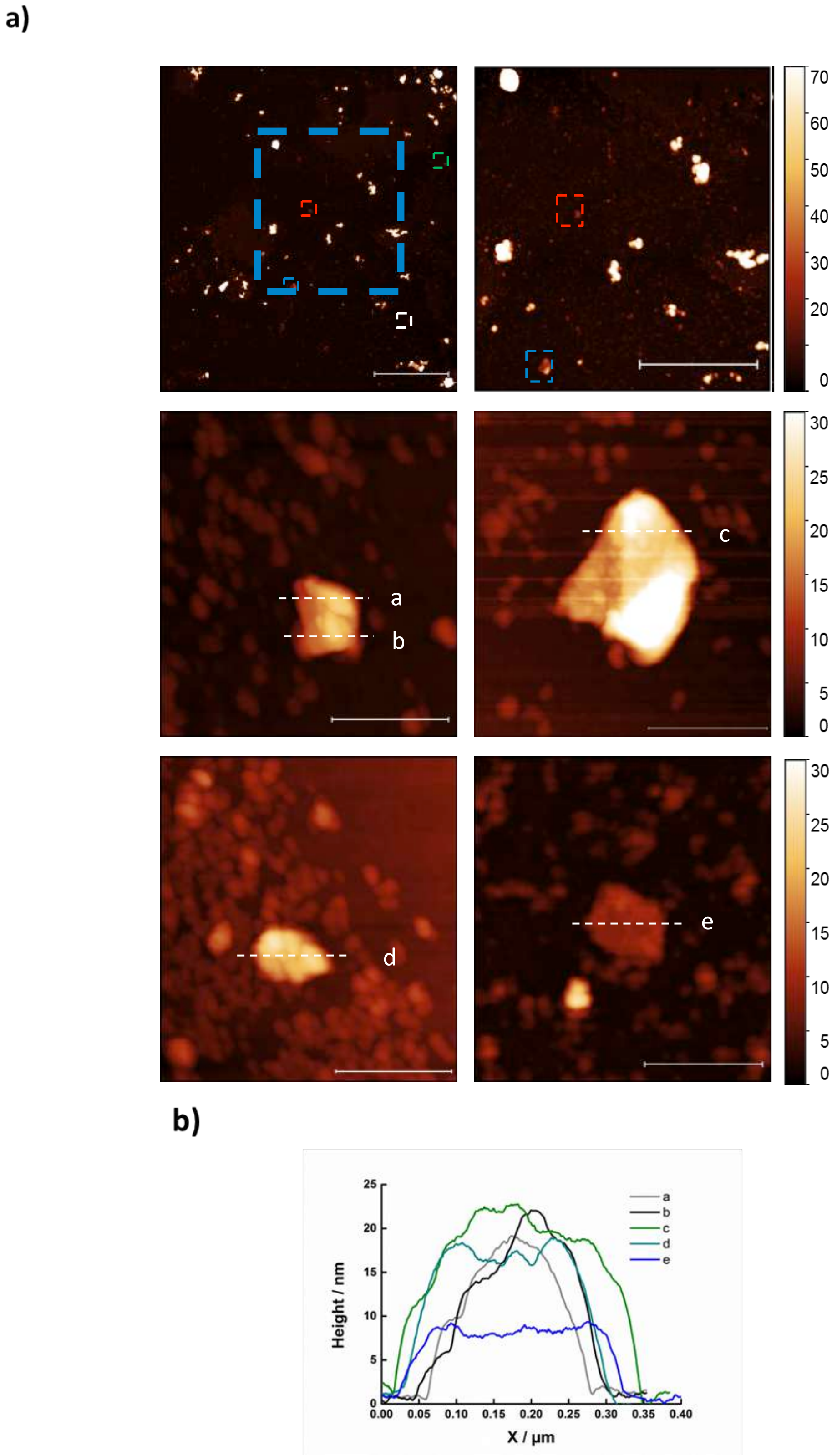

**Supplementary Figure 7.** a) Series of 6 AFM images showing several zooms of areas with thin flakes (scale bars: top right 5 µm, left 2 µm and bottom 200 nm). The first image is the one which was correlated to the Raman map in Supplementary Figure 5 b) Corresponding AFM height profiles across lines a–e.

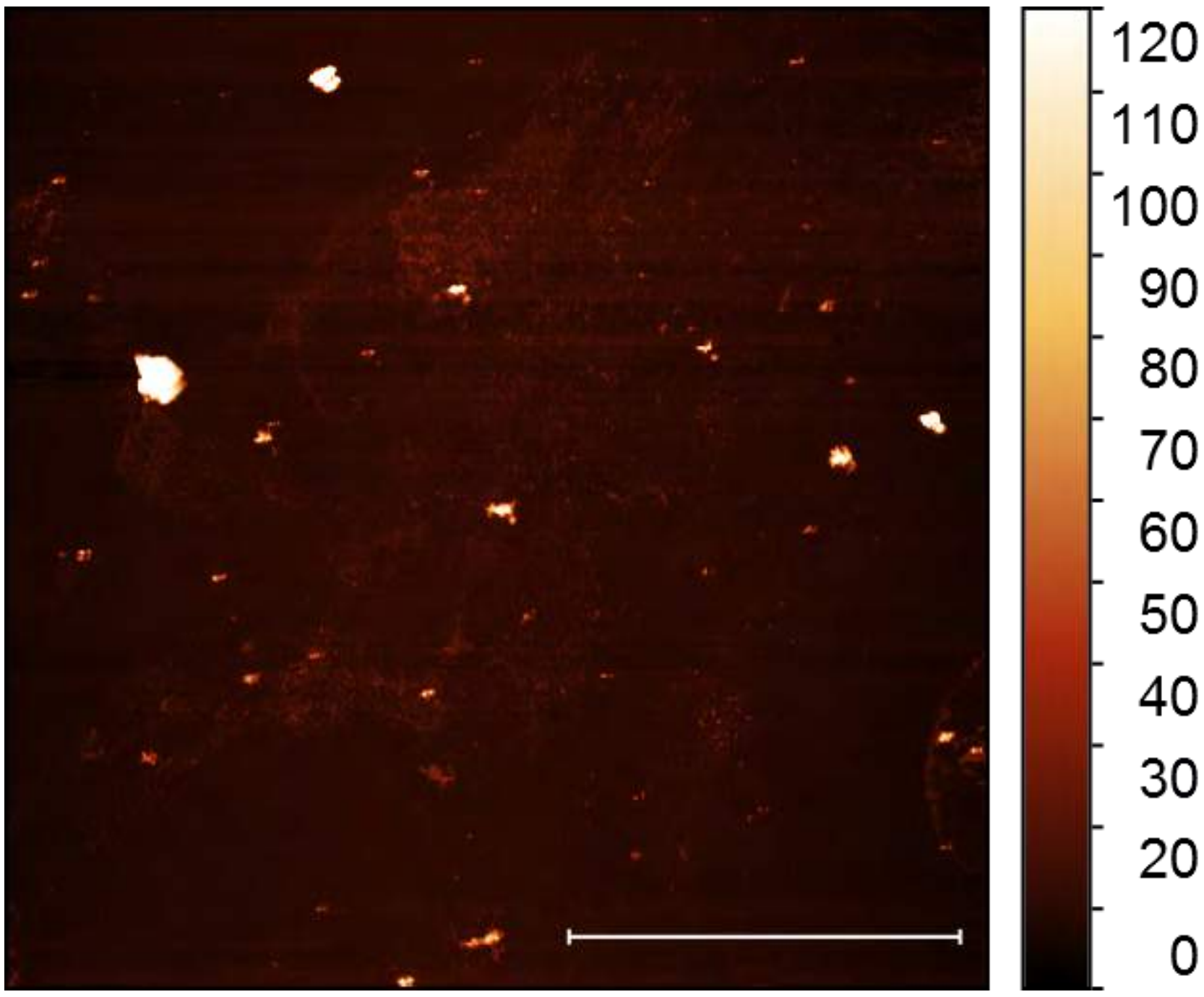

**Supplementary Figure 8**. AFM image of IL deposited on a Si/$SiO_2$ substrate washed with ACN used as reference (scale bar represents 2 μm). Source data are provided as a Source Data file.

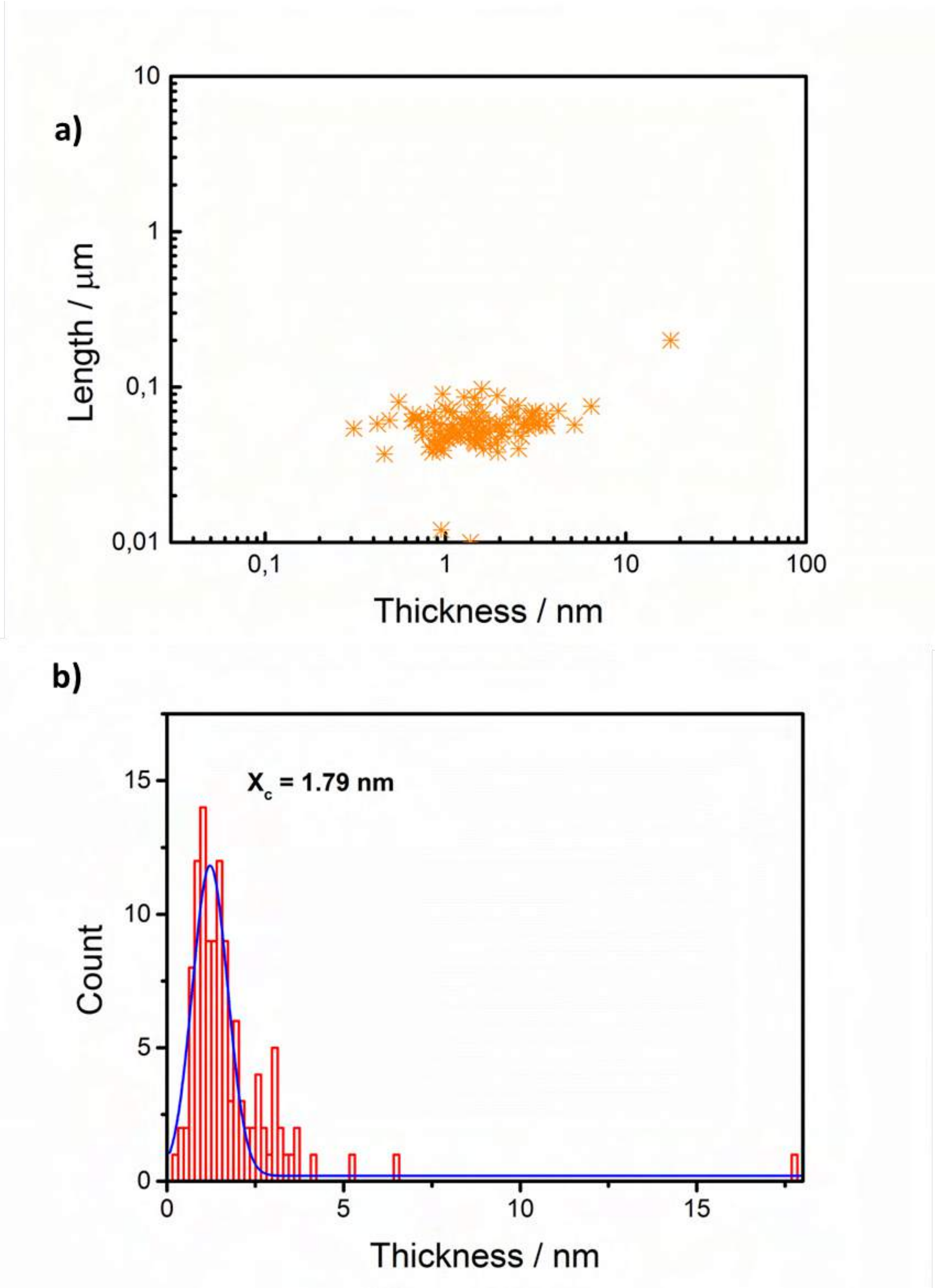

**Supplementary Figure 9**. a) Scatter plot of the AFM values obtained from the length and height of IL deposited on $Si/SiO_2$ substrate (Supplementary Figure 8). b) Represents the histogram of the height values with a mean value of 1.79 nm considering a total amount of 116 replicates. Source data are provided as a Source Data file.

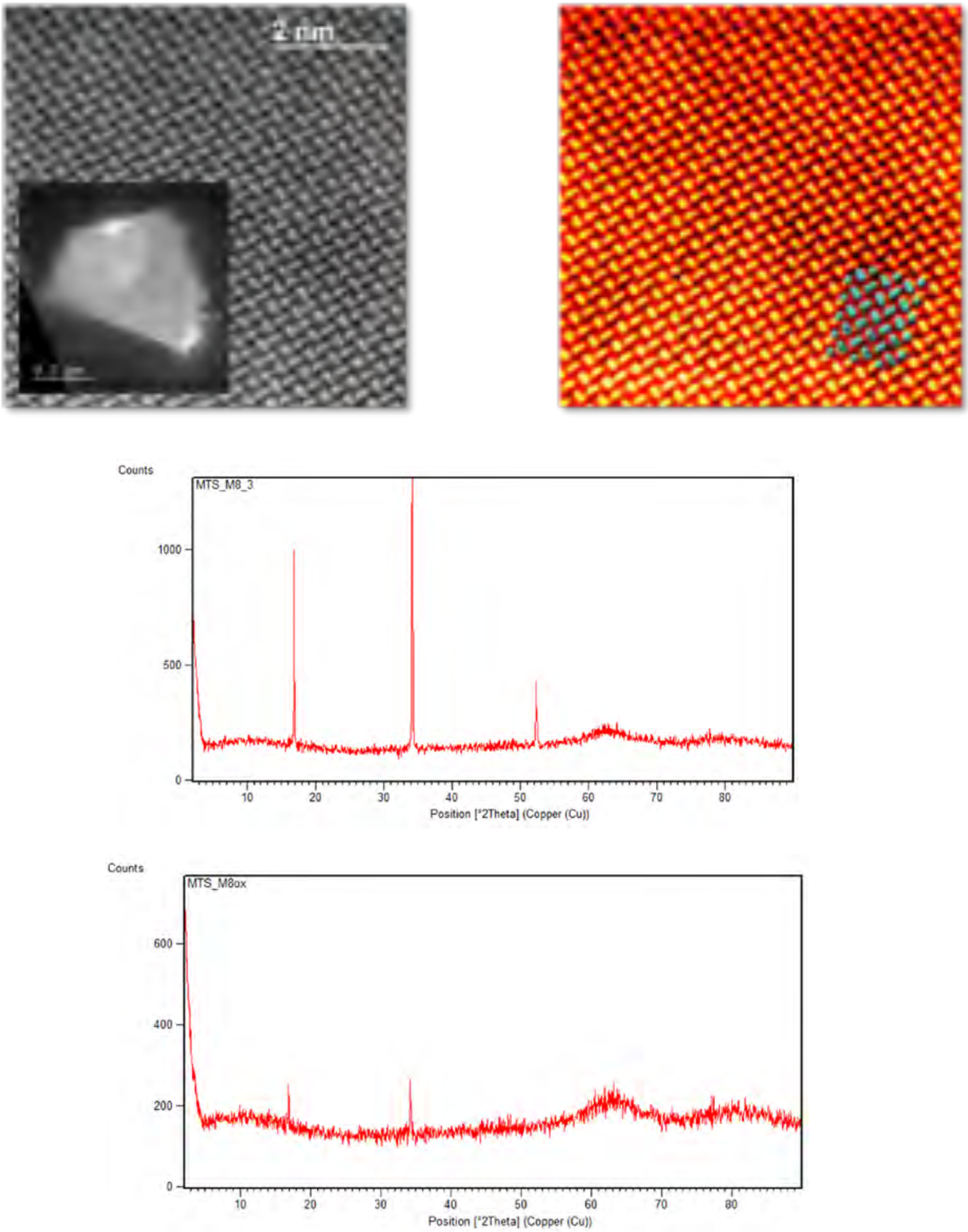

**Supplementary Figure 10.** (Top-left) Atomic resolution HAADF image acquired down the [110] axis, from the edge of a free–standing portion of a flake, including the raw image and a Fourier filtered (FFT) version, in false color (top-right). A sketch of the crystal structure is overlayed. The scale bar represents 2 nm. The inset exhibits a low magnification image of a FL–BP flake, the scale bar is 500 nm. Data acquired at 80 kV. (Middle) XRD of reaggregated FL–BP after washings with tetrahydrofurane solvent, ultracentrifugation and evaporation under inert atmosphere, and the corresponding spectrum after exposure to air for 30 min (bottom).

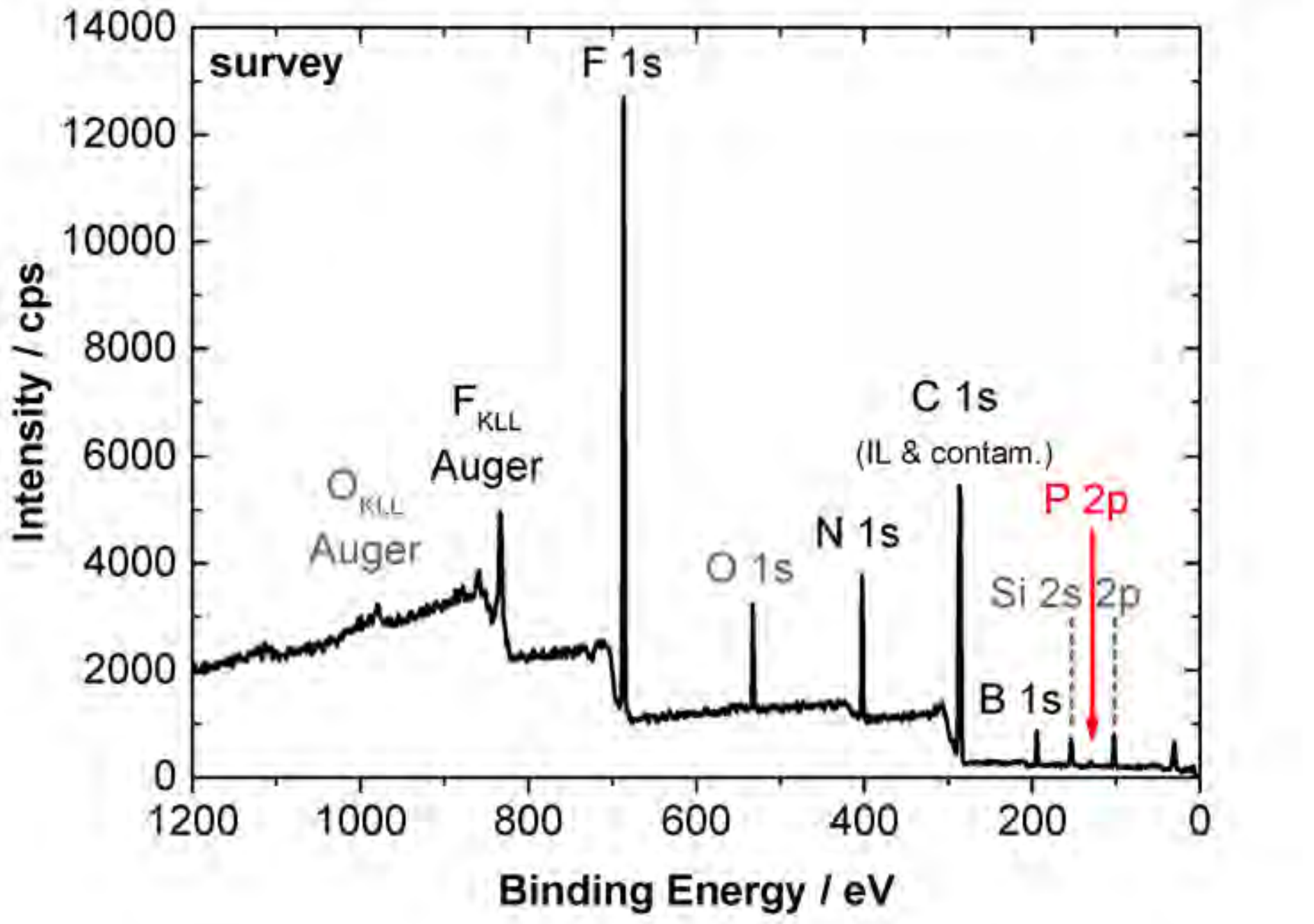

**Supplementary Figure 11.** XPS survey spectrum of FL-BP (Si, O, and C signals originating from an IL contamination are indicated in gray). Source data are provided as a Source Data file.

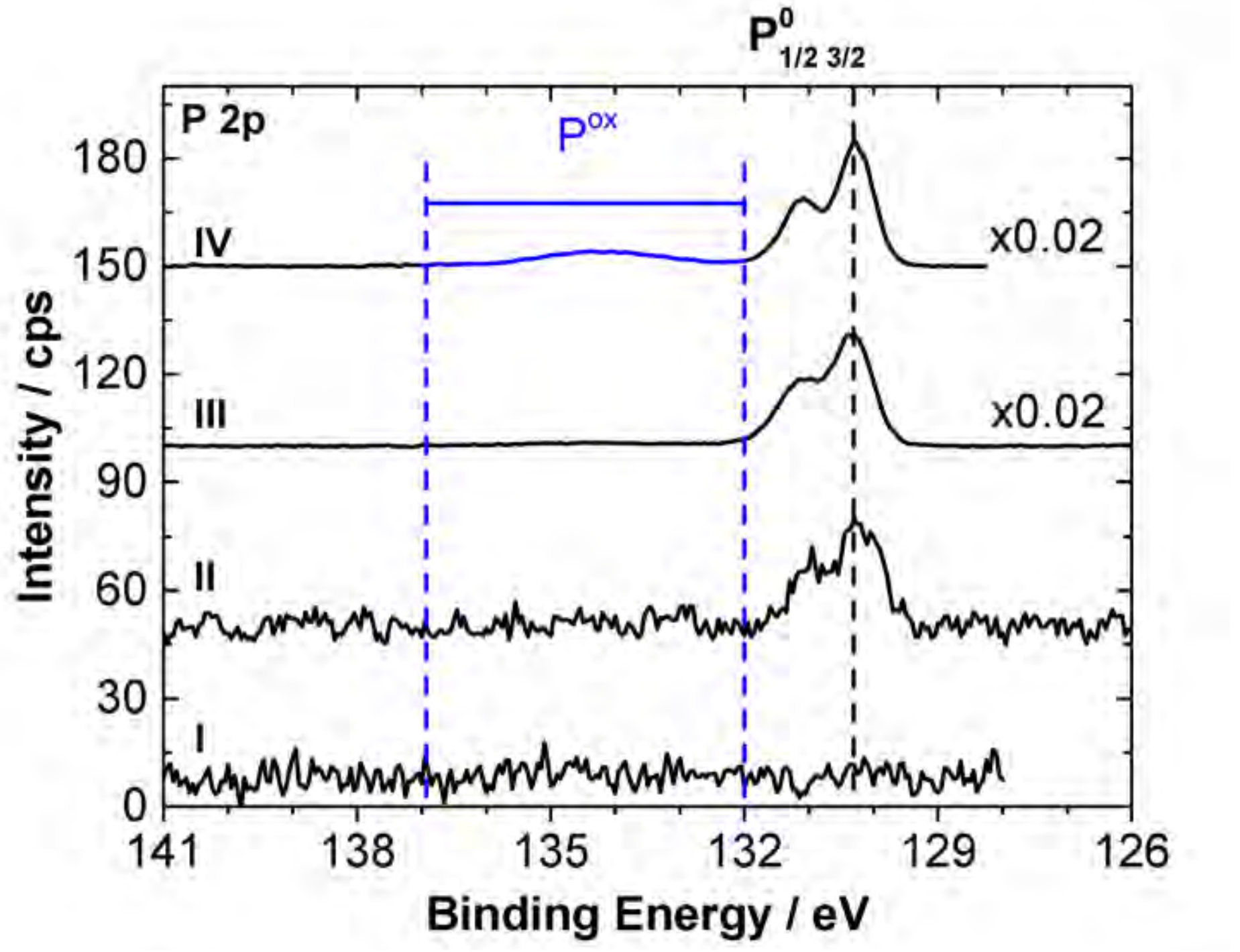

**Supplementary Figure 12.** XPS P 2p region of the neat bmim-$BF_4$ IL (I), the highly-concentrated FL-BP suspension (II) showing only P in oxidation state zero at P $2p_{3/2}$ = 130.2 eV (region for oxidized P species is indicated), after removal of most of the IL by heating in UHV (III), and after having exposed the sample subsequently to environmental conditions for a day, showing the presence of a broad oxide P component around 134 eV (IV). Spectra are offset and re-scaled for sake of clarity. Source data are provided as a Source Data file.

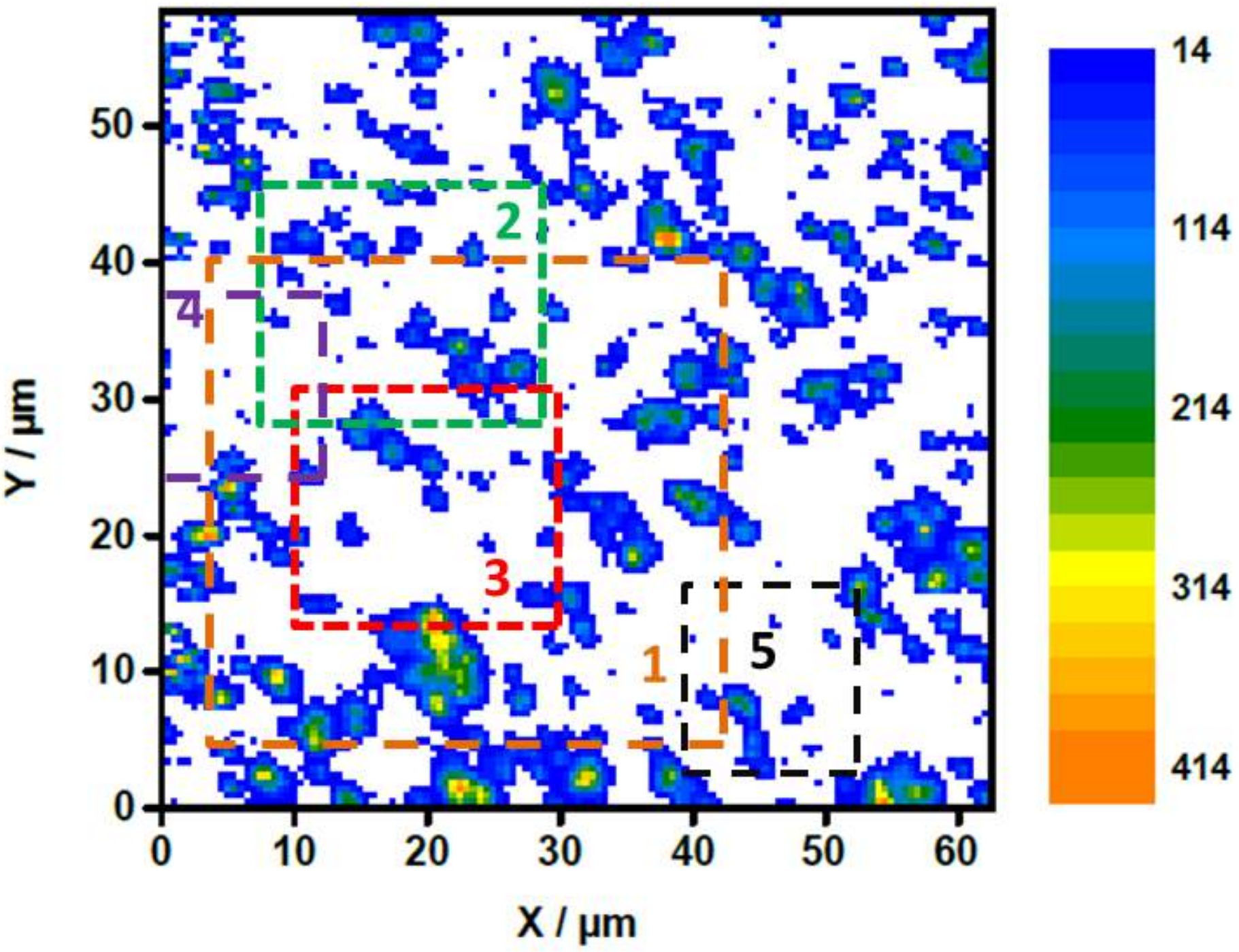

**Supplementary Figure 13.** Raman map of the $A_{1g}$ band of Sb with different squares numbered from 1 to 5 in order to correlate the Sb Raman signals with AFM images. Source data are provided as a Source Data file.

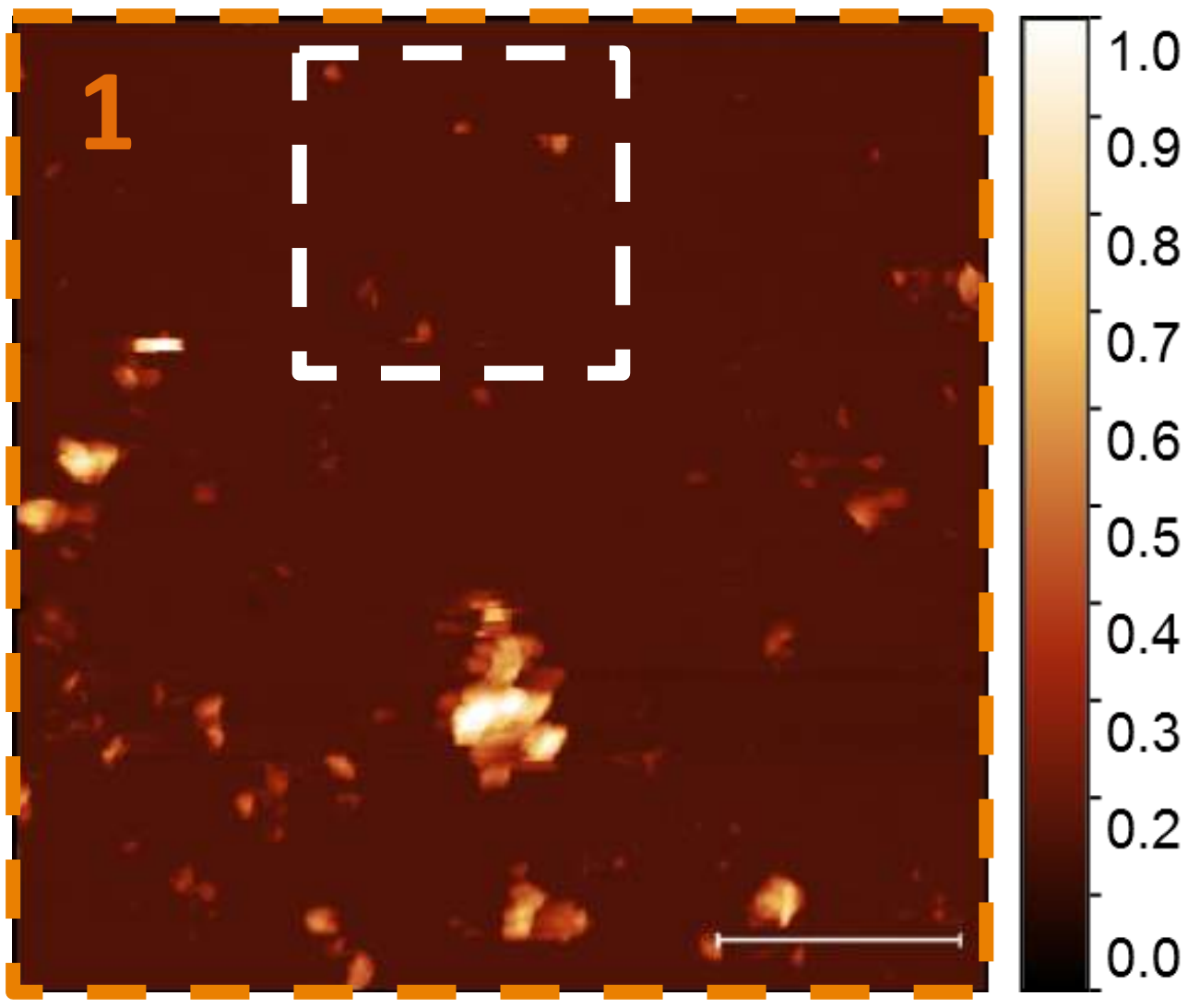

**Supplementary Figure 14.** AFM image showing area 1 (orange) of the corresponding Raman map in Supplementary Figure 13 (scale bar represents10 µm). Source data are provided as a Source Data file.

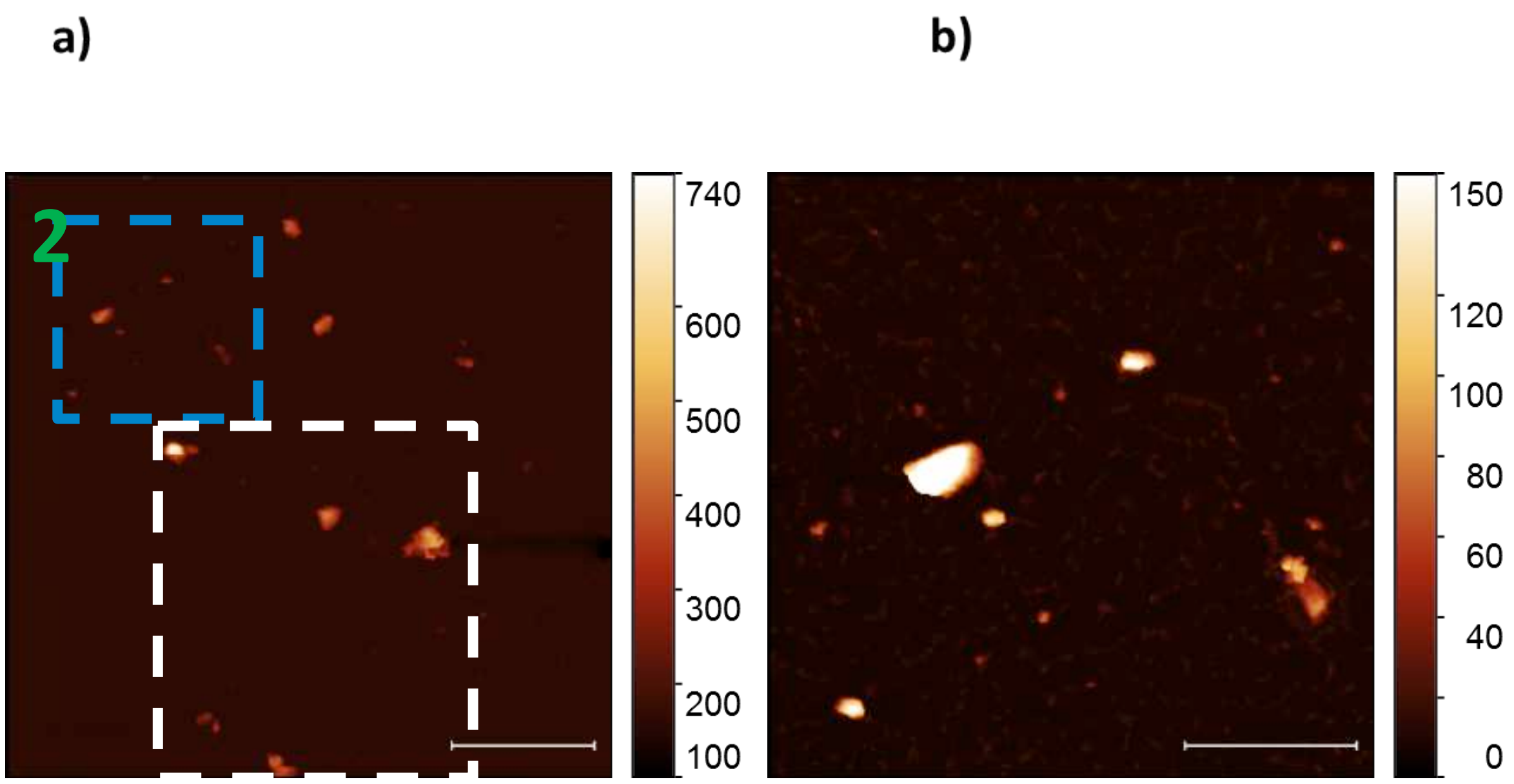

**Supplementary Figure 15.** AFM image showing a Zoom in of area 2 (green) in Supplementary Figure 13, highlighting the white area (a) and a further zoom in b) (scale bars: left 5 µm and right 2 µm). Source data are provided as a Source Data file.

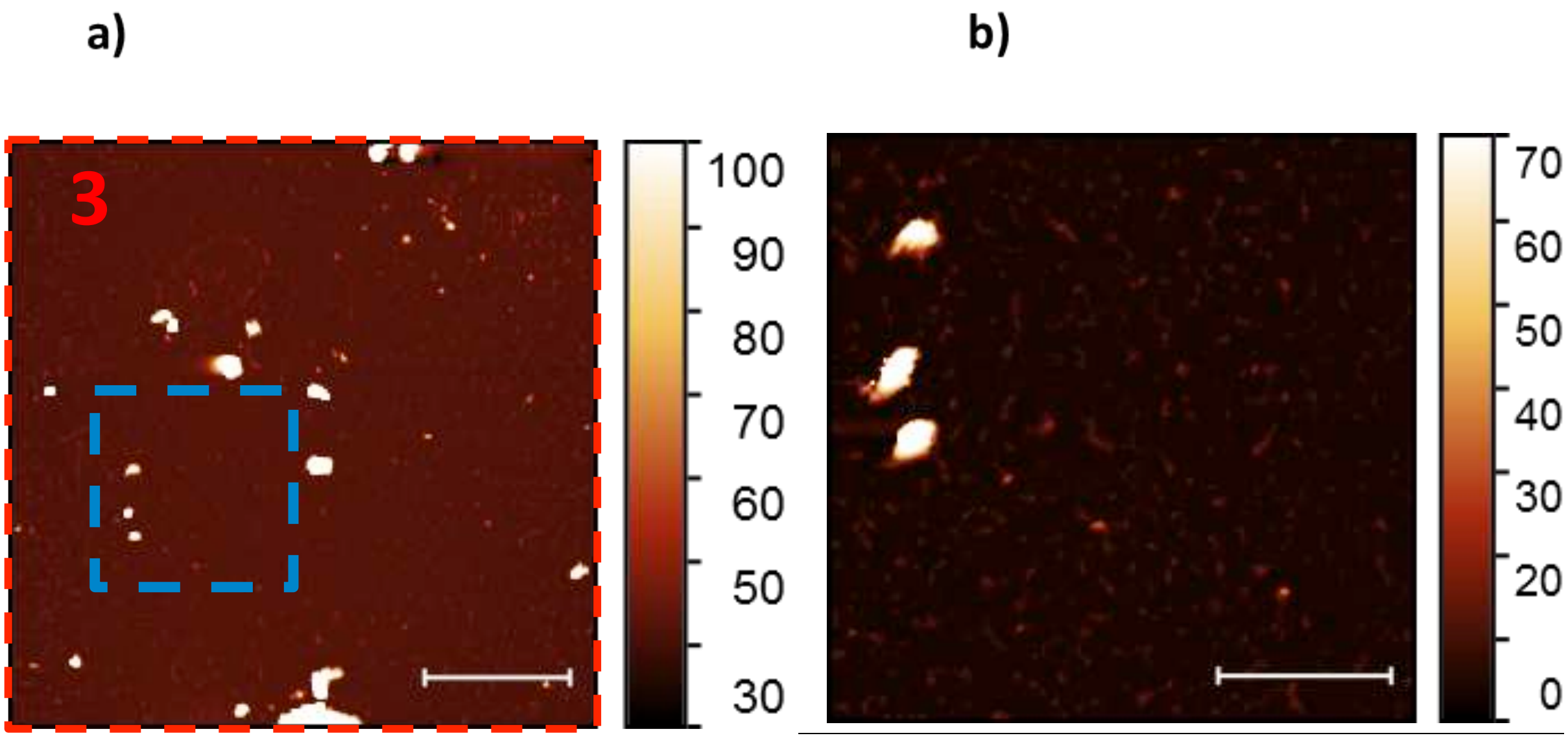

**Supplementary Figure 16.** a) AFM image showing the Zoom in of area 3 (red) in Supplementary Figure 13 and b) the further zoom in of the blue area (scale bars: left 5 µm and right 2 µm). Source data are provided as a Source Data file.

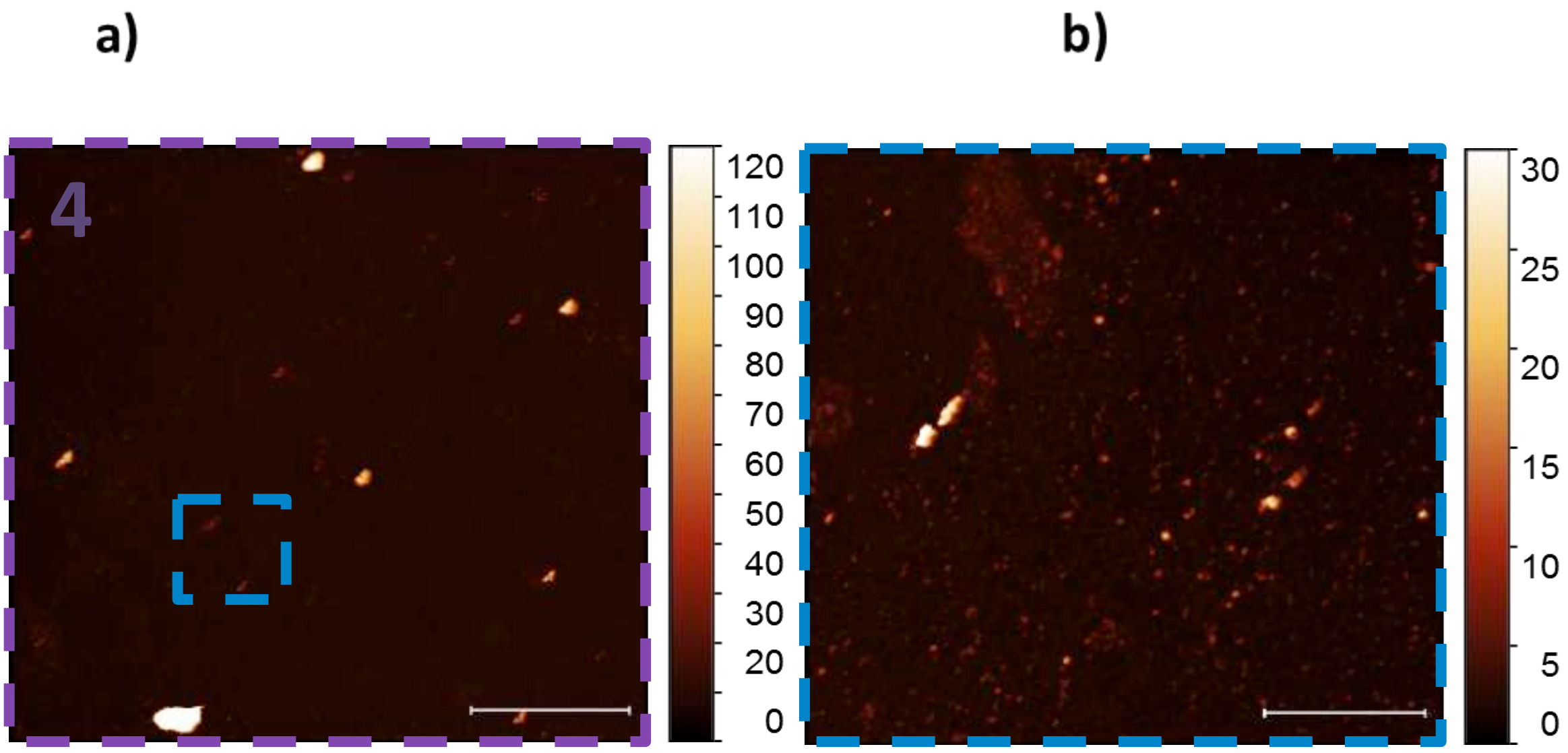

**Supplementary Figure 17.** a) AFM image showing the Zoom in of area 4 (violet) in Supplementary Figure 13 and b) the further zoom in of the blue area (scale bars: left 5 μm and right 1 μm). Source data are provided as a Source Data file.

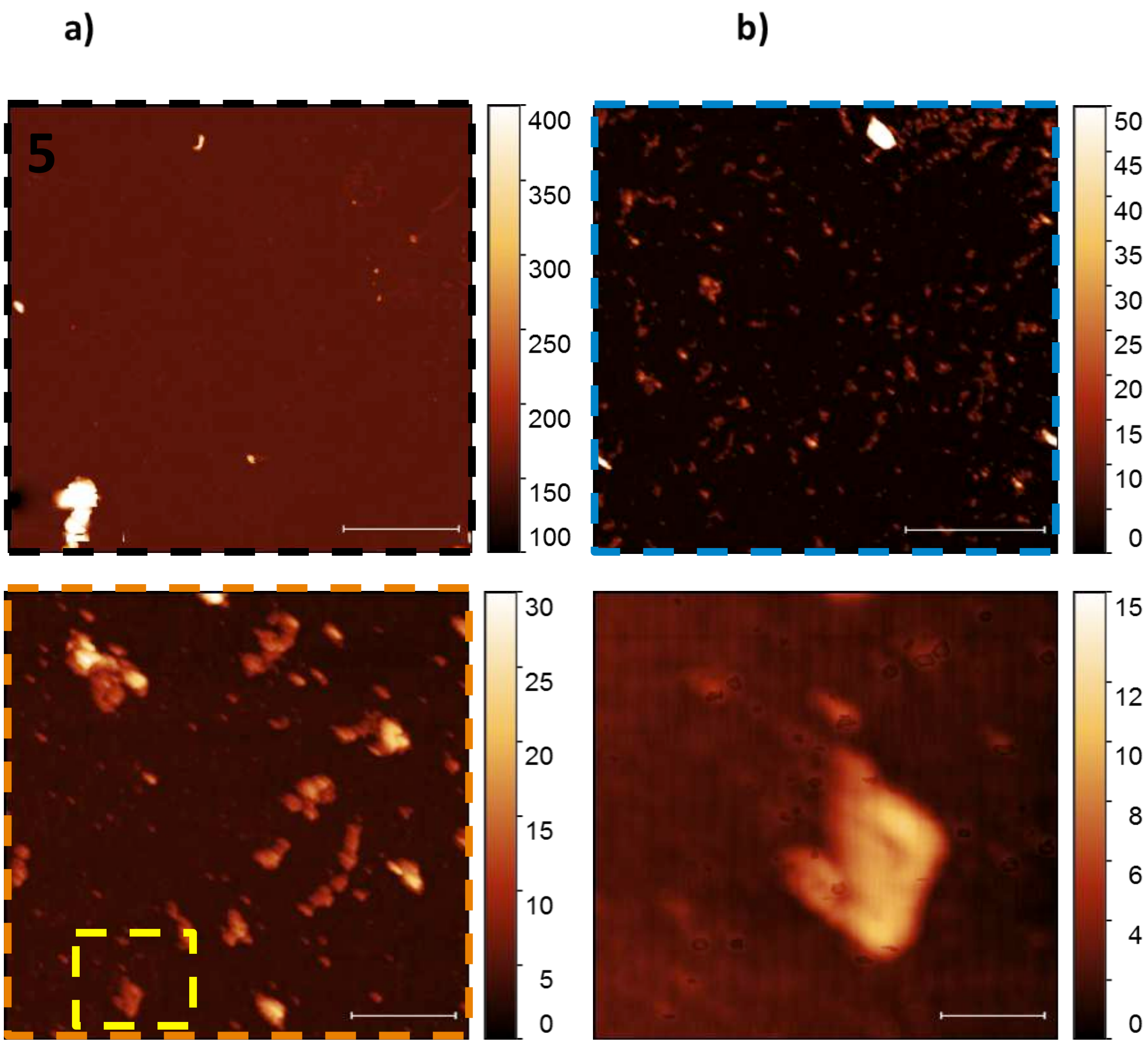

**Supplementary Figure 18.** a) AFM image showing the zoom in of area 5 (black, left; scale bar represents 5 µm) in Supplementary Figure 13 and further zoom in of the selected areas (b–d, scale bars: right 2 µm, bottom left 500 nm and right 100 nm).

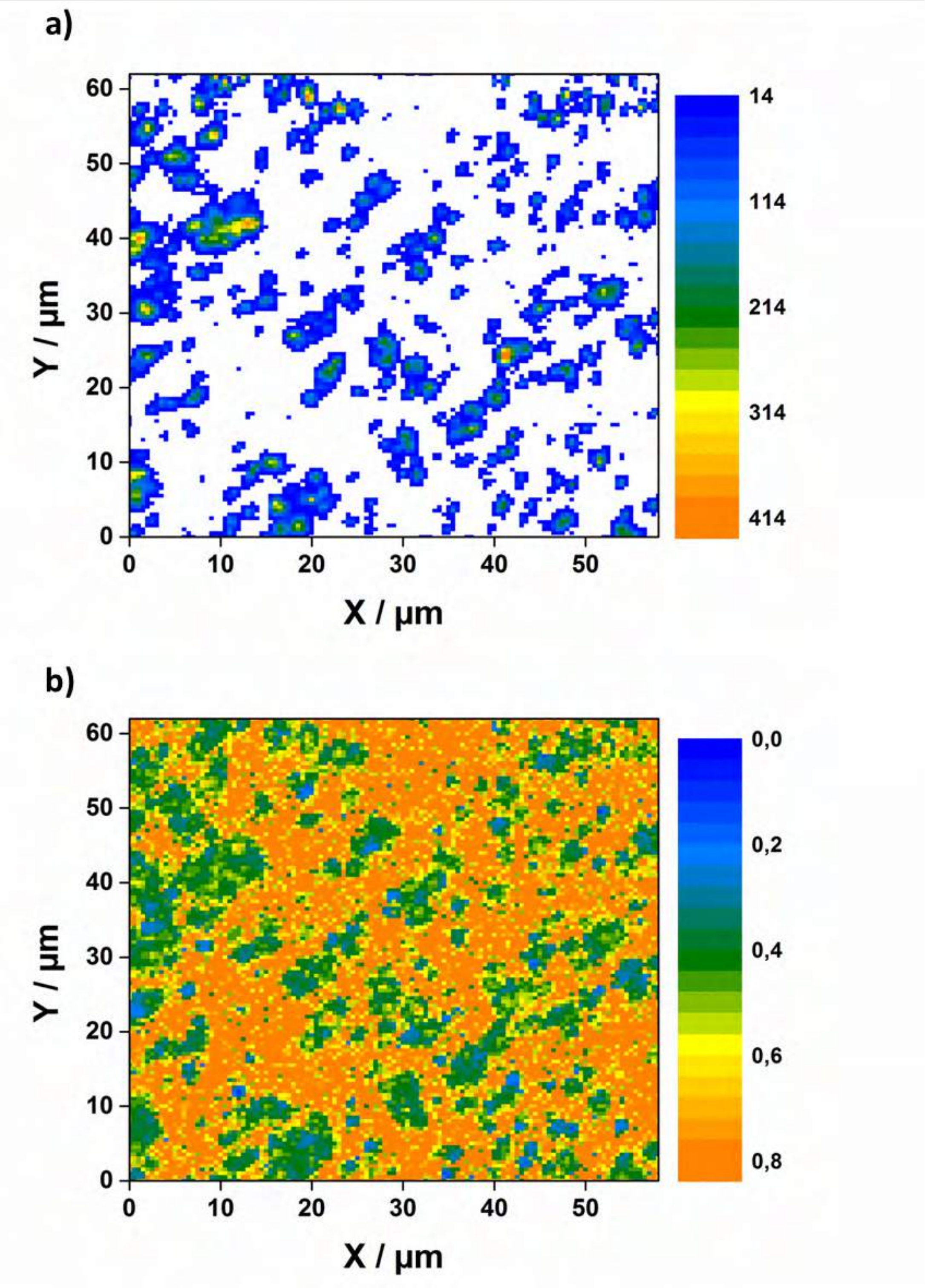

**Supplementary Figure 19**. a) Raman mapping of the $A^1_g$ band of Sb and b) Raman mapping of the Ratio (R) between the $A^1_g$ and $E_g$ Raman mode of Sb. Source data are provided as a Source Data file.

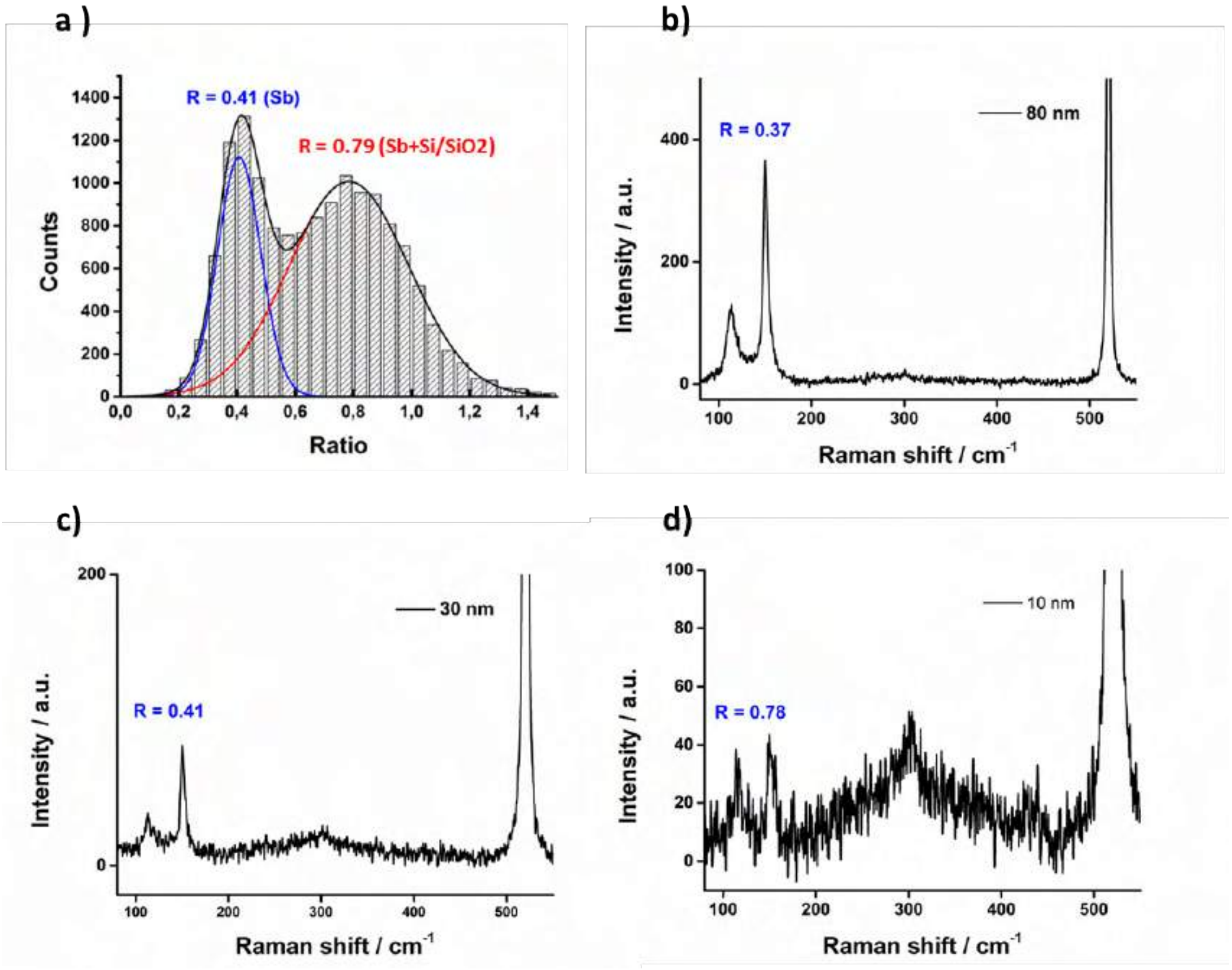

**Supplementary Figure 20.** a) Histogram showing the Ratio $A^1_g/E_g$ of Sb (blue R=0.41). Raman single point spectra of different Sb flakes with thicknesses from 80 nm to 10 nm (b–d) highlighting the ratio. Note that with decreasing thickness the Ratio increases but at a certain point cannot be separated anymore from the $Si/SiO_2$ substrate because of the low counts. Source data are provided as a Source Data file.

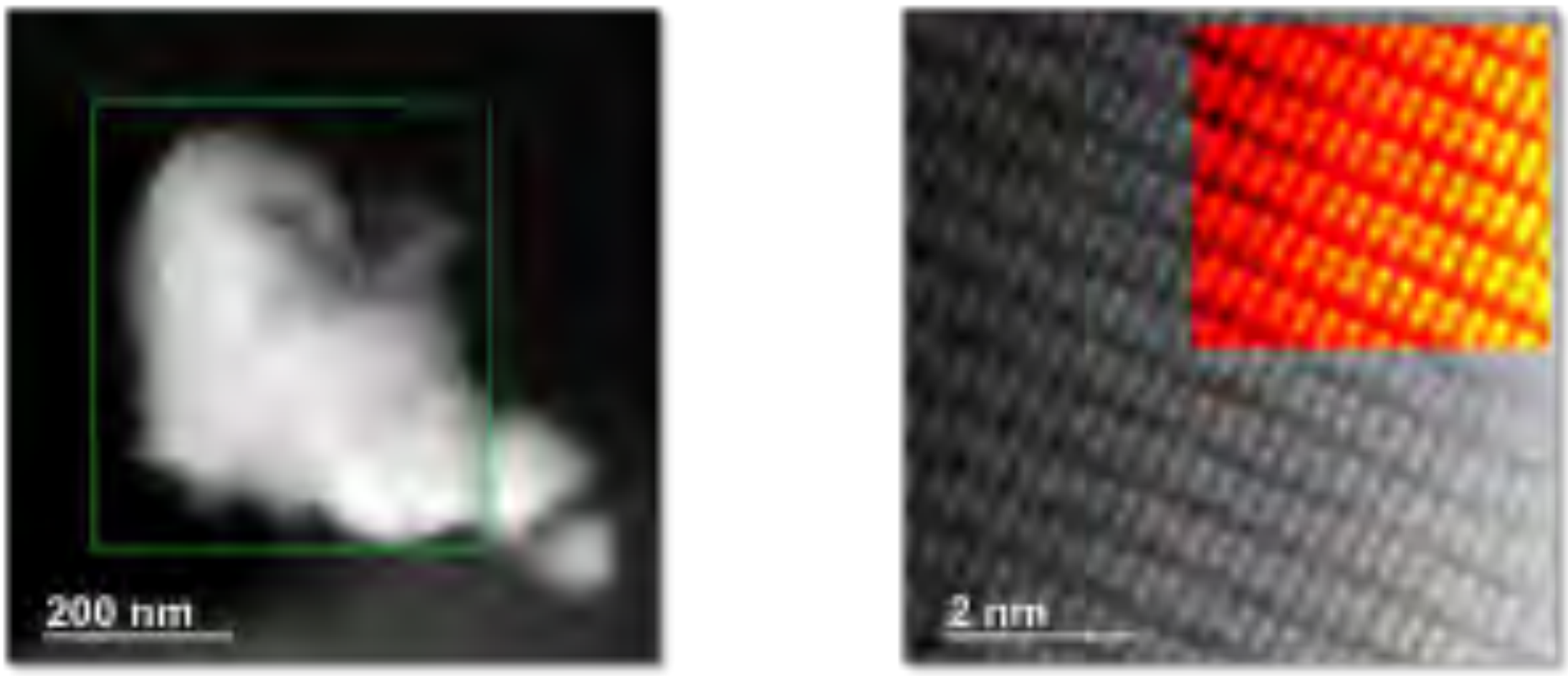

**Supplementary Figure 21.** (left) Low magnification HAADF STEM image of a FL–Sb flake. The scale bar represents 200 nm. Compositional maps derived from EEL spectrum images of the flake shown in Figure 2 of the main text has been acquired from the area highlighted with a green rectangle. (right) Atomic resolution HAADF image acquired on the edge of a free–standing portion of a flake, near the edge, along with a Fourier filtered (FFT) image in the inset, acquired down the [210] orientation. The scale bar is 2 nm. Data acquired at 80 kV.

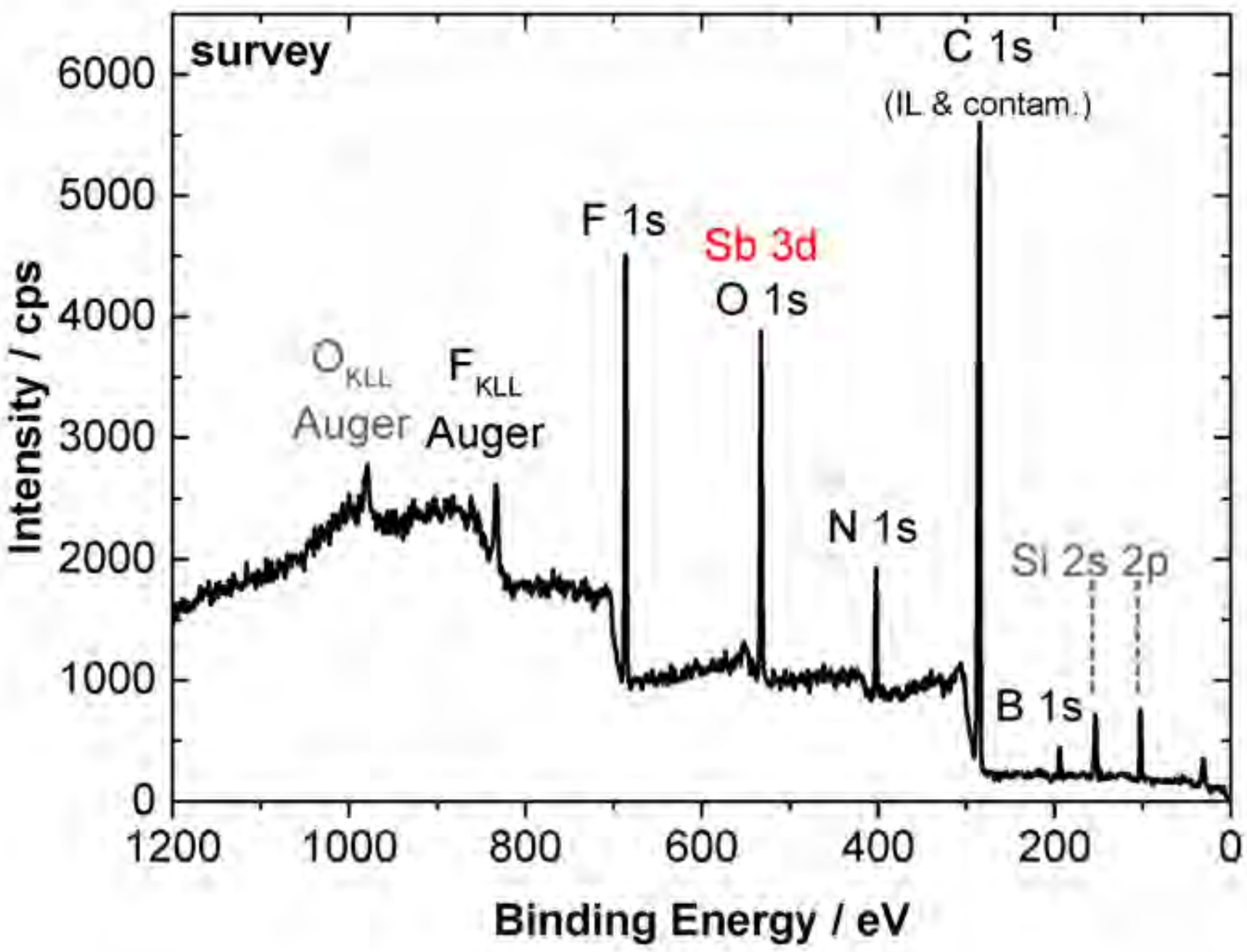

**Supplementary Figure 22.** XPS survey spectrum of FL-Sb. Source data are provided as a Source Data file.

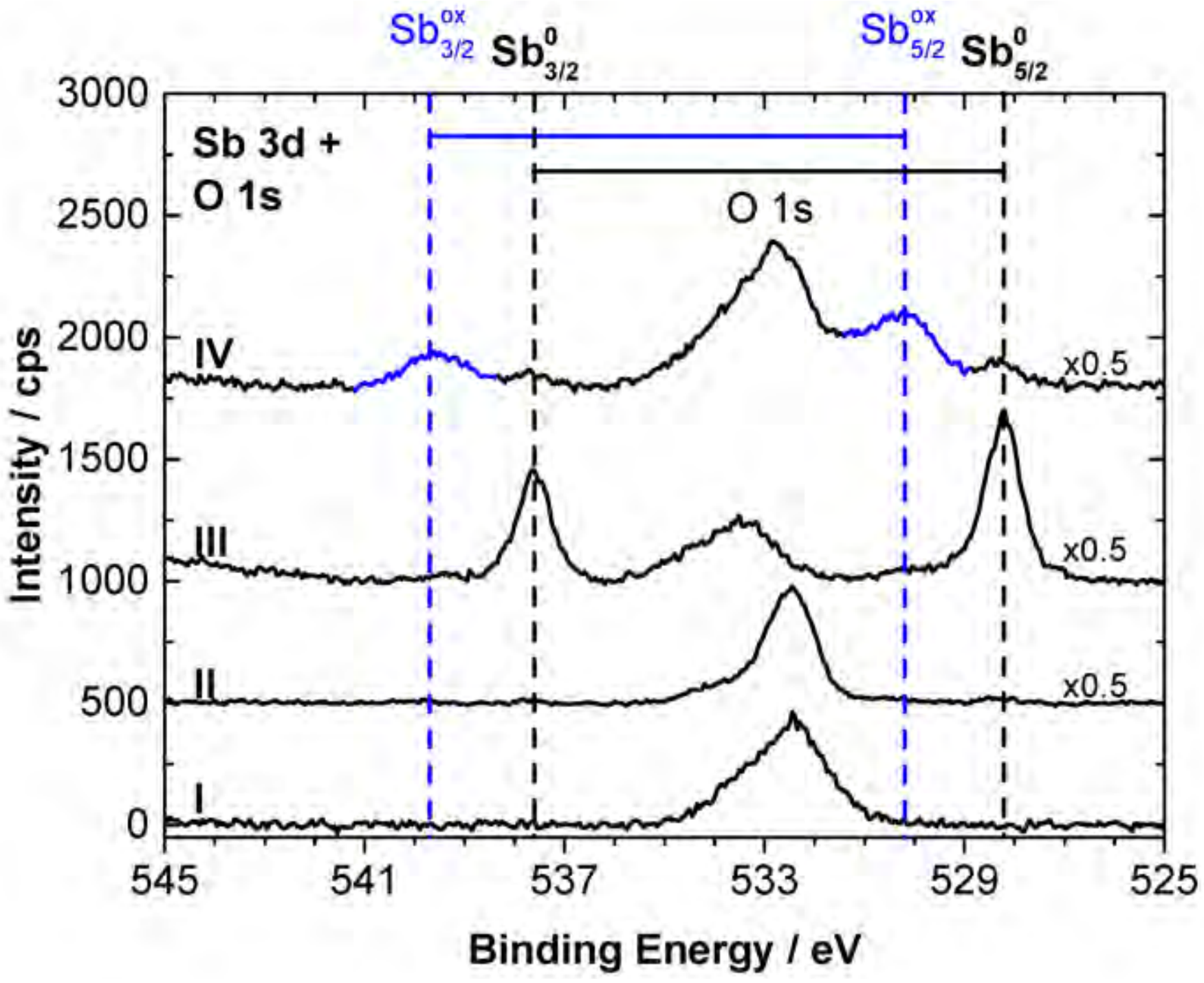

**Supplementary Figure 23.** XPS Sb 3d and O 1s region of the neat bmim-$BF_4$ IL (I) showing oxygen signals from the IL surface contamination layer, of the highly-concentrated FL-Sb suspension (II) showing small signals of non-oxidized (Sb $3d_{5/2}$ at 528.2 eV) and minor contributions from oxidized (530.3 eV) antimony next to the oxygen contamination, after removal of most of the IL by heating in UHV (III), and after submitting the sample to environmental conditions for a day, showing a drastic decrease in Sb(0) and concomitant

increase of the oxidized Sb species. Spectra are offset and re-scaled for sake of clarity. Source data are provided as a Source Data file.

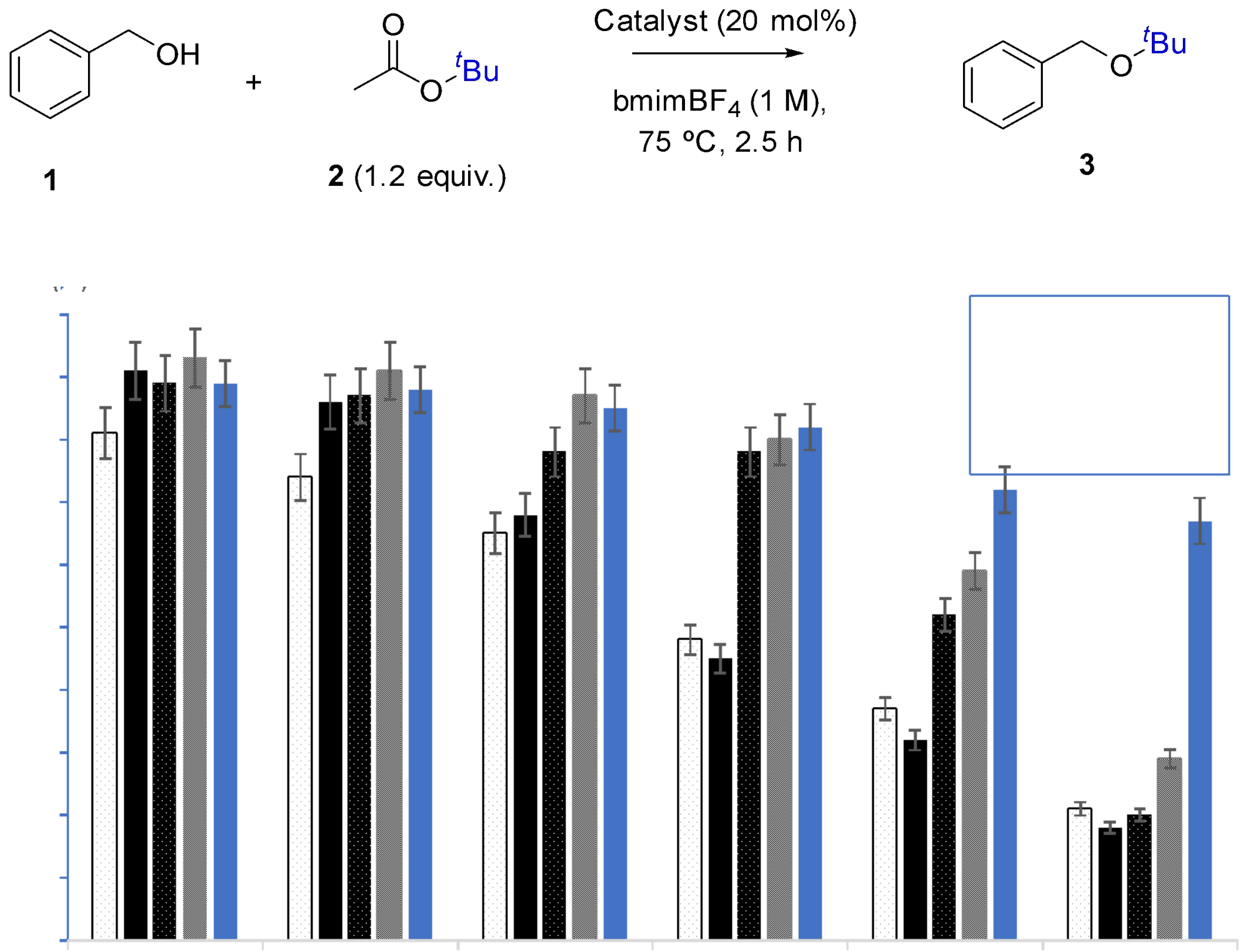

**Supplementary Figure 24.** Reuse tests. The FL-BP catalyst lifetime was prolonged with a sonication treatment after each use (**5**). Error bars account for 5% uncertainty. Source data are provided as a Source Data file.

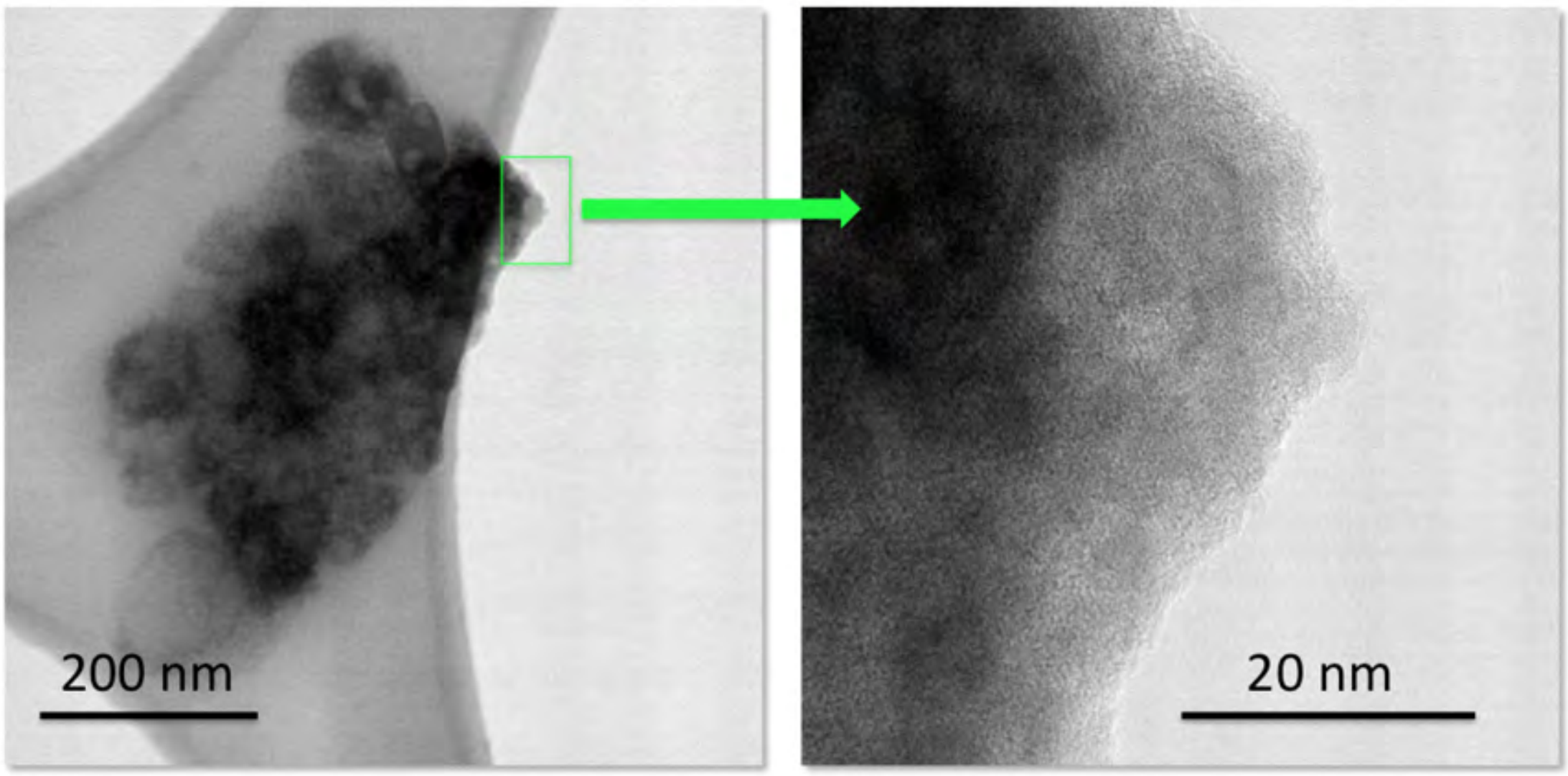

**Supplementary Figure 25.** Low magnification (left) and high magnification (right) annular bright field images of a $N_2$–cycled FL–Sb sample exhibiting significant edge amorphization.

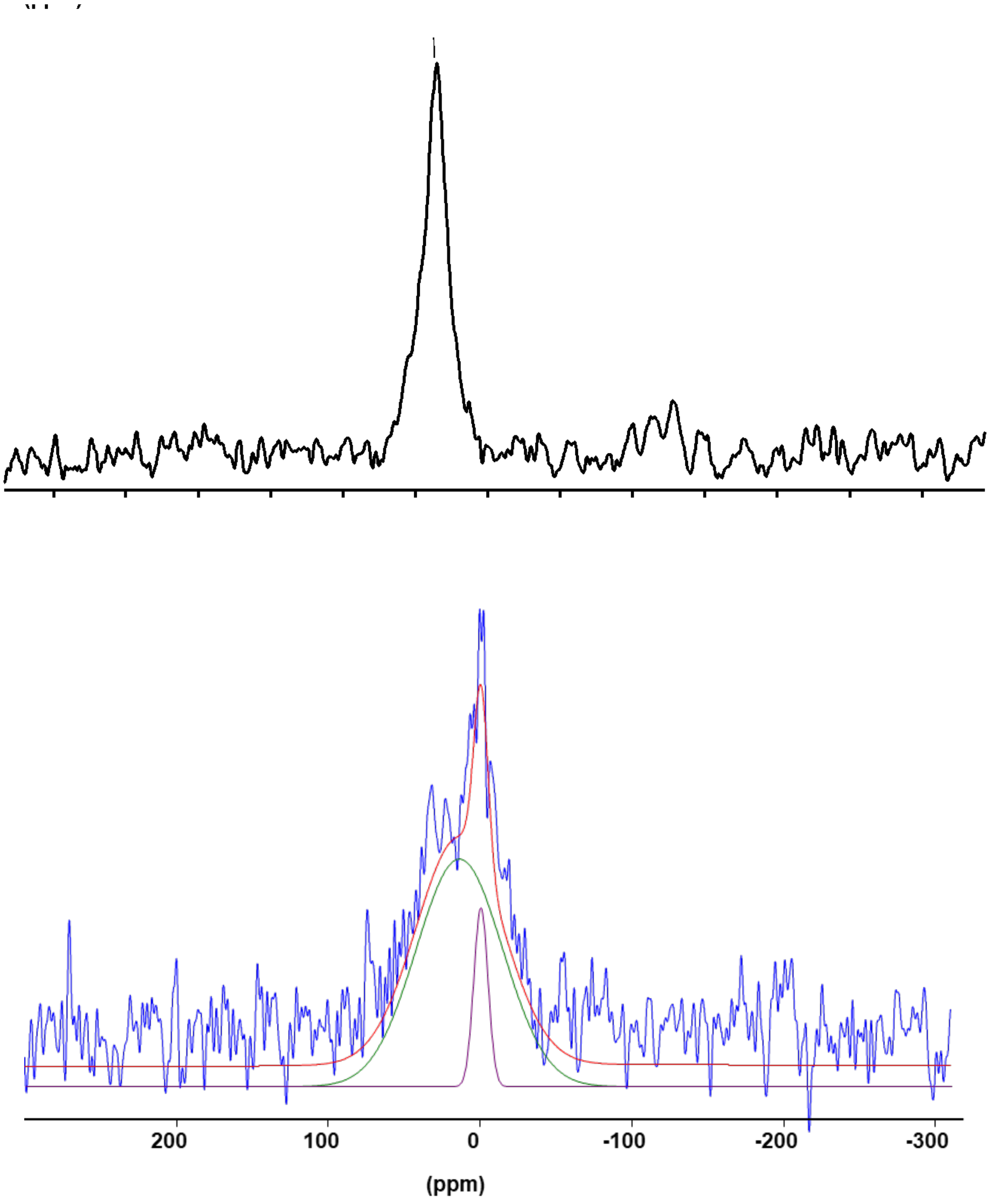

**Supplementary Figure 26.** Top: $^{31}$P MAS NMR spectrum of FL–BP in the presence of ten–fold amounts of **1** and **2**. Bottom: Deconvoluted solid state NMR of FL–BP in statics, without spinning. Blue line shows the raw whole spectrum, red line the sum of components, green line P in zero oxidation state and purple line oxidized P, which accounts for 8% out of the total.

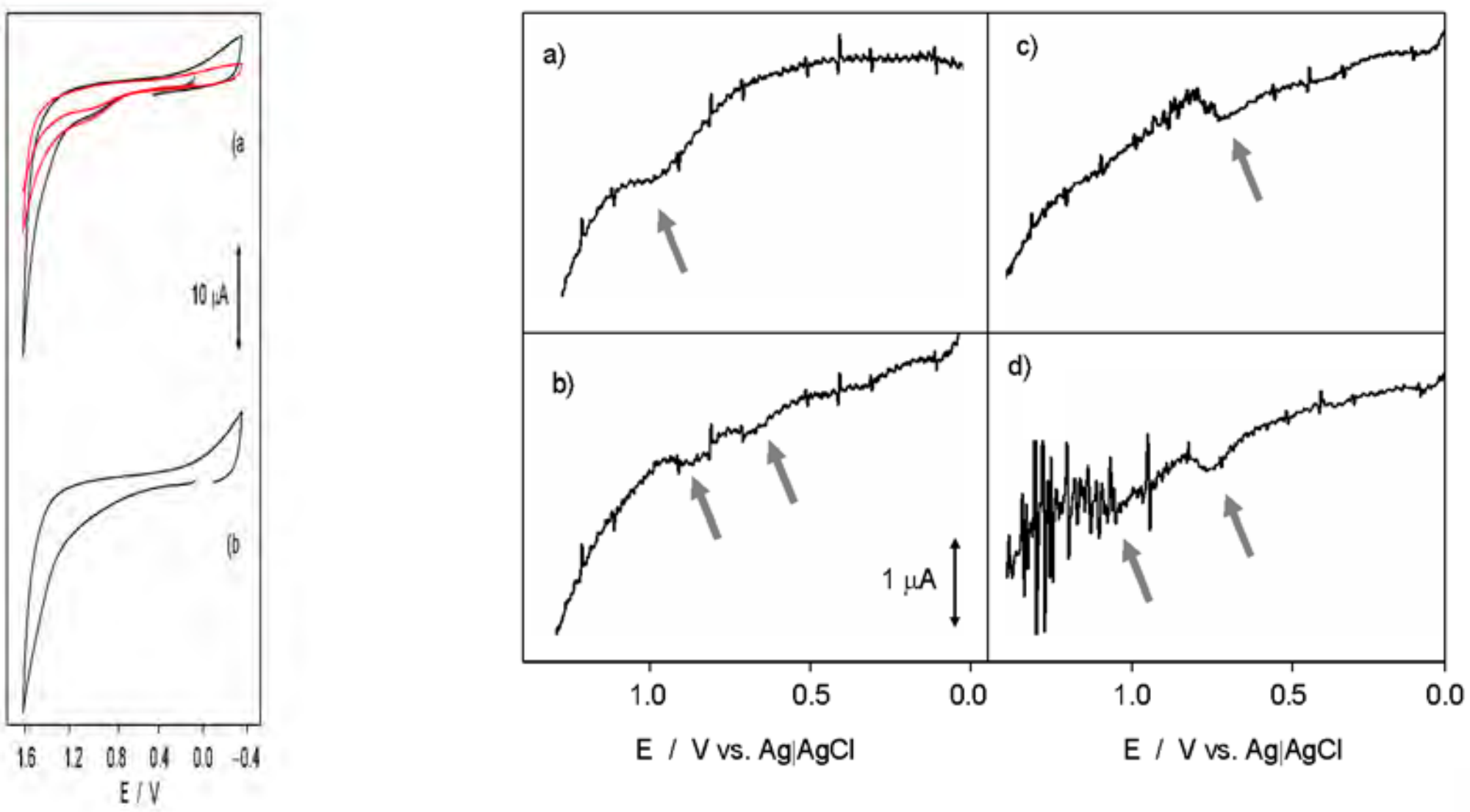

**Supplementary Figure 27** Left: Cyclic voltammograms at glassy carbon electrode of suspensions of a) FL–BP and b) Fl–Sb (black lines) in air-saturated $bmimBF_4$. Red line in a) corresponds to the blank voltammogram in $bmimBF_4$ under the same experimental conditions. Potential scan rate 50 mV $s^{-1}$; potentials relative to Pt pseudo-reference electrode. Right: Linear potential scan voltammograms, after semi-derivative convolution, of a,b) FL-Sb and c,d) FL-BP nanoparticulate deposits on glassy carbon electrode immersed into 0.10 M potassium phosphate aqueous buffer at pH 7.0 extracted from IL solutions a,c) before and b,d) after addition of the BrOAc plus *t*-BuOAc reactants. Potential scan initiated at 0.0 V in the positive direction. The arrows indicate signals of agglomeration/fractioning of the particles.

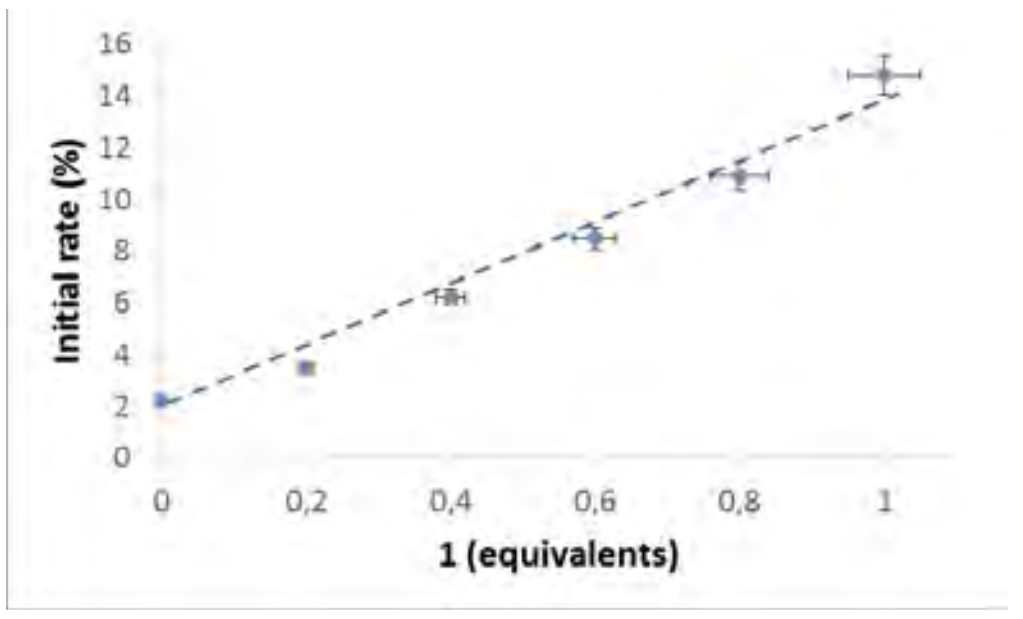

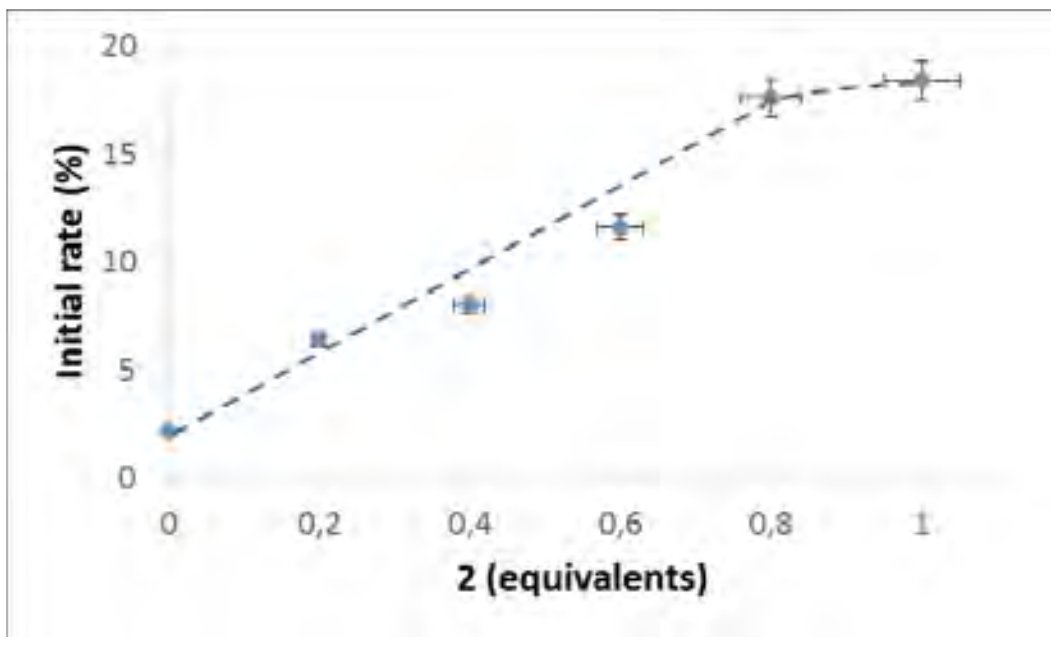

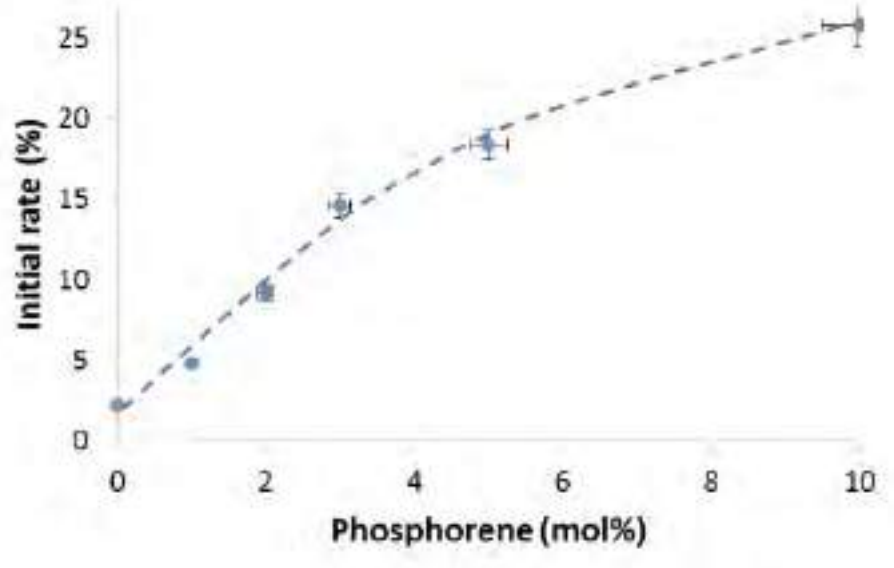

**Supplementary Figure 28.** Reaction order for each reagent in the alkylation reaction (Table 2) with FL–BP catalyst. Similar lines obtained for FL–Sb. Error bars account for 5% uncertainty. Source data are provided as a Source Data file.

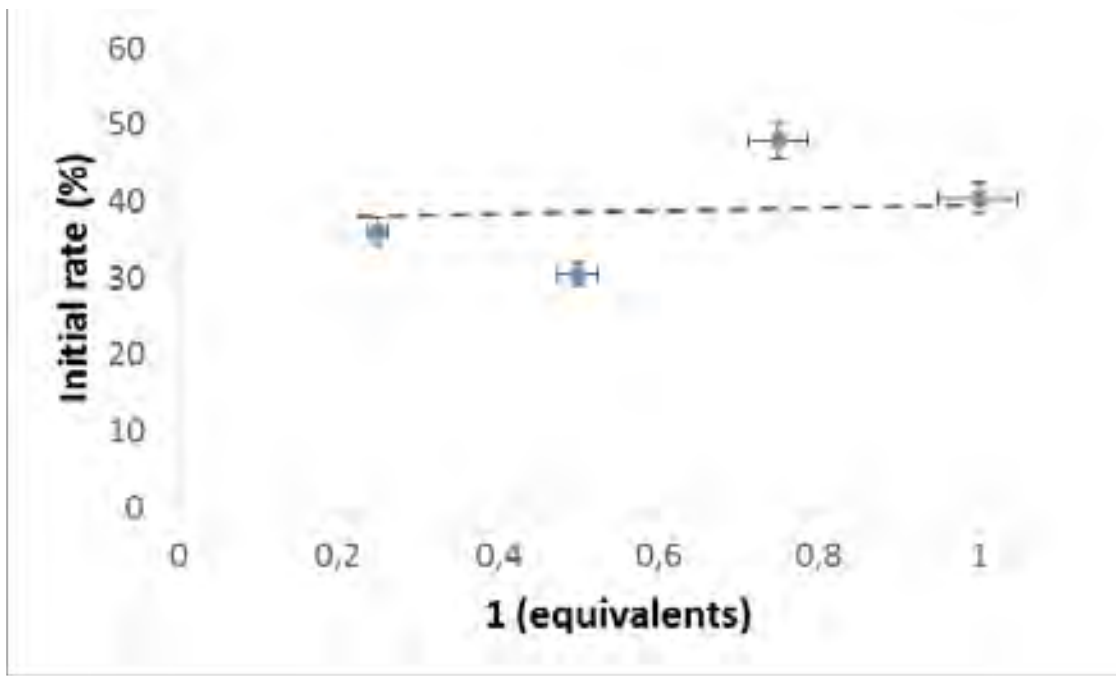

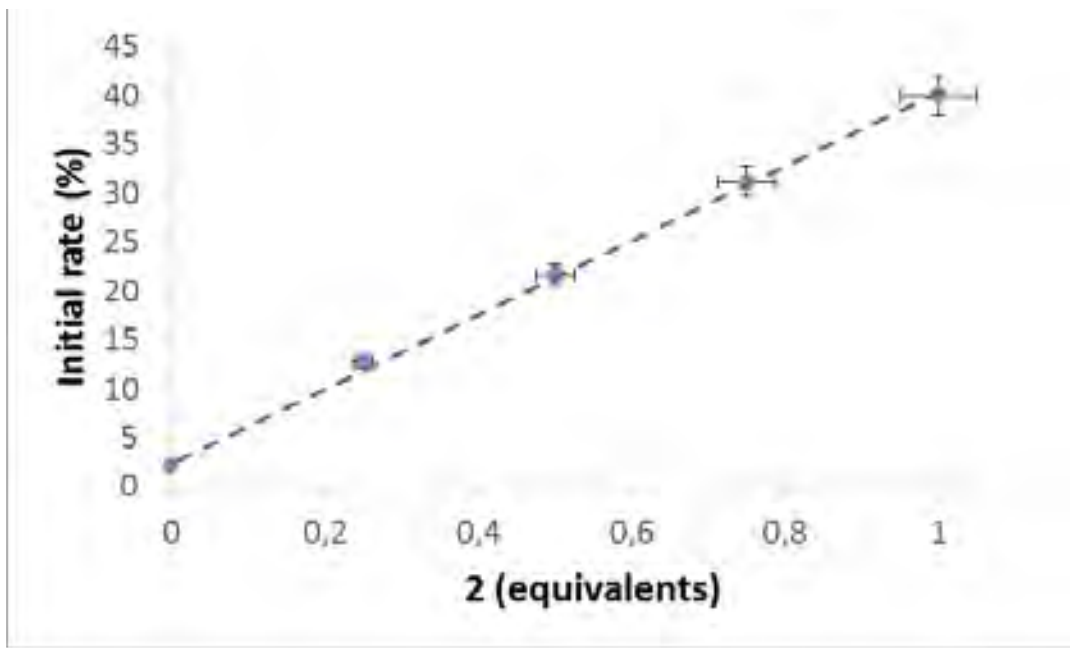

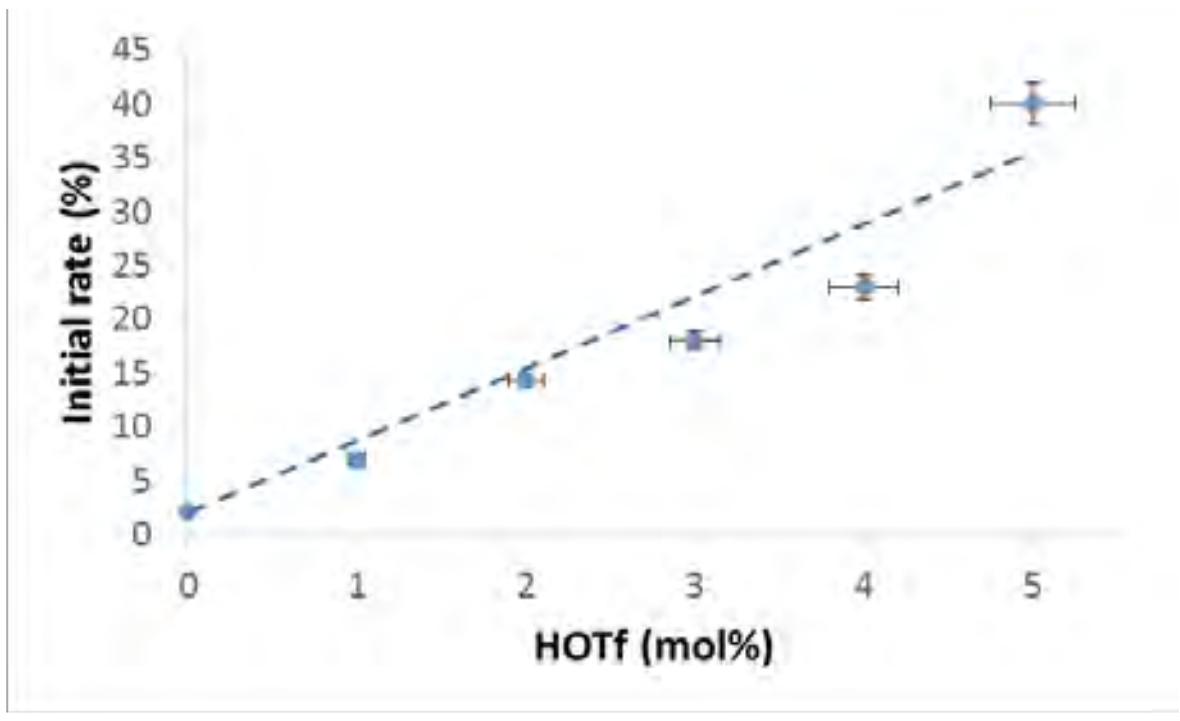

**Supplementary Figure 29.** Reaction order for each reagent in the alkylation reaction of **1** with **2** in the presence of HOTf catalyst at 25 ºC. Error bars account for 5% uncertainty. Source data are provided as a Source Data file.

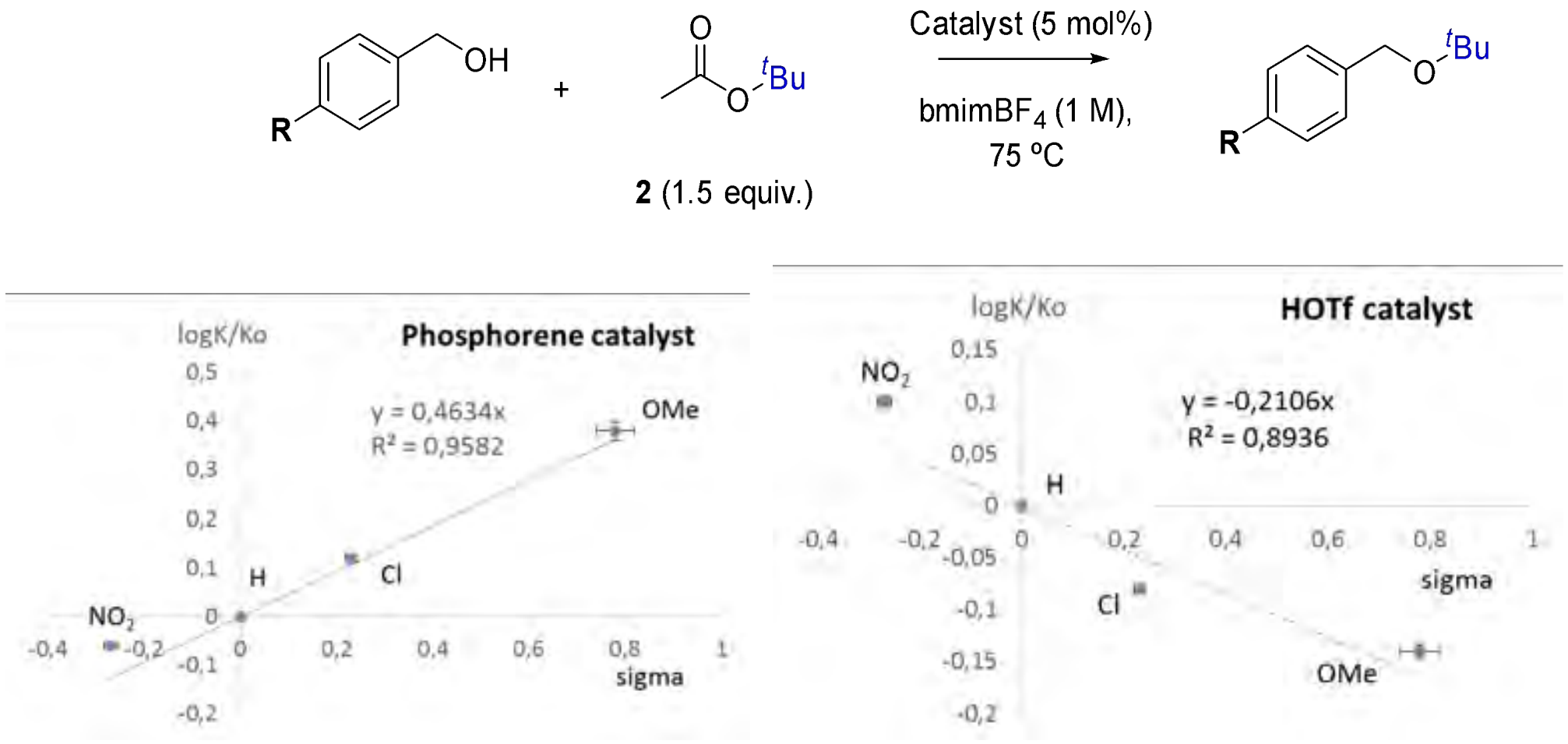

**Supplementary Figure 30.** Hammett–plot for the reaction above with FL–BP (left) and triflic acid (HOTf, right) catalyst. Error bars account for 5% uncertainty. Source data are provided as a Source Data file.

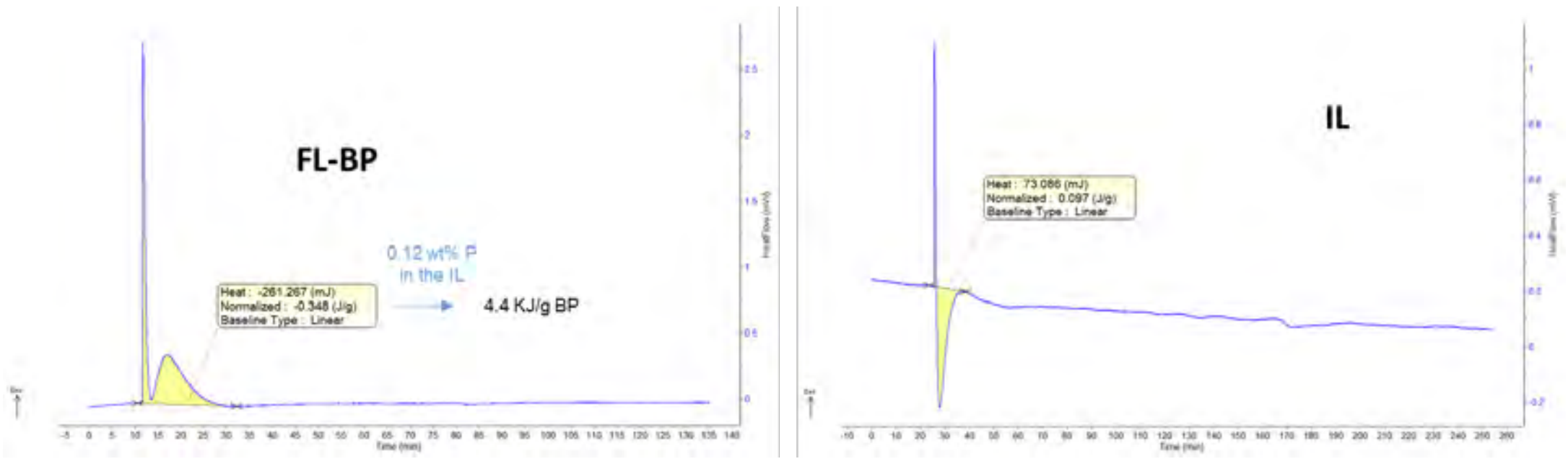

**Supplementary Figure 31** Calorimetry results for the adsorption of **1** on FL-BP (left) and neat IL (right).

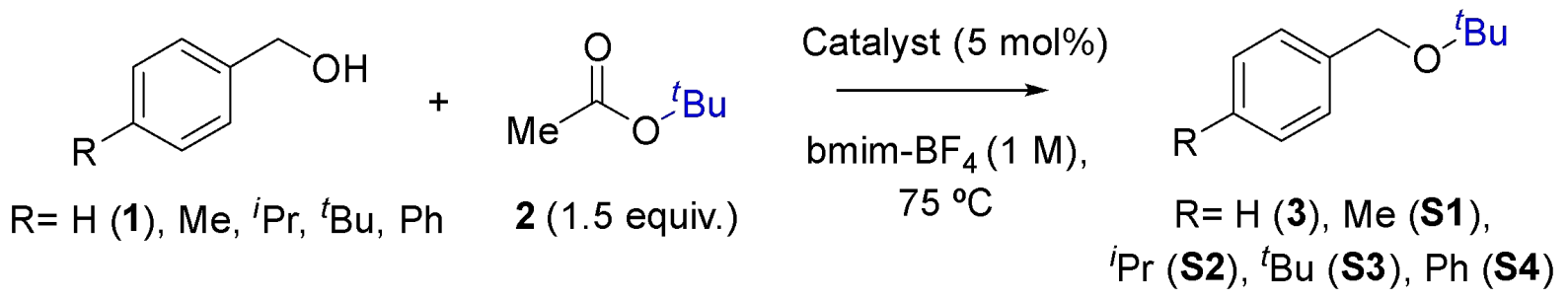

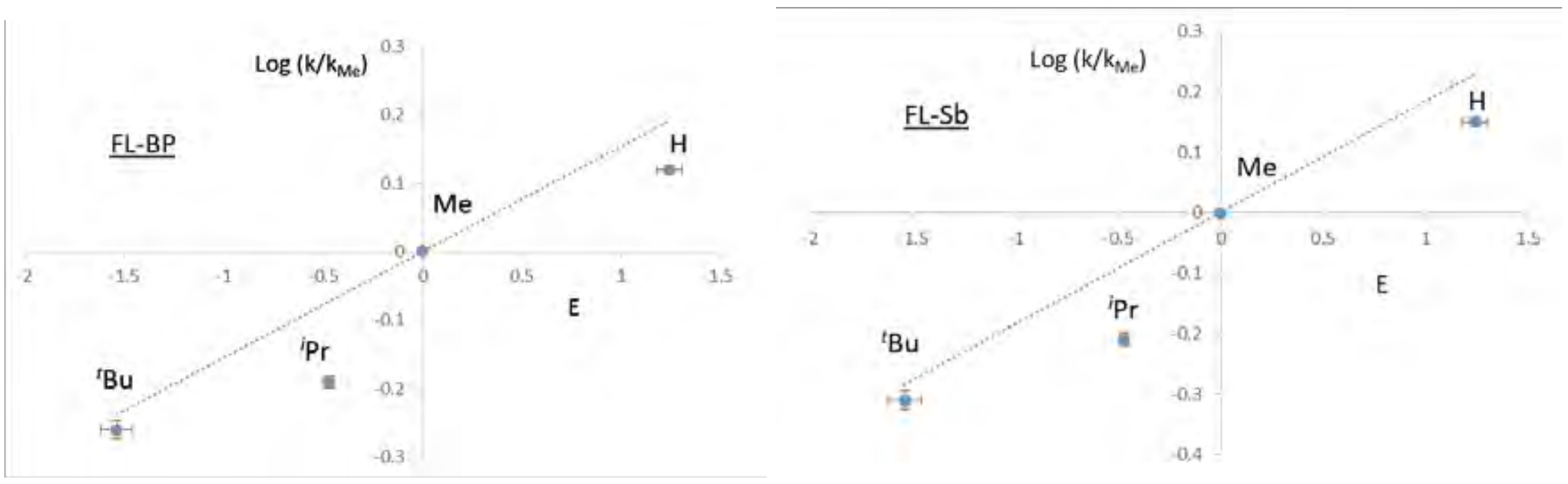

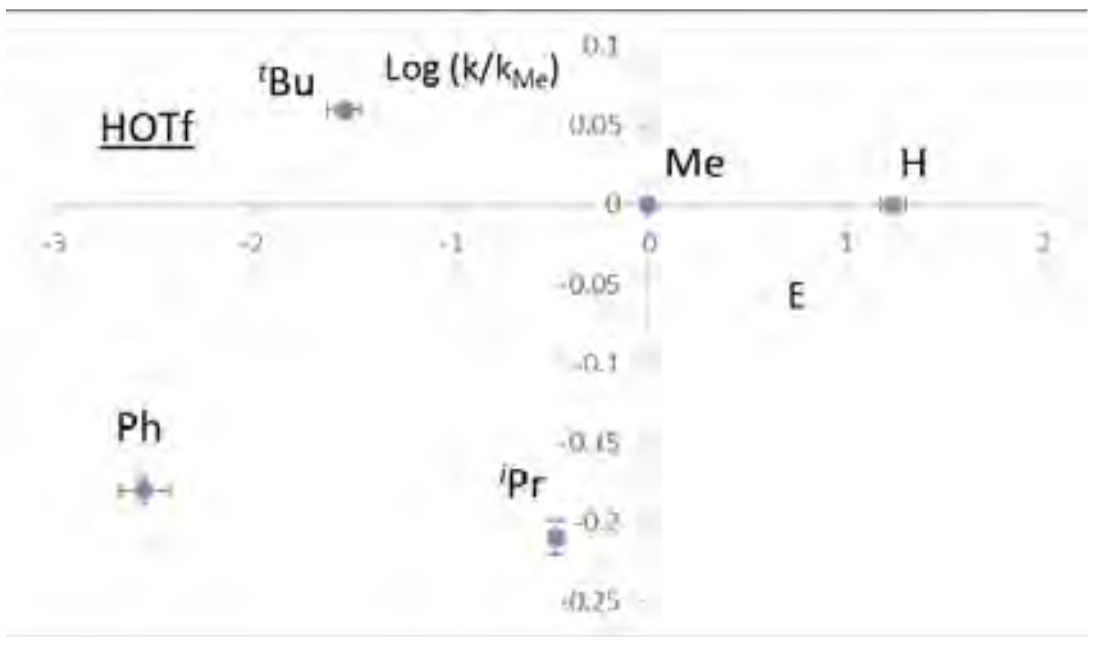

**Supplementary Figure 32.** Taft plots for FL-BP, FL-Sb and HOTf (0.5 mol%), under reaction conditions in Table 2, otherwise indicated. Error bars account for 5% uncertainty. Source data are provided as a Source Data file.

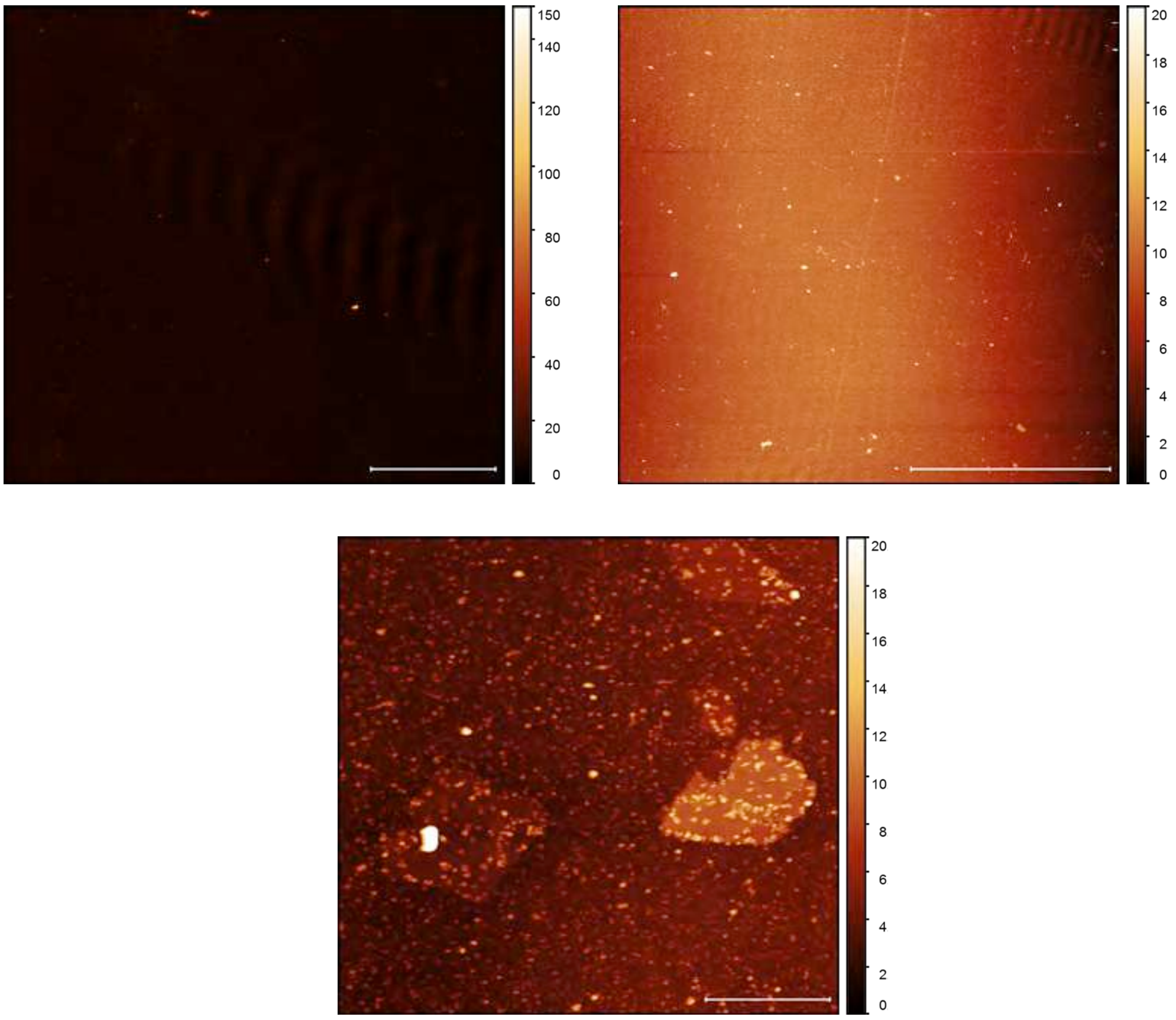

**Supplementary Figure 33.** AFM topography image of the exfoliated FL–BP sample prepared without ultracentrifugation (left, scale bar represents 20 μm ). AFM topography image of the same sample submitted to ultracentrifugation (2 h at 13.000 rpm) (right, scale bar represents 20 μm ), and zoom–in of some thin flakes with minimum thicknesses of *ca*. 1.5 nm (bottom, scale bar represents 500 nm).

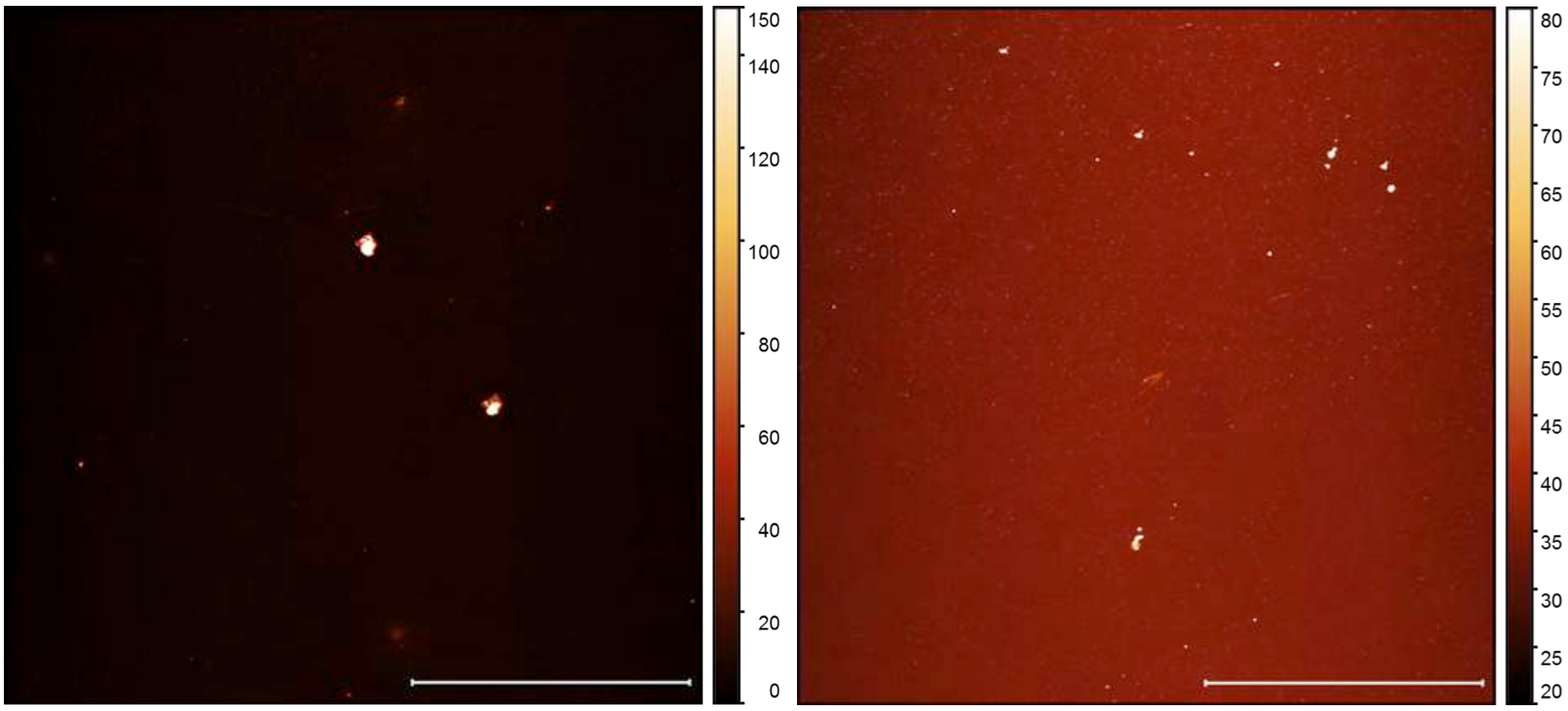

**Supplementary Figure 34.** AFM topography image of the exfoliated FL–Sb sample prepared without ultracentrifugation (left, scale bar represents 20 μm). AFM topography image of the same sample submitted to ultracentrifugation (2h at 13.000 rpm) (right, scale bar represents 20 μm). Source data are provided as a Source Data file.

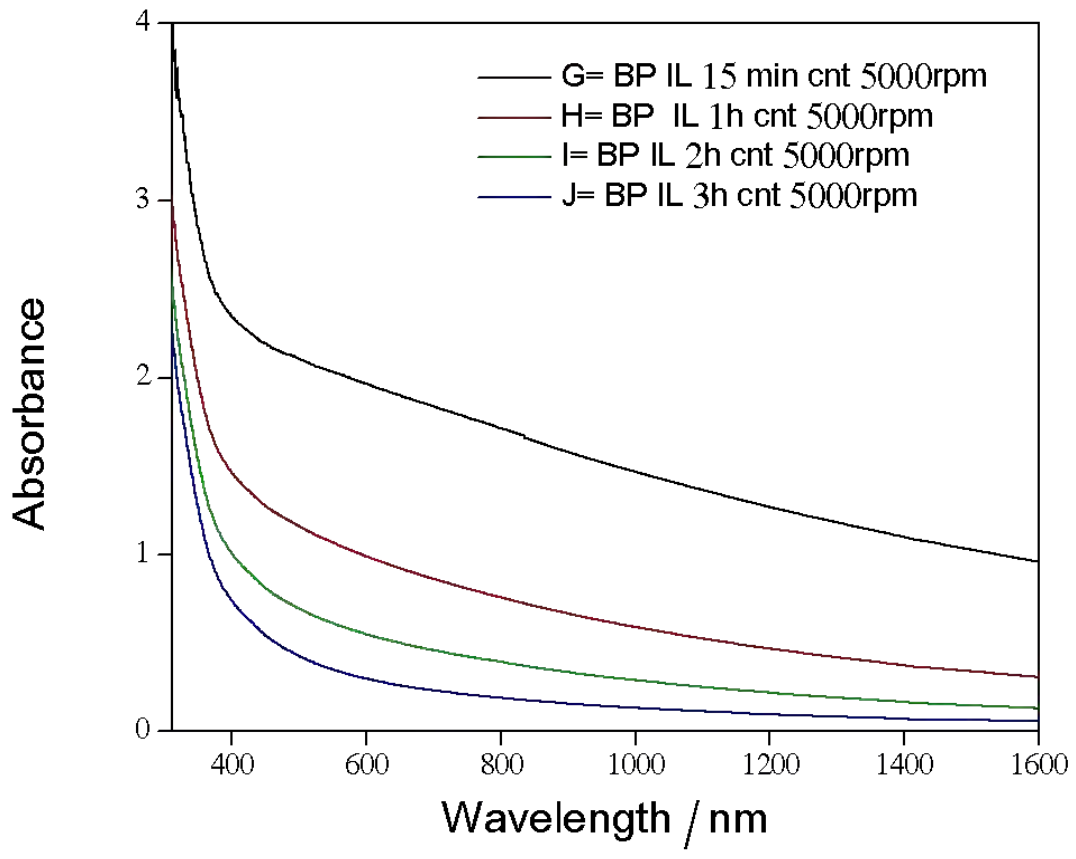

**Supplementary Figure 35.** Absorption visible spectra for FL-BP centrifuged for different times. Black and blue lines correspond to 0.12 and 0.01 mmol P $g^{-1}$, respectively. It is worth to remark the high quality of the solvent used for the dispersions, which is reflected in the absence of water absorption peaks in the NIR spectra at around 1430 nm. Source data are provided as a Source Data file.

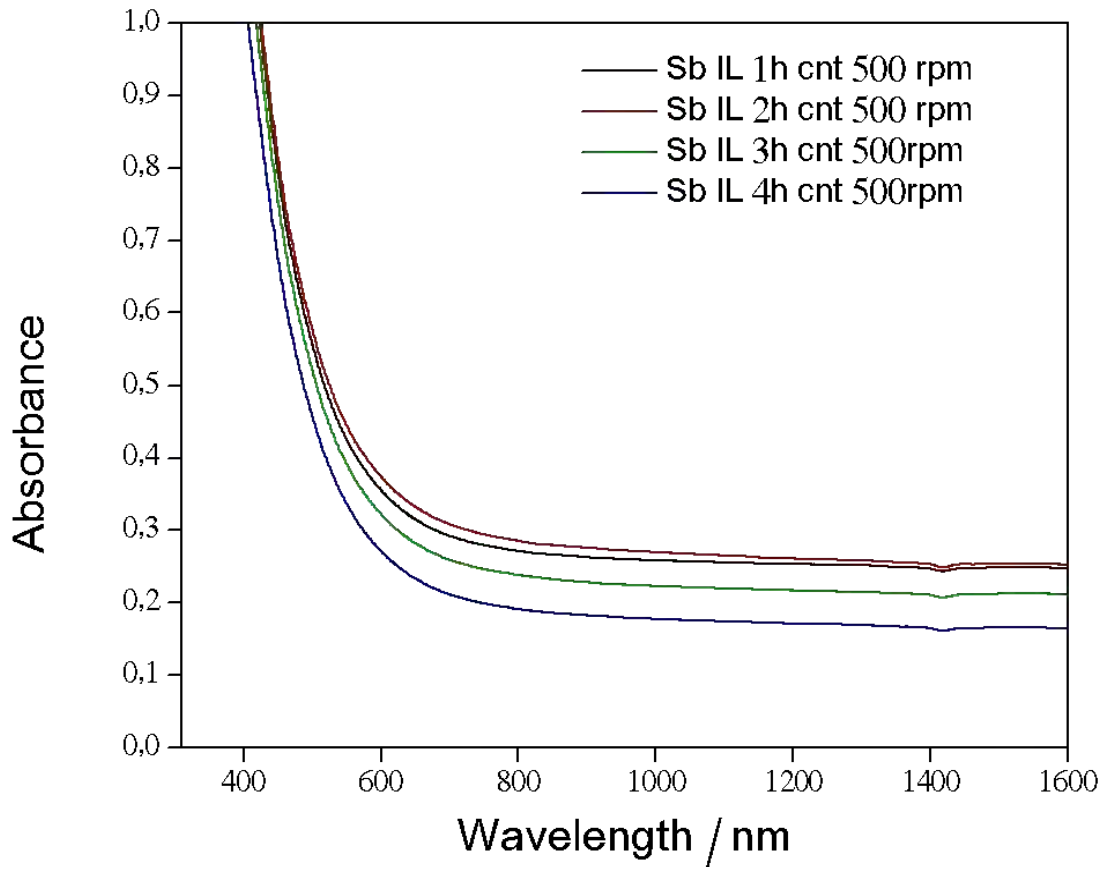

**Supplementary Figure 36.** Absorption visible spectra for FL-Sb centrifuged for different times. Black and blue lines correspond to 0.15 and 0.02 mmol Sb $g^{-1}$, respectively. Source data are provided as a Source Data file.

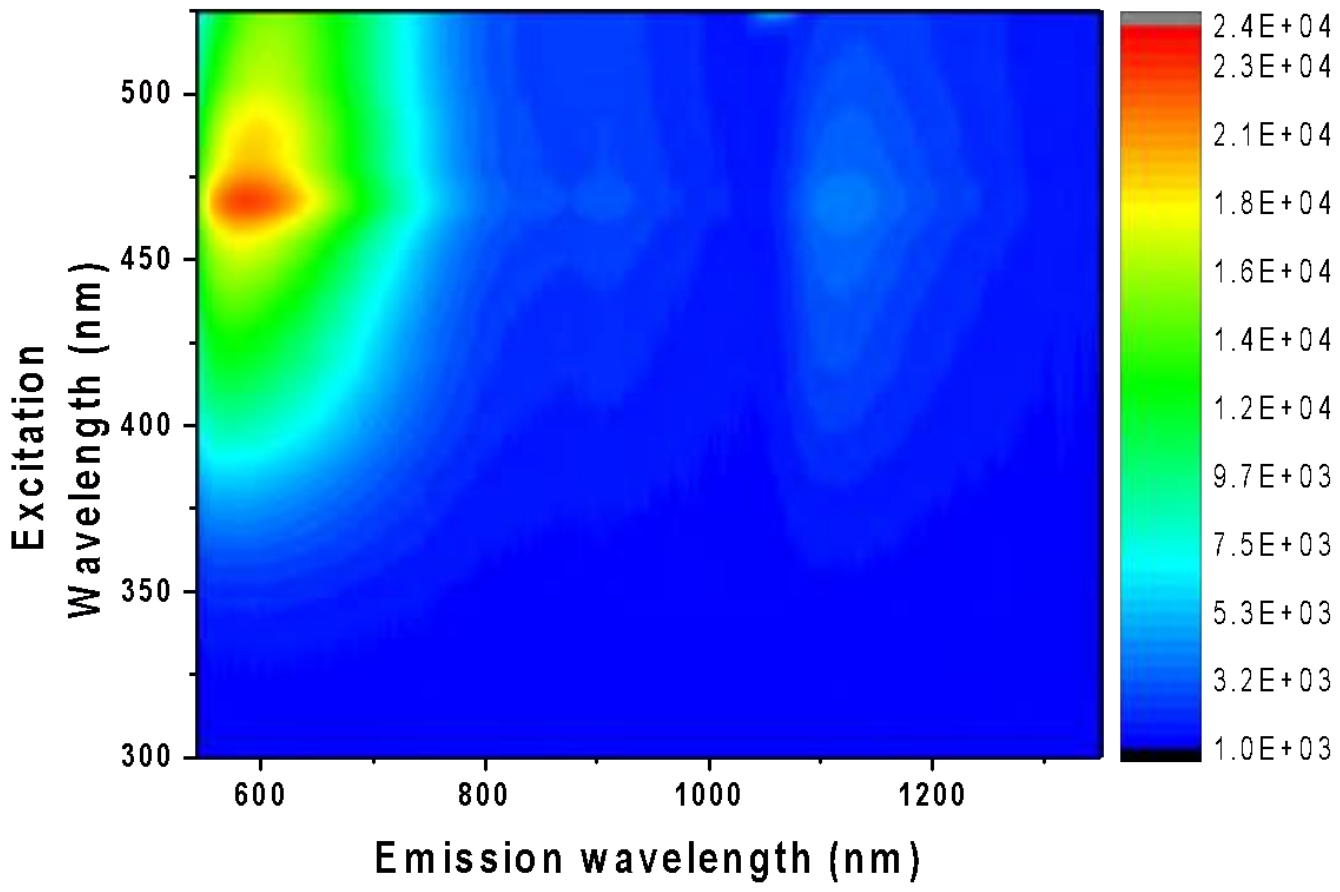

**Supplementary Figure 37.** Photoluminescence emission–excitation contour maps measured on a FL-BP dispersion centrifuged for 3h at 5000 rpm under inert gas conditions (see Methods) measured with a 550 nm cut-off filter in emission. According to Hanlon et al. (*Nature Communications*, 2015, *6*, 8563) the observed emission lines at ~600, 900, and 1125 nm can be associated with PL from 1, 2, and 3 layers of BP, respectively. Source data are provided as a Source Data file.

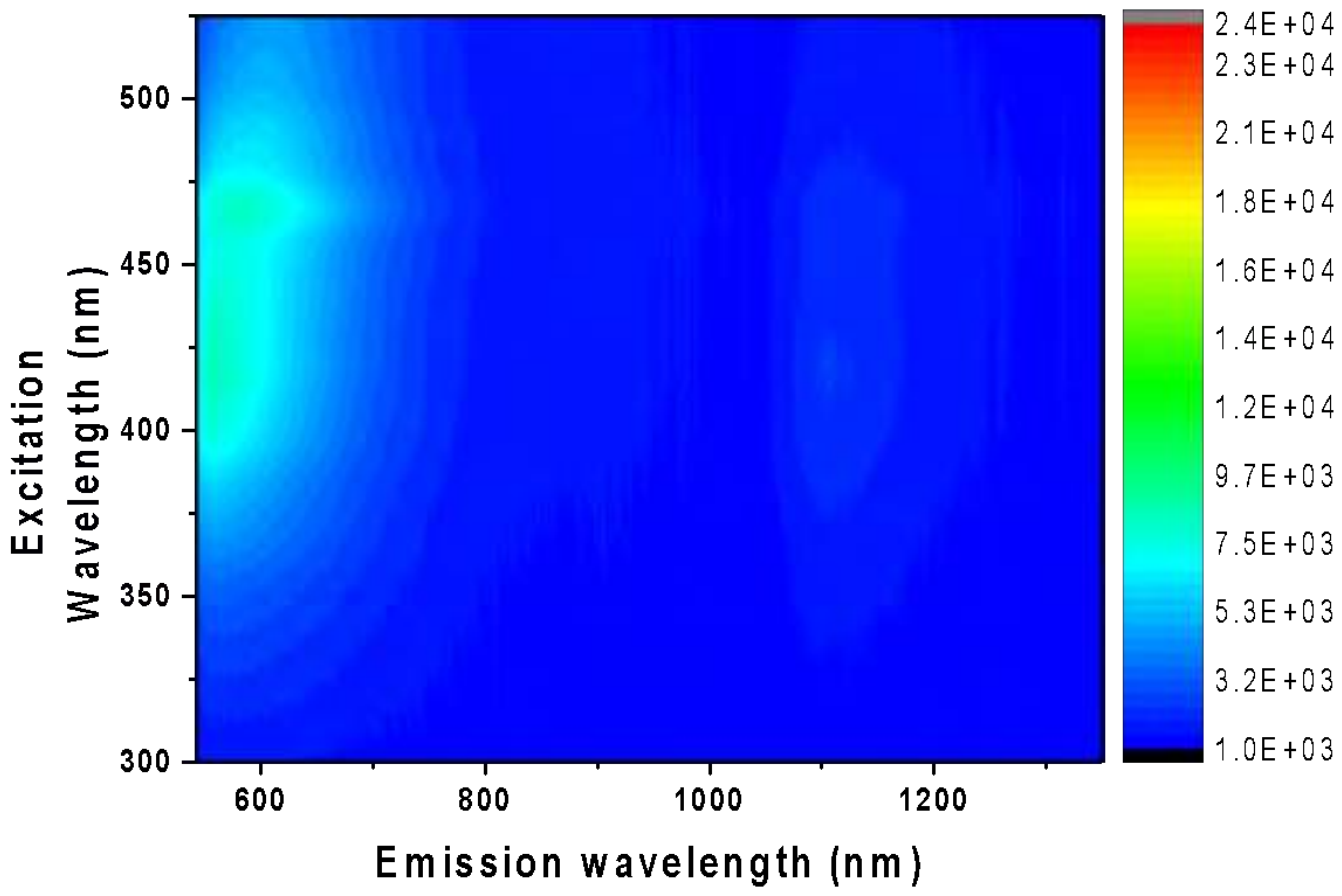

**Supplementary Figure 38.** Photoluminescence emission–excitation contour maps measured on pristine bmim-$BF_4$ under inert gas conditions (see Methods) measured with a 550 nm cut-off filter in emission. The observed weak emissions have been attributed to supramolecular aggregates (Bath et al. *Journal of Molecular Liquids*, 2013, *181*, 142–151 and references therein). Source data are provided as a Source Data file.

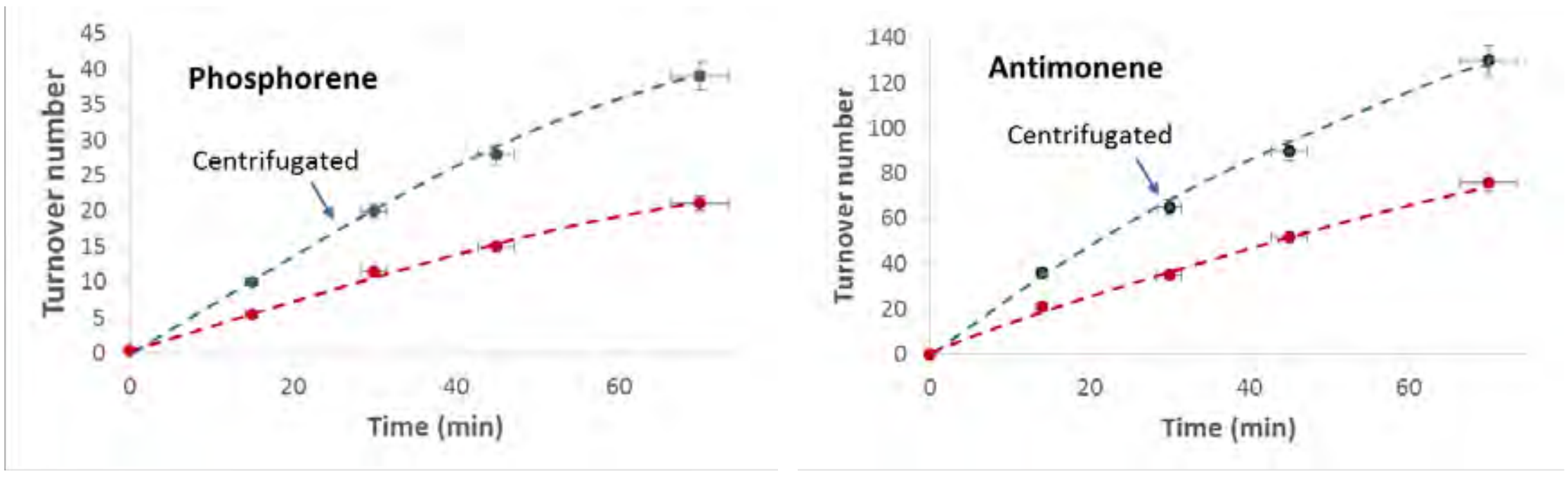

**Supplementary Figure 39.** Turnover number–time plot for FL–BP (left) and FL–Sb (right) before (red circles) and after centrifugation (blue circles). Error bars account for 5% uncertainty. Source data are provided as a Source Data file.

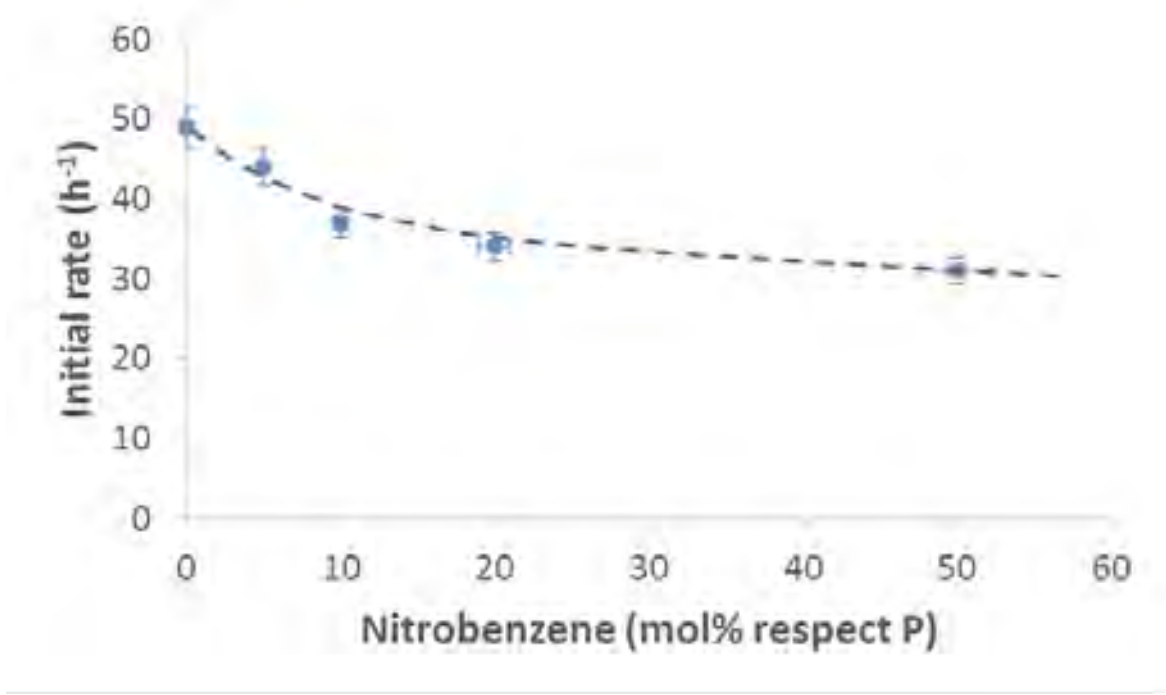

**Supplementary Figure 40.** Inhibition kinetics with nitrobenzene in the presence of FL–BP (10 mol%). Error bars account for 5% uncertainty. Source data are provided as a Source Data file.

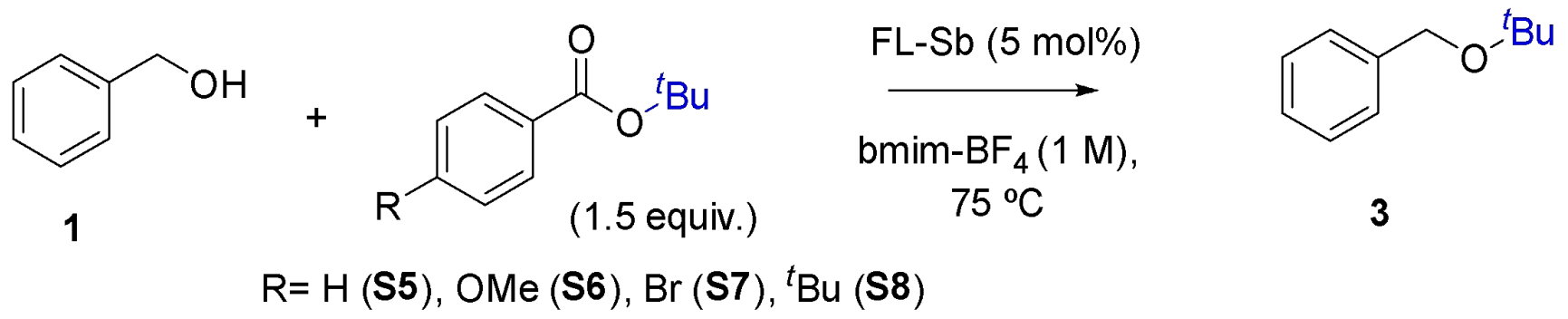

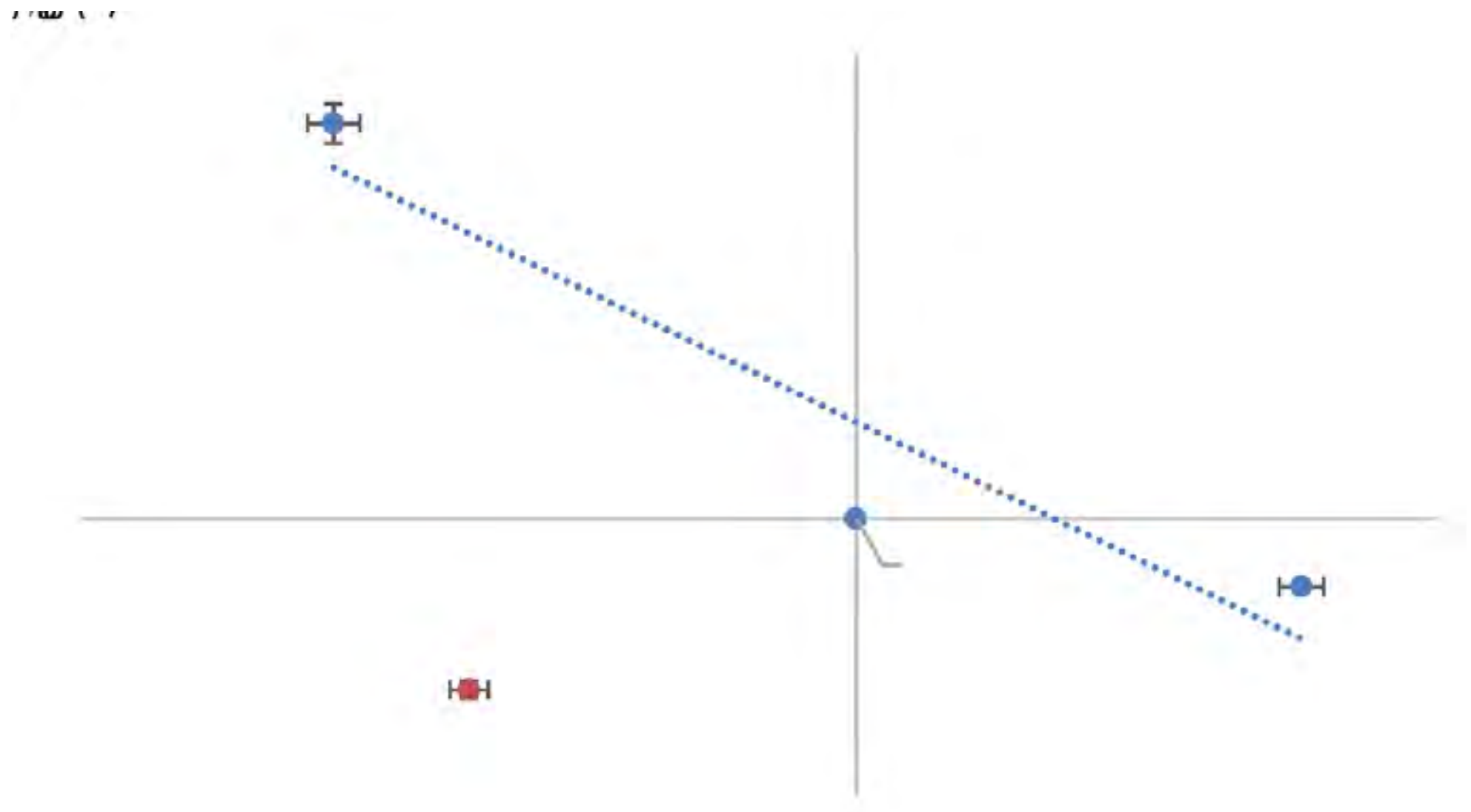

**Supplementary Figure 41.** Hammett plot for the tert-butylation of **1** with different benzoates under FL-Sb catalysis. Error bars account for 5% uncertainty. Source data are provided as a Source Data file.

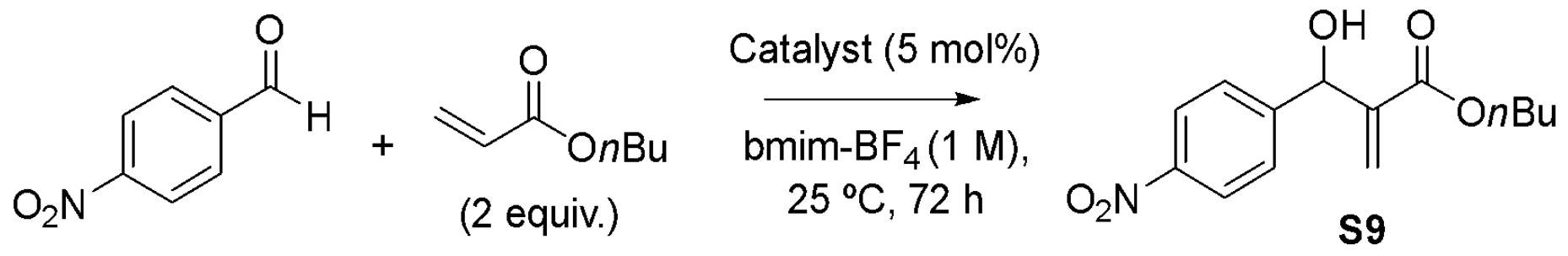

| FL-P / Fl-Sb | S9 (TON) |
| --- | --- |
| Non-centrifugated | 0.6 / 1.0 |
| Centrifugated | 1.8 / 2.3 |
| DABCO | 1.0 |
| Blank | 0 |

**Supplementary Figure 42.** Baylis-Hillman reaction catalysed by FL-P, FL-Sb or DABCO in IL.

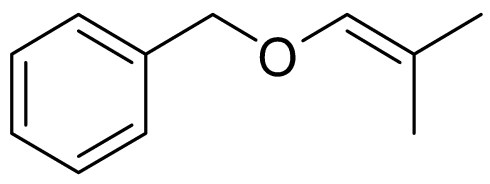

**Supplementary Table 1** Reported procedures to synthesize the allylic benzyl alcohol above (see references below).

| Entry | Reaction | Yield (%) |
|---|---|---|
| 1[1] | 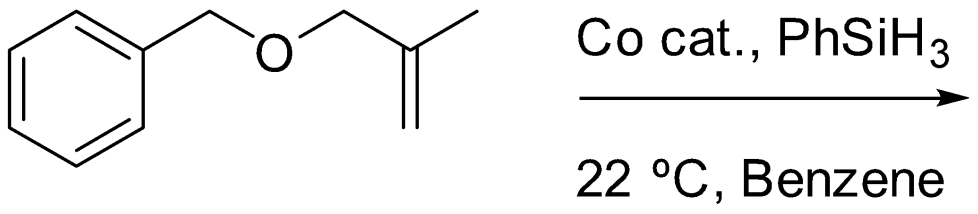  | 81 |
| 2[2] | 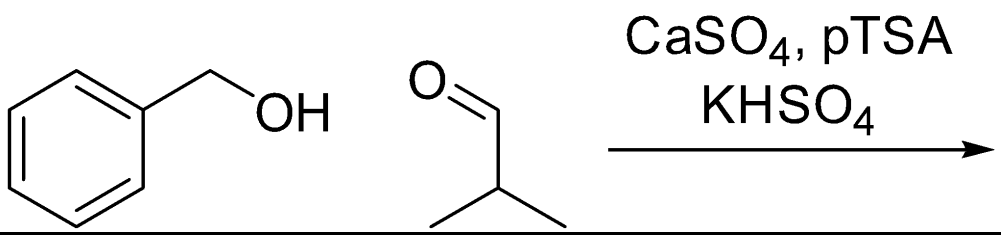  | - |
| 3[2] | 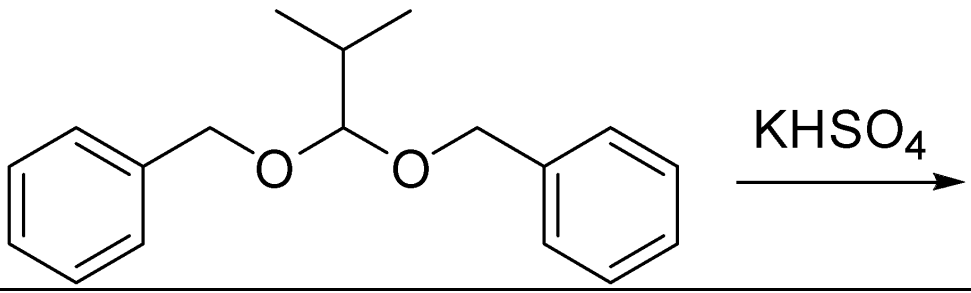  | 70 |
| 4[3] | 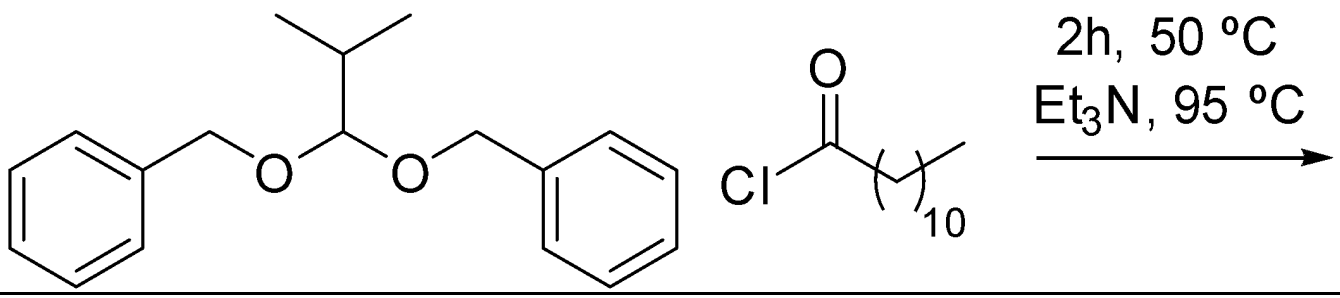  | 69 |
| 5[4] | 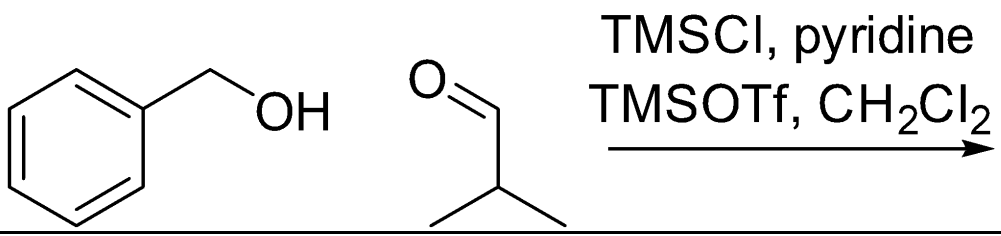  | - |
| 6[4] | 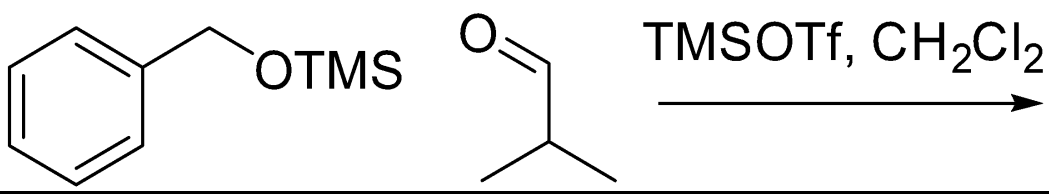  | 42 |
| 7[5] | 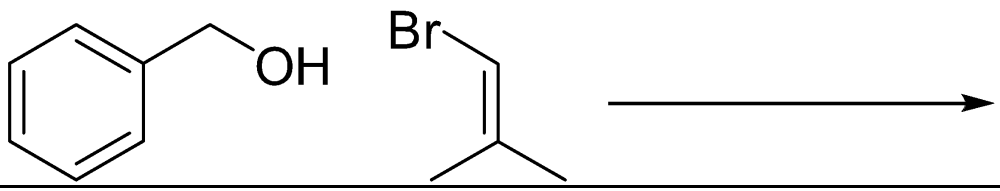  | - |
| 8[6] | 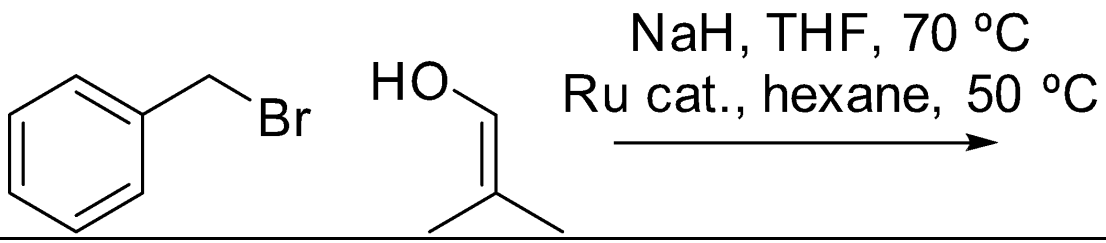  | - |
| 9[6] | 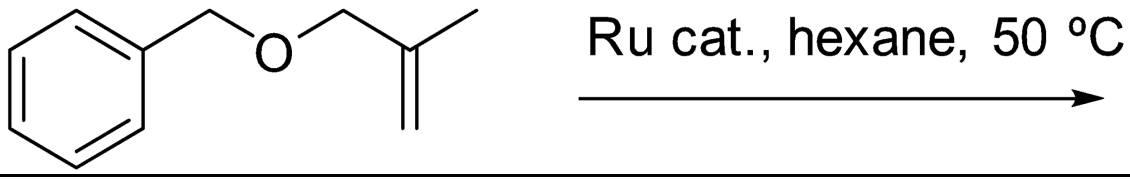  | 95 |
| 10[7] | 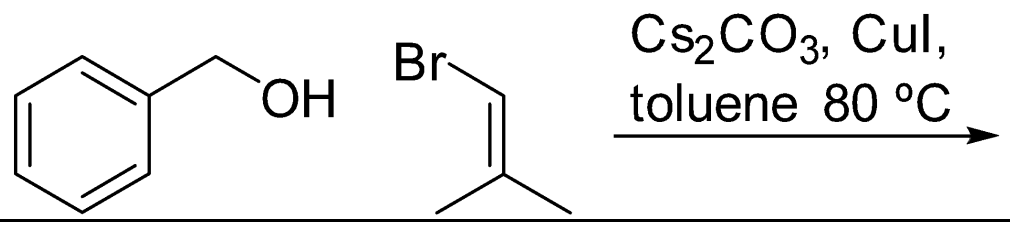  | 64 |
| 11[8] | 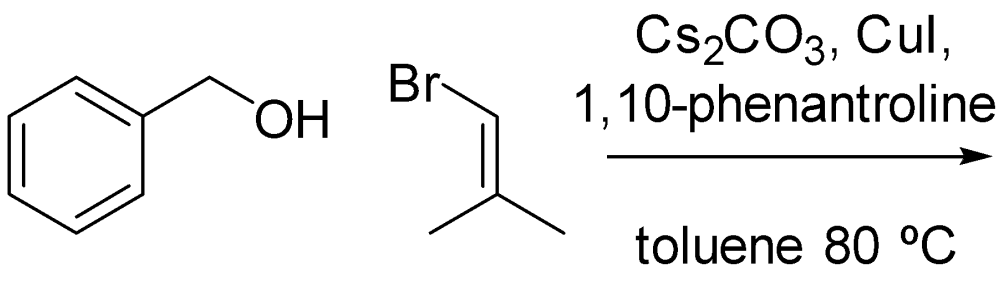  | 50 |
| 12[9] | 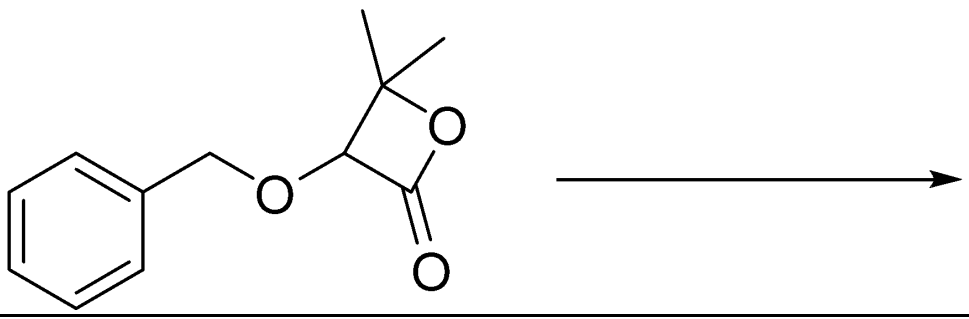 | 85 |

**Supplementary Table 2** Data showing the Taft parameters (Es and σ) for the substituents and log($k/k_{Me}$) for all the catalysts in competitive experiments. For reaction conditions see Table 1.

| Group | Es | σ | HOTf | FL-BP | FL-Sb |
|---|---|---|---|---|---|
| *H* | 1.24 | 0.49 | 0.182 | -0.118 | 0.343 |
| *Me* | 0 | 0 | 0 | 0 | 0 |
| *iPr* | -0.47 | -0.19 | 0.223 | -0.028 | 0.375 |
| *tBu* | -1.54 | -0.3 | -0.980 | -0.639 | -0.383 |
| *Ph* | -2.55 | 0.6 | -1.674 | n.m. | n.m. |

**Supplementary Table 3** Multiple regression analysis of the different catalysts in competitive experiments. S.D. = standard deviation. δ = steric effects. $\rho^*$ = polar effects.

| | Intercept | S.D. | δ | S.D. | ρ* | S.D. |
|---|---|---|---|---|---|---|
| *HOTf* | -0.213 | 0.163 | -0.680 | 0.479 | 2.280 | 1.578 |
| *FL-BP* | -0.038 | 0.029 | 0.825 | 0.084 | -2.229 | 0.278 |
| *FL-Sb* | -0.160 | 0.122 | 0.276 | 0.359 | 0.424 | 1.182 |

**Supplementary Table 4** Multiple regression analysis of the different catalysts in competitive experiments forcing the analysis to intercept the origin of coordinates (0, 0, 0). S.D. = standard deviation. δ = steric effects. $\rho^*$ = polar effects.

| | Intercept | S.D. | δ | S.D. | ρ* | S.D. |
|---|---|---|---|---|---|---|
| *HOTf* | 0 | -- | -0.325 | 0.460 | 1.156 | 1.545 |
| *FL-BP* | 0 | -- | 0.887 | 0.081 | -2.427 | 0.272 |
| *FL-Sb* | 0 | -- | 0.542 | 0.345 | -0.418 | 1.157 |

**Supplementary Table 5** Comparison between graphene and graphene-based catalysts with traditional catalyst in common reactions in organic chemistry. G = graphene; GO = graphene oxide; rGO = reduced graphene oxide; (X)G = graphene doped with X).

| Entry | Reaction | Catalyst | Temperature | Yield (%) |
|---|---|---|---|---|
| 1[10] | 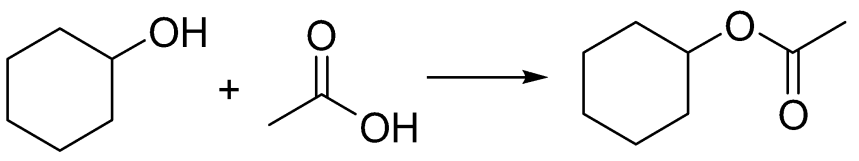 | G-SO3H | 100 ºC | 79.5 |
| 2[10] | | GO | 100 ºC | 11.5 |
| 3[10] | | rGO | 100 ºC | 10.2 |
| 4[11] | | Fe2O3/C | 116 ºC | 97 |
| 5[10] | 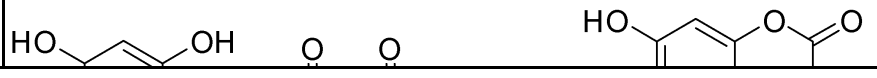 | G-SO3H | 110 ºC | 82.1 |
| 6[10] | | GO | 110 ºC | 12.3 |
| 7[10] | | rGO | 110 ºC | 11.1 |
| 8[12] | | H2SO4 | 25 ºC | 97 |
| 9[10] | 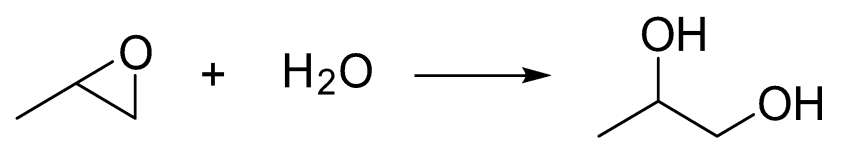 | G-SO3H | 27 ºC | 66.8 |
| 10[10] | | GO | 27 ºC | 0 |
| 11[10] | | rGO | 27 ºC | 0 |
| 12[13] | | Co-salen | 40 ºC | 97 |
| 13[14] | 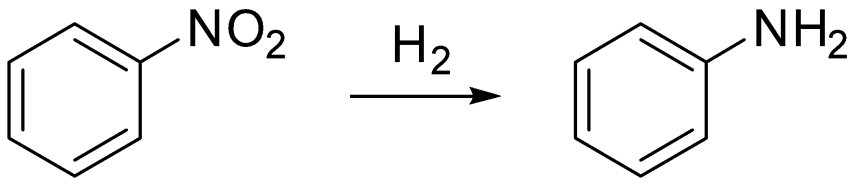 | rGO | 100 ºC | 94.2 |
| 14[14] | | GO | 100 ºC | 91.1 |
| 15[15] | | Pd | 25 ºC | 100 |
| 16[16] | Selective acetylene hydrogenation | G | 110 ºC | 81 |
| 17[16] | | rGO | 110 ºC | 87.5 |
| 18[17] | | PdIn/Al2O3 | 60 ºC | 95 |
| 19[18] | 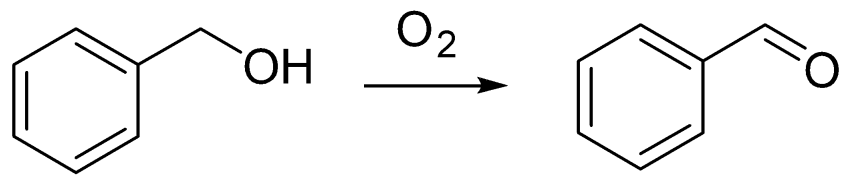 | GO 200 wt% | 100 ºC | 92 |
| 20[19] | | Pt/TiO2 | 25 ºC | 76.7 |
| 21[18] | 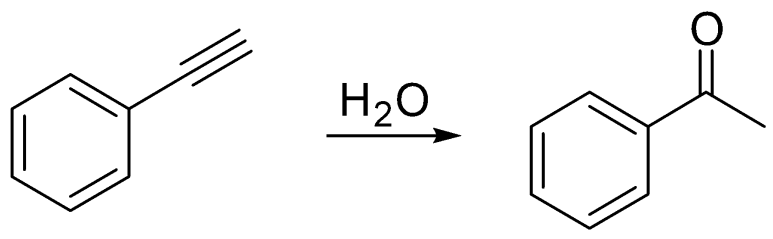 | GO 200 wt% | 100 ºC | 98 |
| 22[20] | | TfOH | 25 ºC | 100 |
| 23[21] | 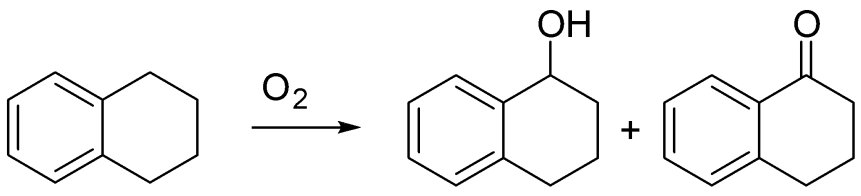 24 h, 1 bar O2 | - | 120 ºC | 8.2 |
| 24[21] | | GO | 120 ºC | 18.4 |
| 25[21] | | G | 120 ºC | 14.7 |
| 26[21] | | (N)G | 120 ºC | 34.0 |
| 27[21] | | (B)G | 120 ºC | 31.1 |
| 28[21] | | (B,N)G | 120 ºC | 40 |

**Supplementary Note 1**

**Compound characterization.**

**(*Tert*–butoxymethyl)benzene 3**. IR (ν, $cm^{-1}$): 2975, 2867, 1455, 1363, 1197, 1068, 913, 745. $^1H$ NMR (™, ppm; J, Hz): 7.27 (5H, m), 4.38 (2H, s), 1.22 (9H, s). $^{13}C$ NMR (™, ppm; J, Hz): 139.9 (C), 128.3 (2 x CH), 127.4 (2 x CH), 127.1 (CH), 73.4 (C), 64.1 ($CH_2$), 27.7 (3 x $CH_3$). HRMS (ESI) [$M+H^+$, major peak; calculated for $C_{11}H_{16}O$: 165.1280] found 165.1282 *m/z*.

**1–(*Tert*–butoxymethyl)–2–iodobenzene 4**. IR (ν, $cm^{-1}$): 3062, 2973, 2869, 1468, 1363, 1195, 1085, 1012, 746. $^1H$ NMR (™, ppm; J, Hz): 7.71 (1H, d, J = 7.9), 7.43 (1H, d, J = 7.7), 7.27 (1H, t, J = 7.5), 6.88 (1H, t, 7.6). $^{13}C$ NMR (™, ppm; J, Hz): 142.0 (C), 138.8 (CH), 128.7 (CH), 128.6 (CH), 128.2 (CH), 97.3 (C), 73.8 (C), 68.4 ($CH_2$), 27.7 (3 x $CH_3$). HRMS (ESI) [$M^+$, major peak; calculated for $C_{11}H_{15}IO$: 290.0168] found 290.0163 *m/z*.

**((*E*)–3–(((*E*)–3–Iodobut–2–en–1–yl)oxy)prop–1–en–1–yl)benzene 5**. IR (ν, $cm^{-1}$): 3457, 2921, 1677, 1626, 1125, 974, 750, 689. $^1H$ NMR (™, ppm; J, Hz): 7.36–7.16 (5H, m), 6.54 (1H, d, J = 15.9), 6.31 (1H, t, J = 6.9), 6.20 (1H, dt, J = 15.9, 6.1), 4.07 (2H, dd, J = 6.1, 1.3), 3.91 (2H, d, J = 6.9), 2.95 (3H, s). $^{13}C$ NMR (™, ppm; J, Hz): 193.7 (C), 152.7 (C), 129.1 (CH), 128.6 (CH), 128.5 (2 x CH), 127.8 (CH), 126.5 (2 x CH), 125.6 (CH), 70.7 ($CH_2$), 66.6 ($CH_2$), 28.2 ($CH_3$). HRMS (ESI) [M+, major peak; calculated for $C_{13}H_{15}IO$: 314.0168] found 314.0171 *m/z*.

**9–((1phenylethyl)thio)nonan–1–ol, 6**. $^1H$ NMR (™, ppm; J, Hz): 7.21 (5H, m), 3.87 (1H, q, J = 7.1), 3.21 (2H, t, J = 6.7), 2.22 (2H, m), 1.61–1.45 (8H, m), 1.36 (3H, d, 6.5), 1.30–1.10 (6H, m). $^{13}C$ NMR (™, ppm; J, Hz): 144.3 (C), 128.3 (2 x CH), 127.2 (CH), 126.1 (2 x CH), 68.8 ($CH_2$), 44.1 (CH), 39.2 ($CH_2$), 31.3 ($CH_2$), 29.9 ($CH_2$), 29.3 (2 x $CH_2$), 29.1 ($CH_2$), 28.9 ($CH_2$), 26.1 ($CH_2$), 24.2 ($CH_3$).

**4–(*Tert*–butylthio)phenol 7**. IR (ν, $cm^{-1}$): 3378, 2965, 1598, 1494, 1363, 1166, 832, 529. $^1H$ NMR (™, ppm; J, Hz): 7.32 (2H, d, J = 8.7), 6.72 (2H, d, J = 8.7), 1.18 (9H, s). $^{13}C$ NMR (™, ppm; J, Hz): 156.4 (C), 139.1 (2 x CH), 123.7 (C), 115.5 (2 x CH), 45.6

(C), 30.7 (3 x $CH_3$). HRMS (ESI) [M–H$^+$, major peak; calculated for $C_{10}H_{13}OS$: 181.0687] found 181.0680 *m/z.*

**3–(*Tert*–butyl)–1*H*–indole 8a**. IR (ν, cm$^{-1}$): 3417, 3063, 2962, 1702, 1461, 1361, 743. $^1$H NMR (™, ppm; J, Hz): 7.79 (1H, s), 7.75 (1H, d, J = 7.8), 7.29 (1H, d, J = 8.1), 7.10 (1H, t, J = 7.5), 7.02 (1H, t, J = 7.5), 6.87 (1H, s), 1.78 (2H, s), 1.39 (9H, s). $^{13}$C NMR (™, ppm; J, Hz): 137.2 (C), 126.8 (C), 126.0 (C), 121.4 (CH), 121.3 (CH), 119.2 (CH), 118.8 (CH), 111.3 (CH), 31.6 (C), 30.7 (3 x $CH_3$). HRMS (ESI) [M $^+$, major peak; calculated for $C_{12}H_{16}N$: 174.1283] found 174.1281 *m/z*.

**1,3–Di–*tert*–butyl–1*H*–indole 8b**. IR (ν, cm$^{-1}$): 2963, 1711, 1481, 1222, 914, 743. $^1$H NMR (™, ppm; J, Hz): 7.75 (1H, d, J = 7.8), 7.54 (1H, d, J = 8.4), 7.06 (1H, t, J = 7.6), 6.98 (1H, t, J = 8.0), 6.91 (1H, s), 1.64 (9H, s), 1.35 (9H, s). HRMS (ESI) [M $^+$, major peak; calculated for $C_{16}H_{24}N$: 230.1909] found 230.1901 *m/z*.

**3–(*Tert*–butyl)–5–methoxy–1*H*–indole 9a**. IR (ν, cm$^{-1}$): 3417, 2962, 1702, 1482, 1224, 1042, 796. $^1$H NMR (™, ppm; J, Hz): 7.68 (1H, s), 7.18 (2H, m), 6.86 (1H, s), 6.78 (1H, dd, J = 8.8, 2.4), 3.81 (3H, s), 1.37 (9H, s). $^{13}$C NMR (™, ppm; J, Hz): 153.2 (C), 132.5 (C), 126.5 (C), 126.3 (C), 120.2 (CH), 111.8 (CH), 111.2 (CH), 104.1 (CH), 56.1 ($CH_3$), 31.5 (C), 30.6 (3 x $CH_3$). HRMS (ESI) [M+H $^+$, major peak; calculated for $C_{13}H_{17}NO$: 204.1389] found 204.1385 *m/z*.

**1,3–Di–*tert*–butyl–5–methoxy–1*H*–indole 9b**. IR (ν, cm$^{-1}$): 2966, 1709, 1479, 1227, 1038, 668. $^1$H NMR (™, ppm; J, Hz): 7.43 (1H, d, J = 9.1), 7.19 (1H, s), 6.88 (1H, s), 6.74 (1H, dd, J = 9.1, 2.6), 3.80 (3H, s), 1.61 (9H, s), 1.36 (9H, s). HRMS (ESI) [M+H $^+$, major peak; calculated for $C_{17}H_{25}NO$: 260.2014] found 260.2003 *m/z*.

***N*–(2–(5–(*tert*–butoxy)pentanoyl)–4,5–dimethoxyphenethyl)–2,2,2–trifluoroacetamide 11**. IR (ν, cm$^{-1}$): 3325, 2968, 1720, 1518, 1360, 1198. $^1$H NMR (™, ppm; J, Hz): 7.13 (1H, s), 6.66 (1H, s). HRMS (ESI) [M+H$^+$, major peak; calculated for $C_{21}H_{31}FNO_5$: 434.2155] found 434.2159 *m/z*.

**(R)–1–phenylethyl acetate 13**. $^1$H NMR (™, ppm; J, Hz): 7.20 (5H, m), 5.81 (1H, q, J = 6.6), 1.98 (3H, s), 1.45 (3H, d, J = 6.6). $^{13}$C NMR (™, ppm; J, Hz): 170.3 (C), 141.7 (C), 128.5 (2 x CH), 127.9 (CH), 126.1 (2 x CH), 72.3 (CH), 22.2 ($CH_3$), 21.3 ($CH_3$).

**1-(*tert*-butoxymethyl)-4-methylbenzene S1**. IR (ν, $cm^{-1}$): 2974, 2926, 2867, 1517, 1472, 1389, 1362, 1197, 1082, 1021, 894, 802, 474. GC-MS [$M^+$; calculated for $C_{12}H_{18}O$: 178.1358] (m/z, $M^+$ 178.2), major peaks found (m/z, relative intensity): 57.2 (10%), 77.2 (15%), 93.2 (17%), 105.2 (100%), 163.2 (6%), 178.2 (7%). $^1$H NMR (™, ppm; J, Hz): 7.15 (2H, d, 7.5), 7.05 (2H, d, 7.8), 4.32 (2H, s), 2.25 (2H, s), 1.21 (9H, s). $^{13}$C NMR (™, ppm; J, Hz): 136.9 (C), 136.7 (C), 129.0 (2 x CH), 127.5 (2 x CH), 73.3 (C), 64.0 ($CH_2$), 27.7 (3 x $CH_3$), 21.1 ($CH_3$).

**1-(*tert*-butoxymethyl)-4-isopropylbenzene S2**. IR (ν, $cm^{-1}$): 2968, 2930, 2871, 1514, 1467, 1388, 1362, 1235, 1196, 1082, 1019, 895, 818, 542. GC-MS [$M^+$; calculated for $C_{14}H_{22}O$: 206.1671] (m/z, $M^+$ 206.2), major peaks found: 43.2 (11%), 57.3 (13%), 79.2 (12%), 87.3 (12%), 107.2 (23%), 117.2 (19%), 133.2 (100%), 163.2 (8%), 206.2 (9%). $^1$H NMR (™, ppm; J, Hz): 7.19 (2H, d, 8.0), 7.10 (2H, d, 8.1), 4.32 (2H, s), 2.80 (1H, m), 1.21 (9H, s), 1.15 (6H, d, 6.9). $^{13}$C NMR (™, ppm; J, Hz): 147.8 (C), 137.2 (C), 127.6 (2 x CH), 126.4 (2 x CH), 73.3 (C), 64.0 ($CH_2$), 33.9 (CH), 27.7 (3 x $CH_3$), 24.1 (2 x $CH_3$).

**1-(*tert*-butoxymethyl)-4-(*tert*-butyl)benzene S3**. IR (ν, $cm^{-1}$): 2968, 2904, 2868, 1516, 1473, 1389, 1362, 1268. 1197, 1082, 1019, 896, 815, 568. GC-MS [$M^+$; calculated for $C_{15}H_{24}O$: 220.1827] (m/z, $M^+$ 220.3), major peaks found: 57.3 (31%), 91.2 (17%), 107.2 (30%), 117.2 (19%), 132.2 (13%), 147.2 (100%), 149.2 (30%), 163.2 (19%), 205.2 (8%), 220.3 (8%). $^1$H NMR (™, ppm; J, Hz): 7.27 (2H, d, 8.3), 7.20 (2H, d, 8.3), 4.33 (2H, s), 1.23 (9H, s), 1.20 (9H, s). $^{13}$C NMR (™, ppm; J, Hz): 150.0 (C), 136.8 (C), 127.3 (2 x CH), 125.2 (2 x CH), 73.3 (C), 64.0 ($CH_2$), 34.5 (C), 31.4 (3 x $CH_3$), 27.7 (3 x $CH_3$).

***tert*-butyl 4-methoxybenzoate S6**. $^1$H NMR (™, ppm; J, Hz): 7.86 (2H, d, 9.0), 6.82 (2H, d, 9.0), 3.77 (3H, s), 1.50 (9H, s). $^{13}$C NMR (™, ppm; J, Hz): 165.6 (C), 163.0 (C), 131.4 (2 x CH), 124.6 (C), 113.4 (2 x CH), 80.5 (C), 35.0 (C), 55.4 ($CH_3$), 28.3 (3 x $CH_3$).

***tert*-butyl 4-bromobenzoate S7**. $^1$H NMR (™, ppm; J, Hz): 7.86 (2H, d, 8.5), 7.56 (2H, d, 8.5), 1.60 (9H, s). $^{13}$C NMR (™, ppm; J, Hz): 165.0 (C), 131.5 (2 x CH), 131.0 (2 x CH), 130.9 (C), 127.4 (C), 81.4 (C), 28.2 (3 x $CH_3$).

***tert*-butyl 4-(*tert*-butyl)benzoate S8**. [1]H NMR (™, ppm; J, Hz): 7.85 (2H, d, 8.4), 7.36 (2H, d, 8.4), 1.51 (9H, s), 1.26 (9H, s). [13]C NMR (™, ppm; J, Hz): 165.8 (C), 156.0 (C), 129.3 (C), 129.2 (2 x CH), 125.1 (2 x CH), 80.6 (C), 35.0 (C), 31.1 (3 x $CH_3$), 28.2 (3 x $CH_3$).

**Butyl 2-(hydroxy(4-nitrophenyl)methyl)acrylate S9**. GC-MS [$M^+$; calculated for $C_{14}H_{17}NO_5$: 279.2920] (m/z, $M^+$ ), major peaks found: 279.1 (8%), 223.1 (27%), 206.1 (100%). [1]H NMR (™, ppm; J, Hz): 8.00 (4H, m), 6.32 (1H, s), 5.77 (1H, s), 5.55 (1H, s), 4.07 (2H, t, 6.6), 2.10 (1H, s), 1.56 (2H, m), 1.18 (2H, m), 0.84 (3H, t, 7.4).

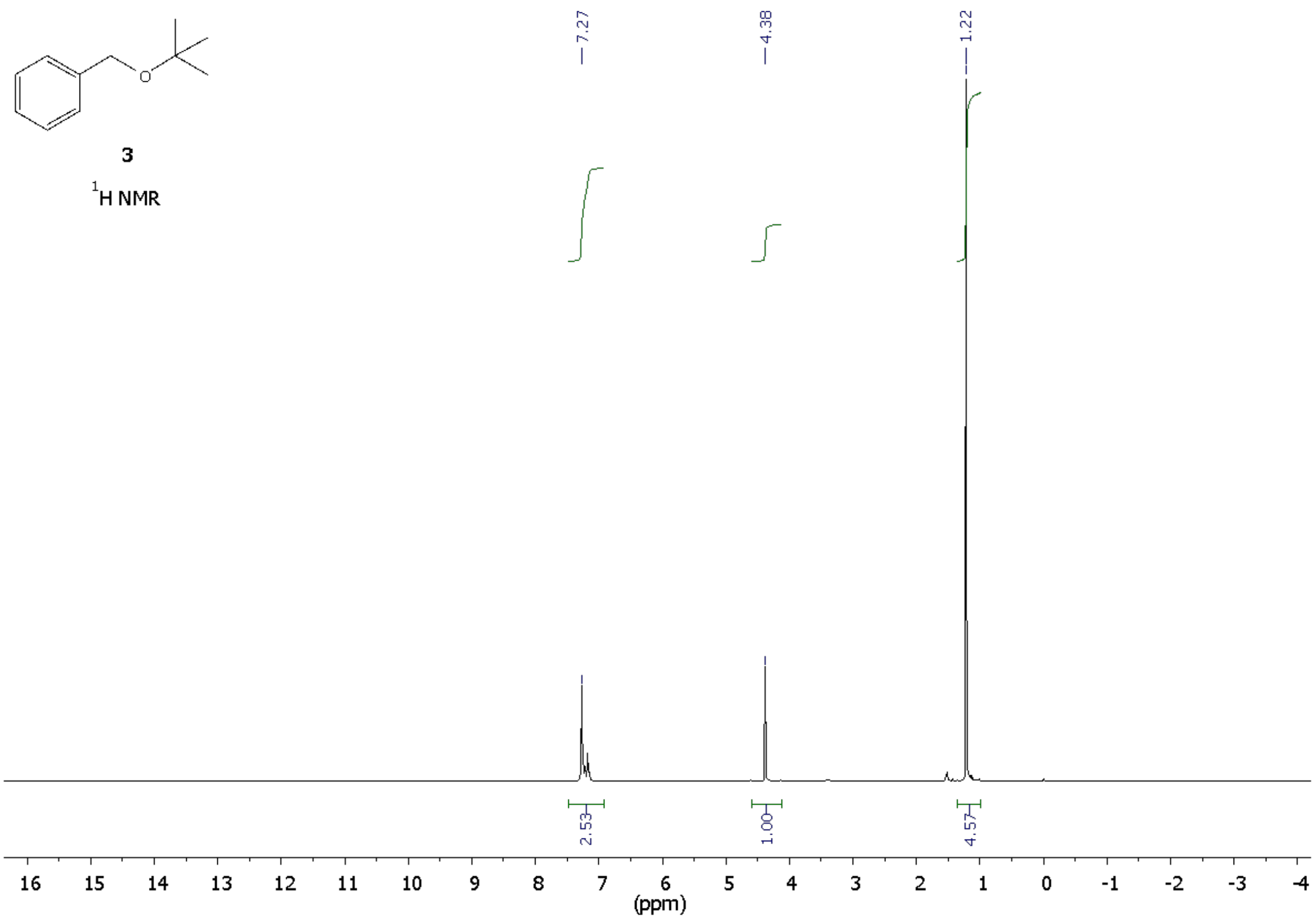

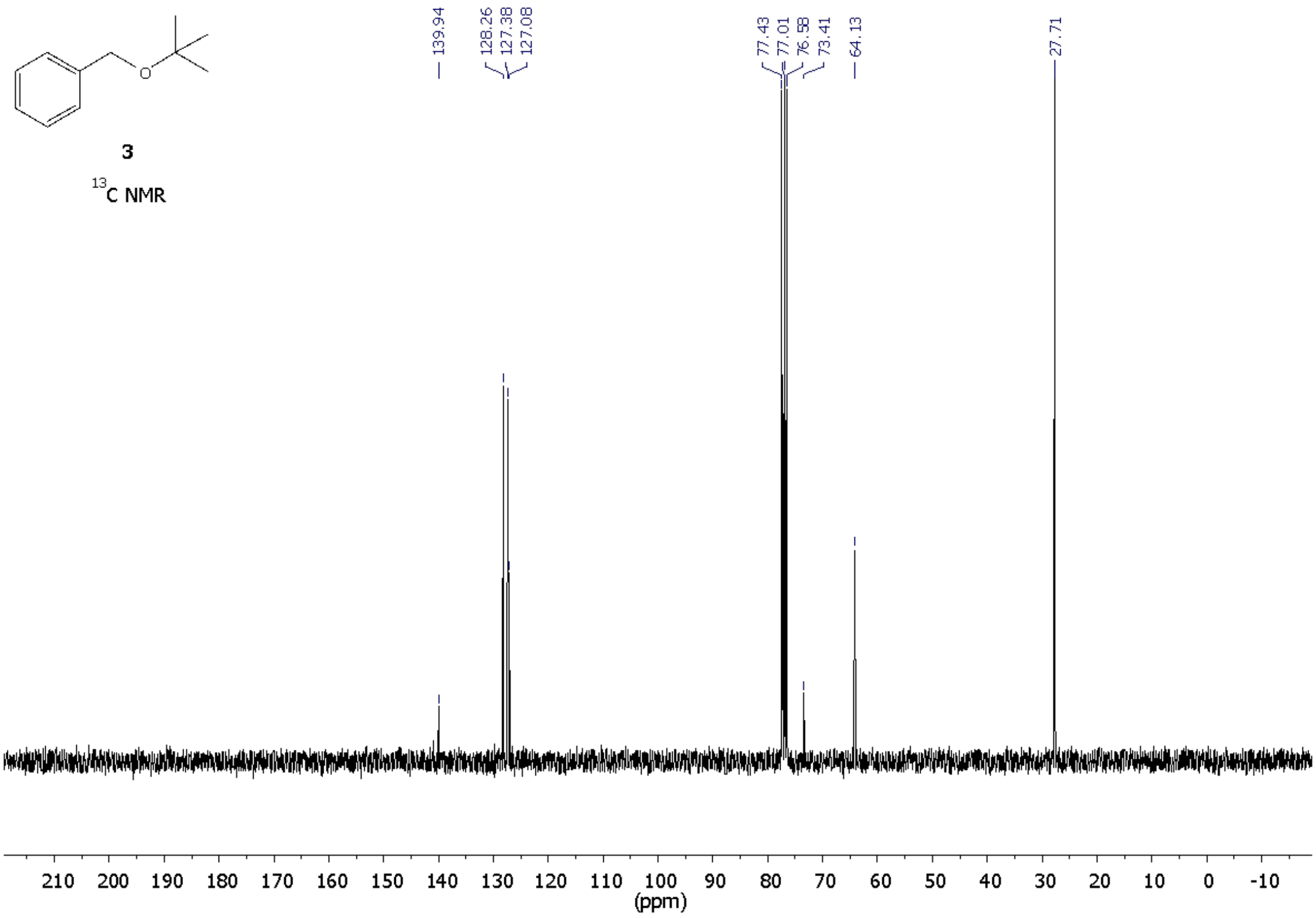
3
$^{13}$C NMR
139.94
128.26
127.38
127.08
77.43
77.01
76.58
73.41
64.13
27.71
210 200 190 180 170 160 150 140 130 120 110 100 90 80 70 60 50 40 30 20 10 0 -10
(ppm)

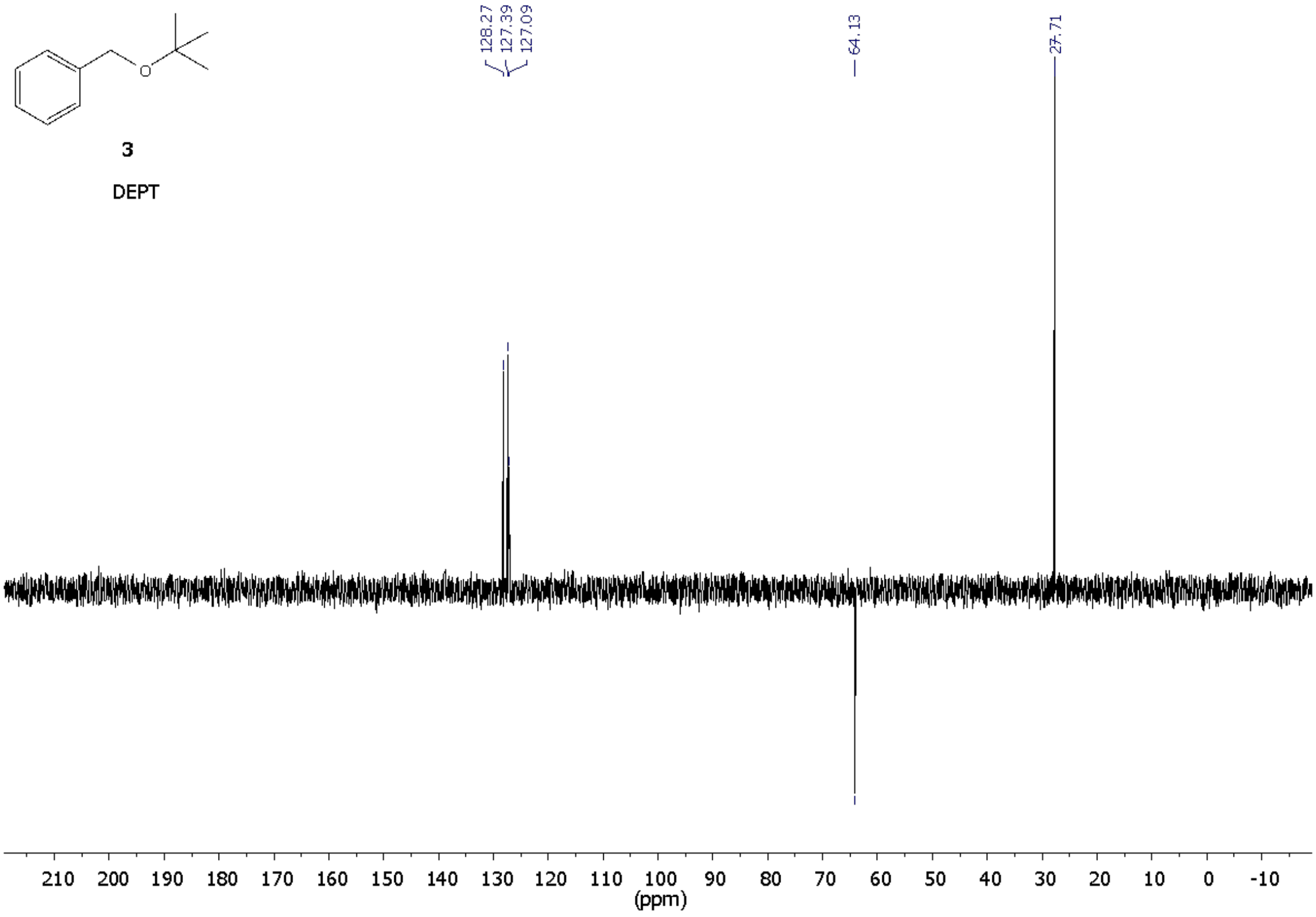
3
DEPT
128.27
127.39
127.09
64.13
27.71
210 200 190 180 170 160 150 140 130 120 110 100 90 80 70 60 50 40 30 20 10 0 -10
(ppm)

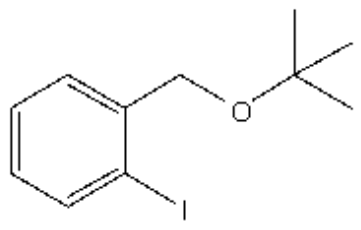

**4**
$^{1}$H NMR

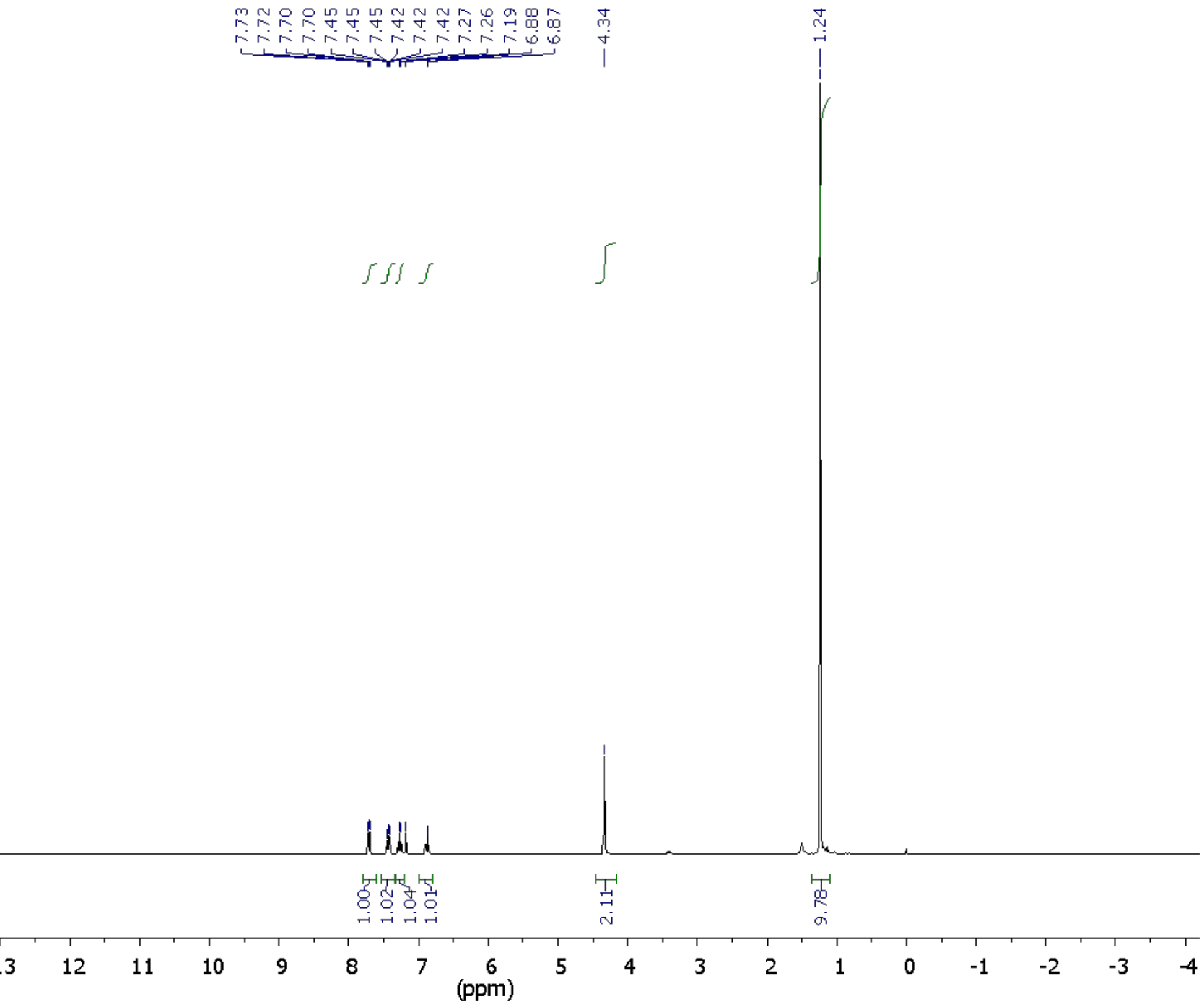

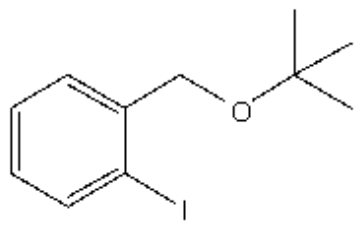

**4**
$^{13}$C NMR

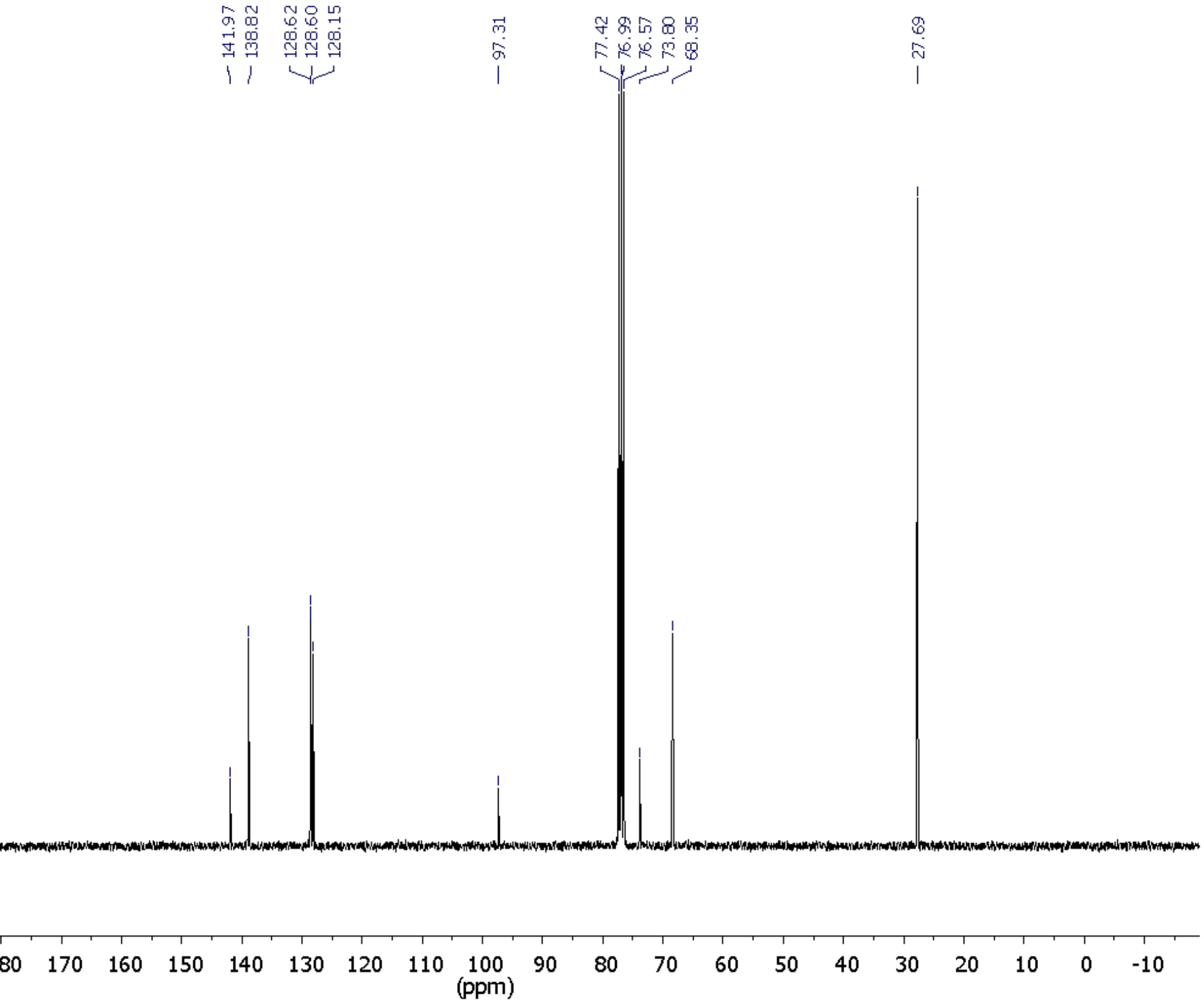

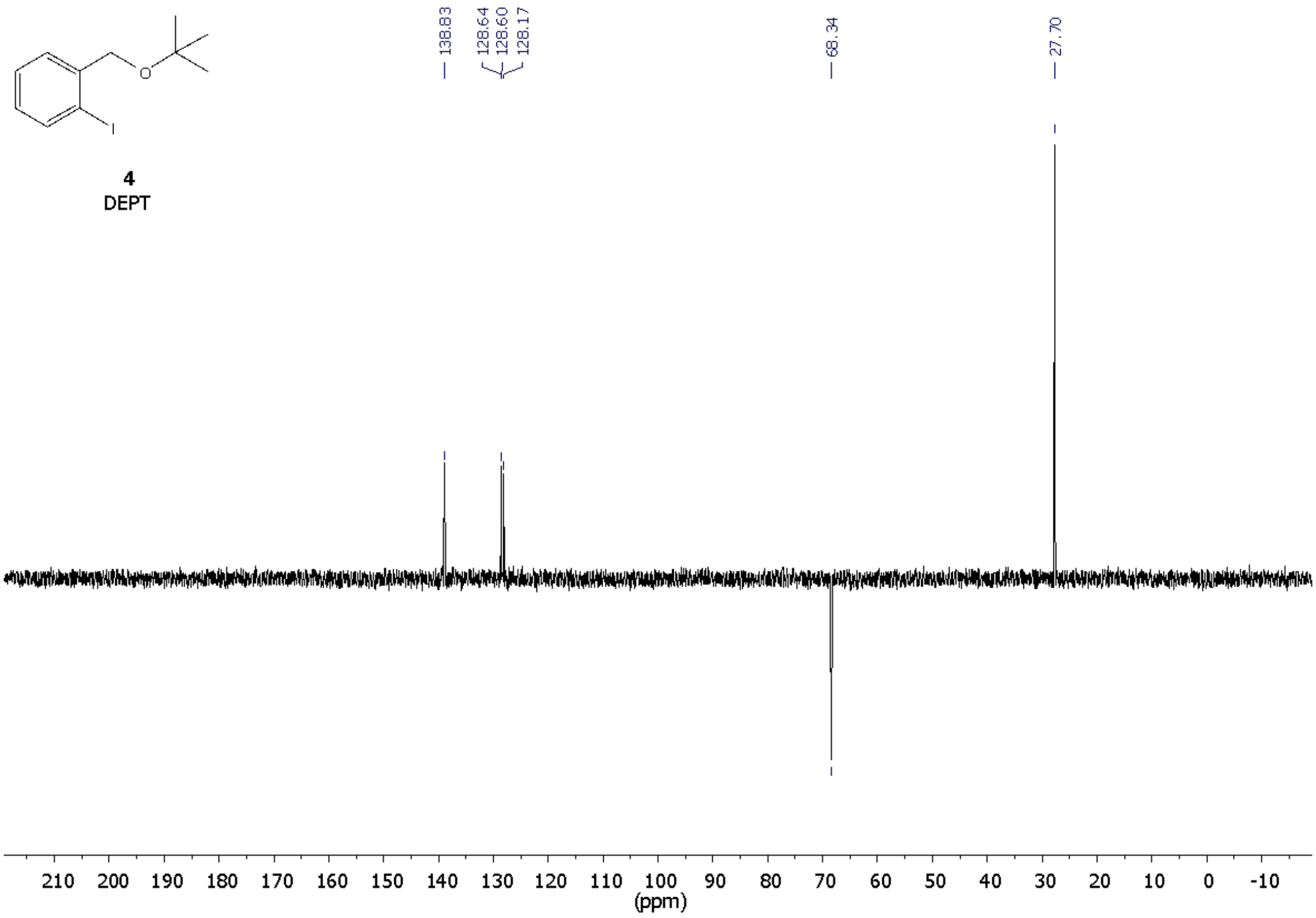
4
DEPT
138.83
128.64
128.60
128.17
68.34
27.70
210 200 190 180 170 160 150 140 130 120 110 100 90 80 70 60 50 40 30 20 10 0 -10
(ppm)

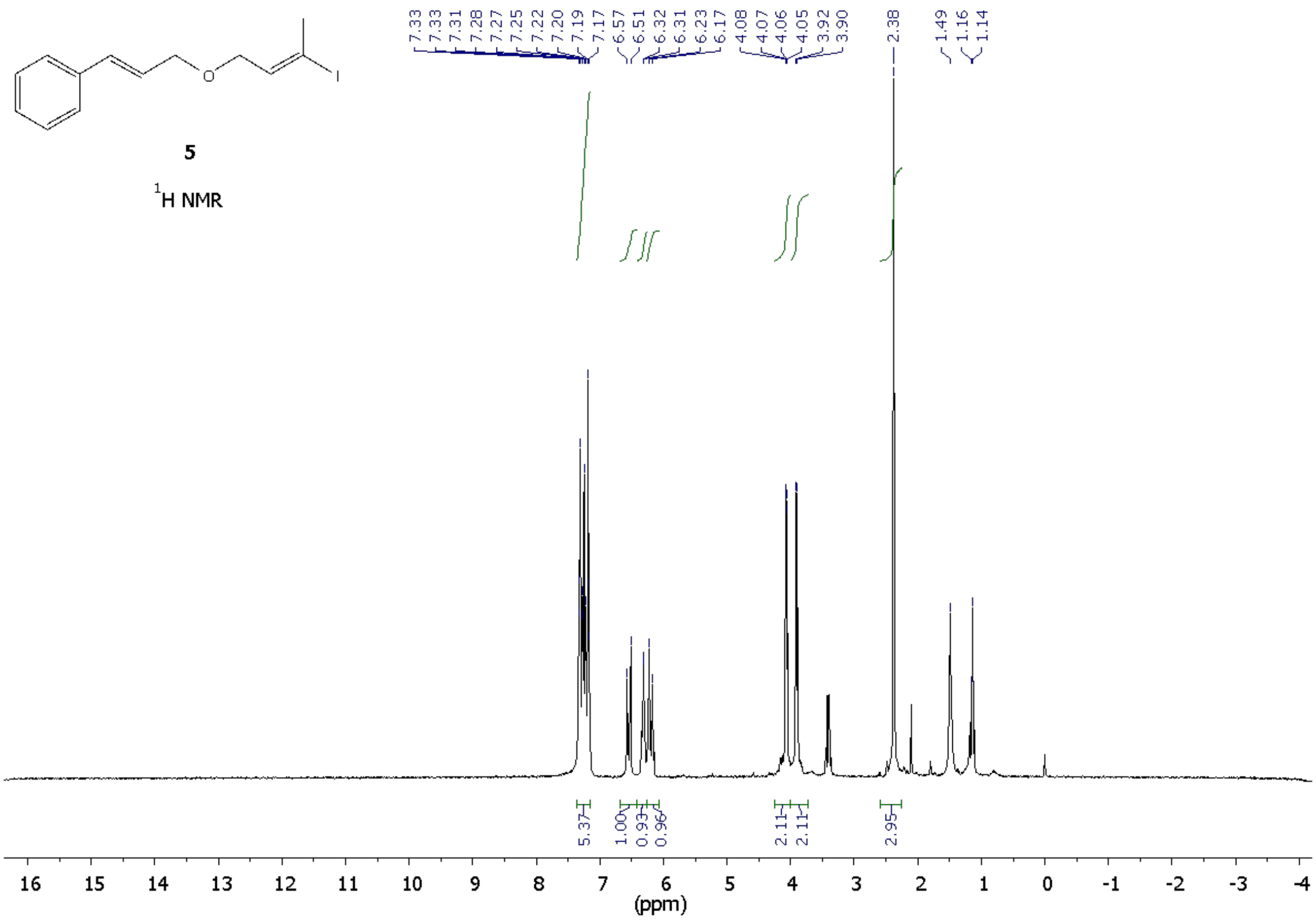
5
1H NMR
7.33
7.33
7.31
7.28
7.27
7.25
7.22
7.20
7.19
7.17
6.57
6.51
6.32
6.31
6.23
6.17
4.08
4.07
4.06
4.05
3.92
3.90
2.38
1.49
1.16
1.14
5.37
1.00
0.93
0.96
2.11
2.11
2.95
16 15 14 13 12 11 10 9 8 7 6 5 4 3 2 1 0 -1 -2 -3 -4
(ppm)

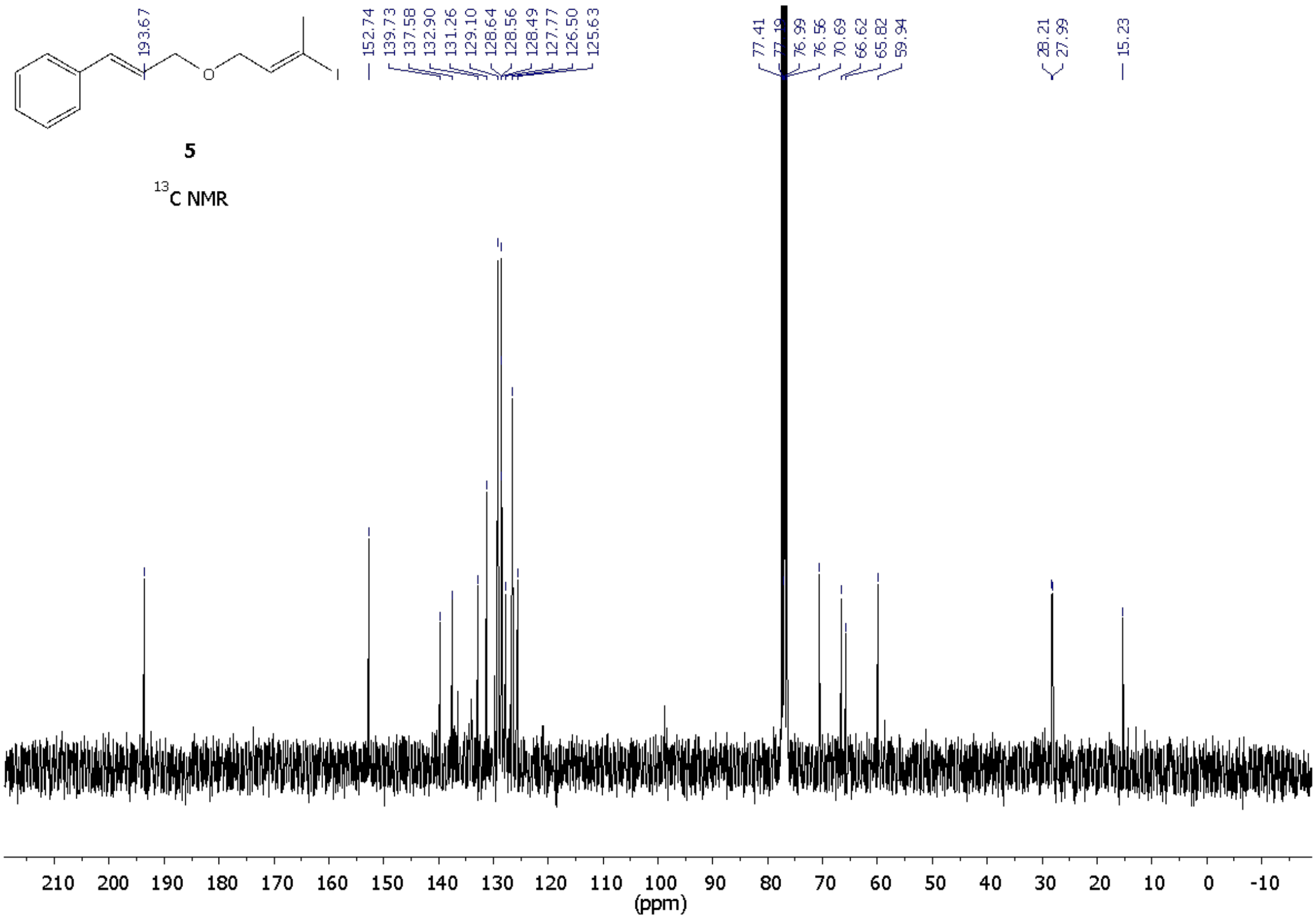
193.67
152.74 139.73 137.58 132.90 131.26 129.10 128.64 128.56 128.49 127.77 126.50 125.63
77.41 77.19 76.99 76.56 70.69 66.62 65.82 59.94
28.21 27.99
15.23
5
13C NMR
210 200 190 180 170 160 150 140 130 120 110 100 90 80 70 60 50 40 30 20 10 0 -10
(ppm)

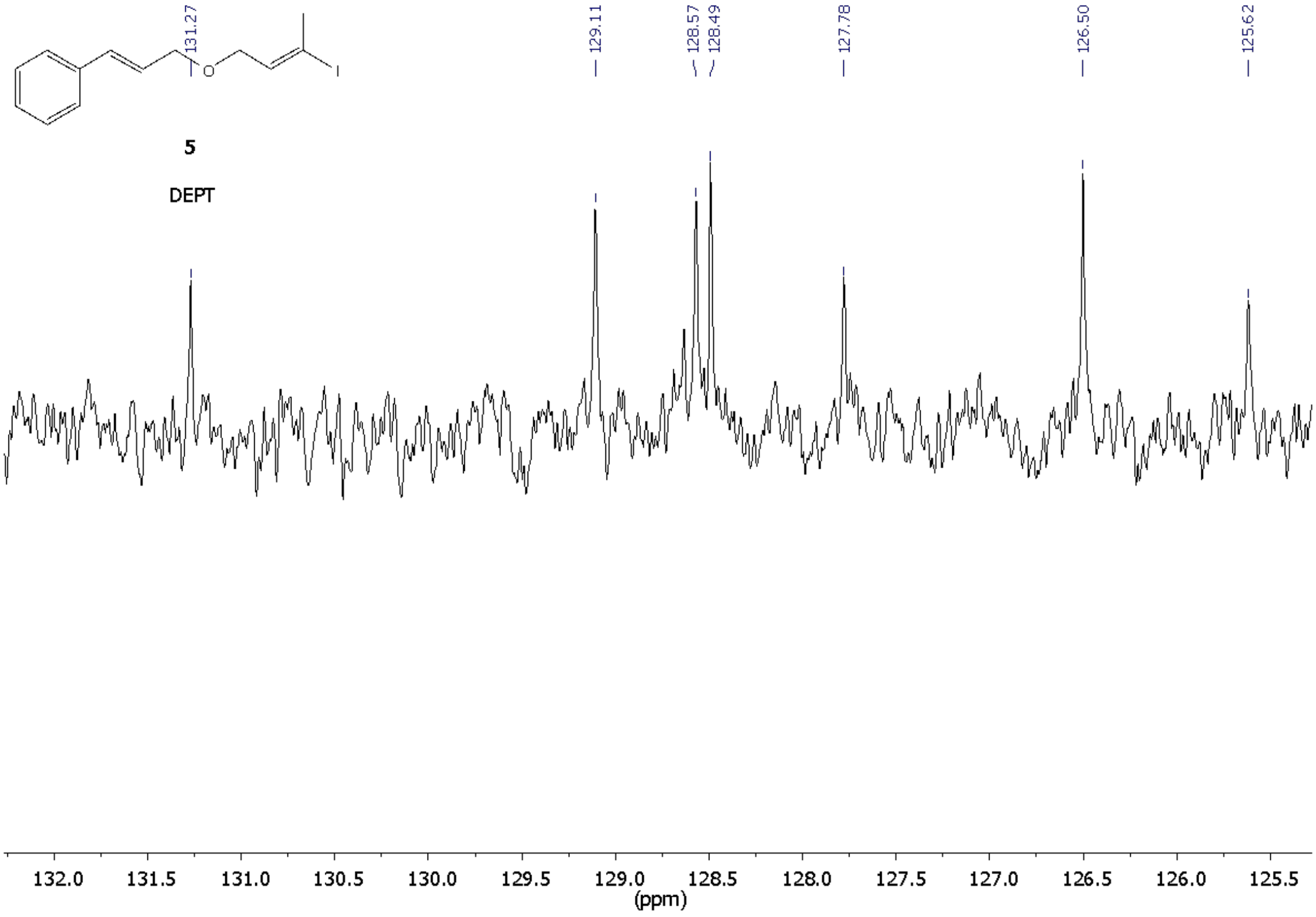
131.27
129.11
128.57 128.49
127.78
126.50
125.62
5
DEPT
132.0 131.5 131.0 130.5 130.0 129.5 129.0 128.5 128.0 127.5 127.0 126.5 126.0 125.5
(ppm)

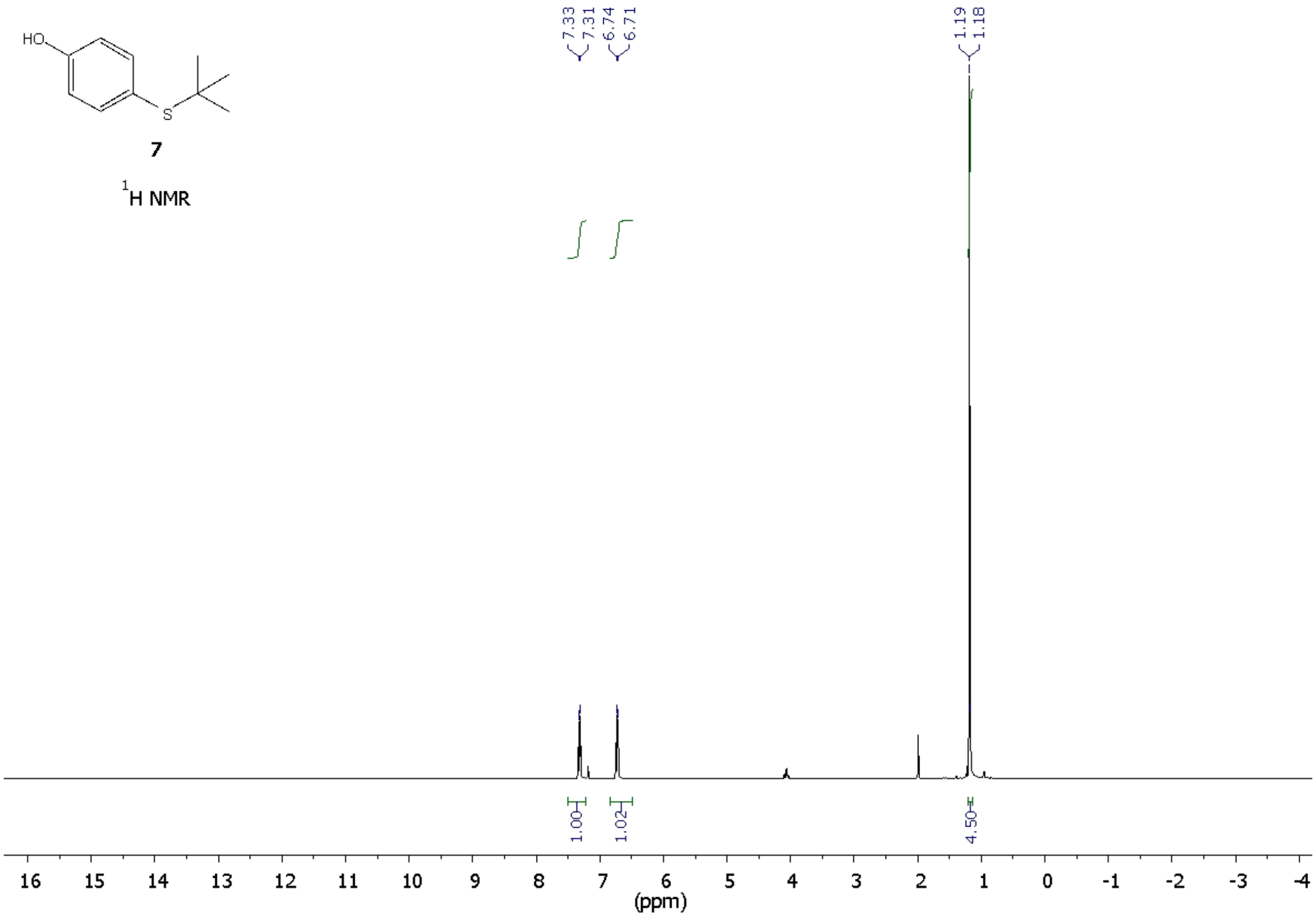
HO
S
7
$^{1}$H NMR
7.33
7.31
6.74
6.71
1.19
1.18
1.00
1.02
4.50
16 15 14 13 12 11 10 9 8 7 6 5 4 3 2 1 0 -1 -2 -3 -4
(ppm)

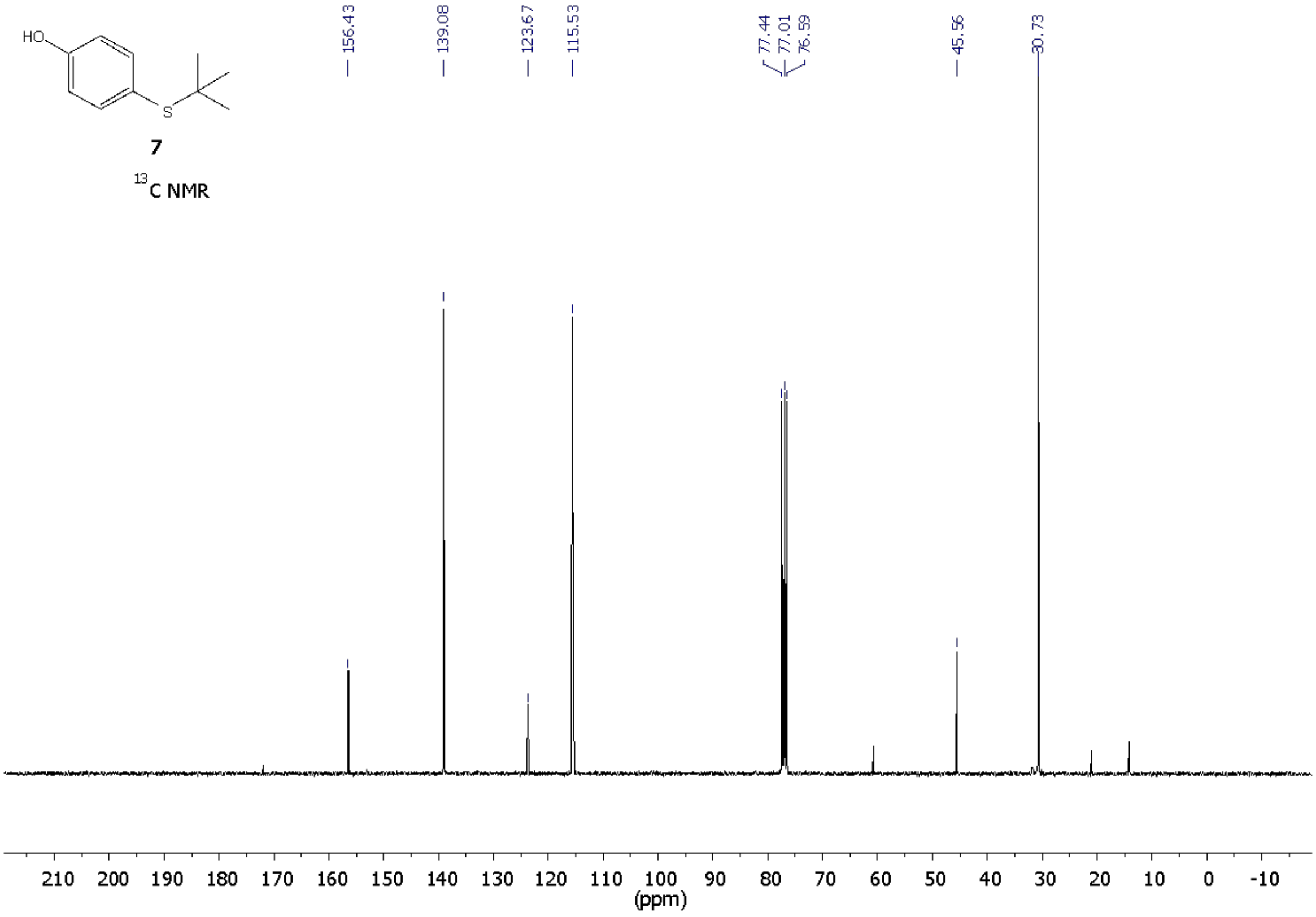
HO
S
7
$^{13}$C NMR
156.43
139.08
123.67
115.53
77.44
77.01
76.59
45.56
30.73
210 200 190 180 170 160 150 140 130 120 110 100 90 80 70 60 50 40 30 20 10 0 -10
(ppm)

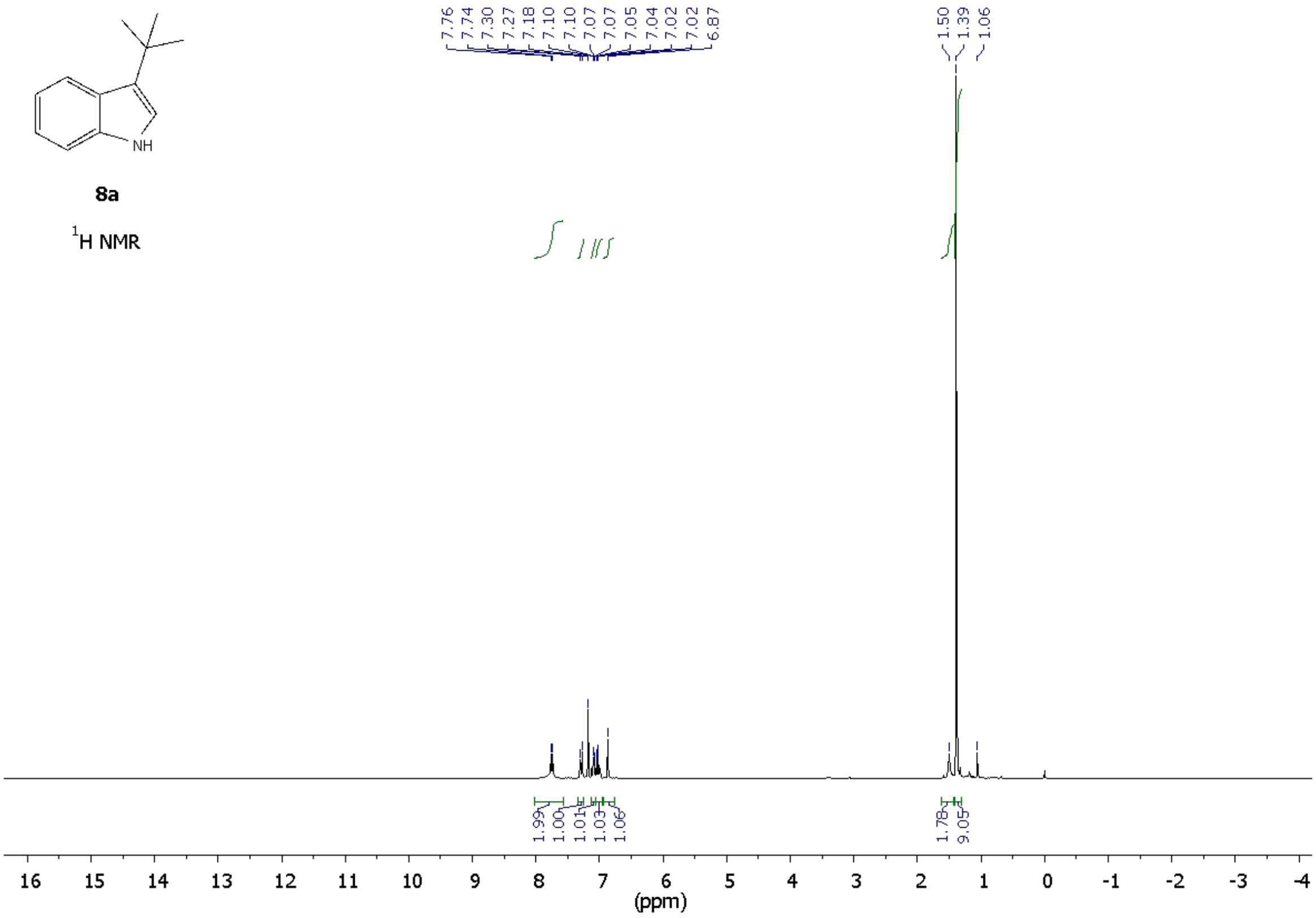

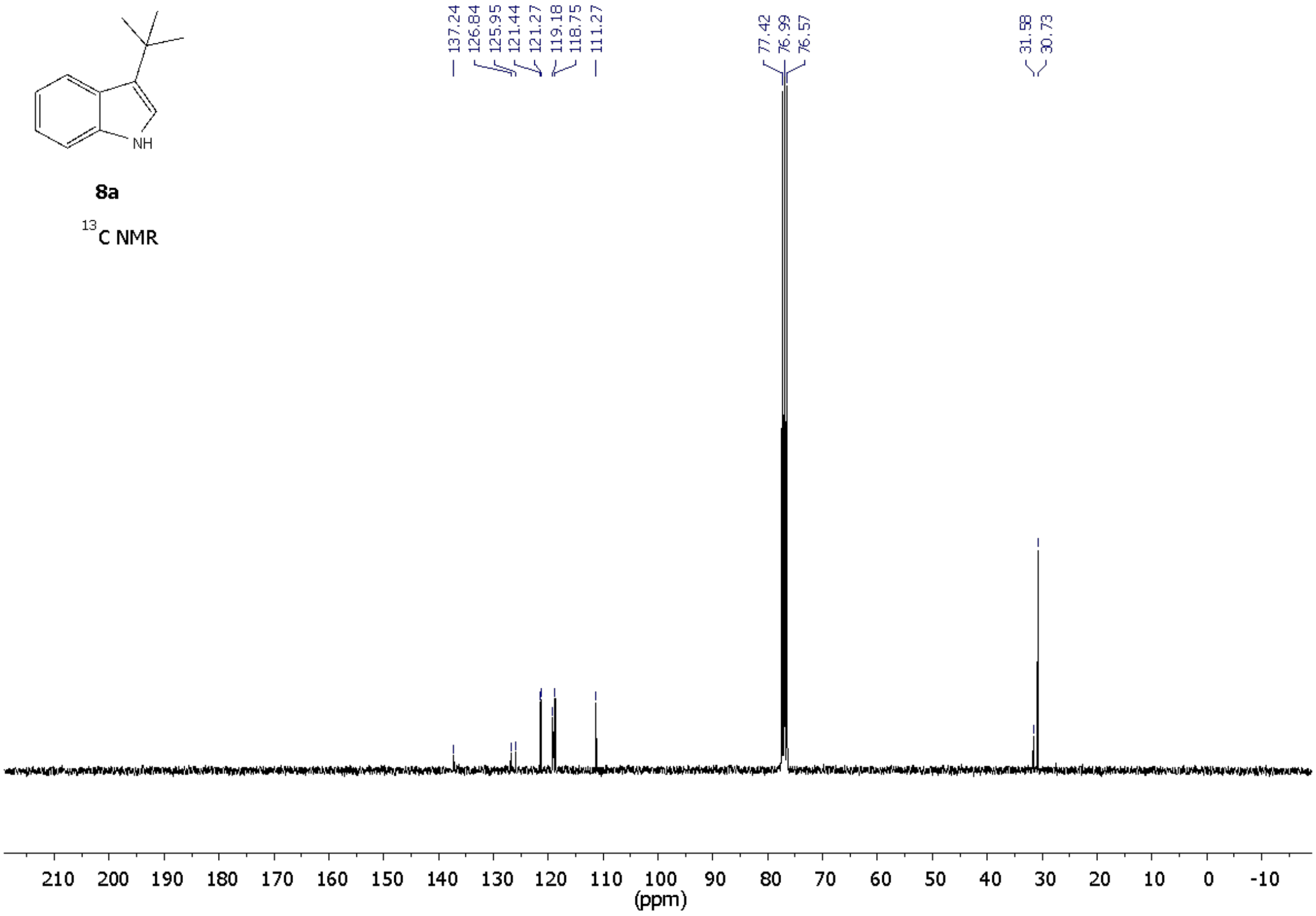

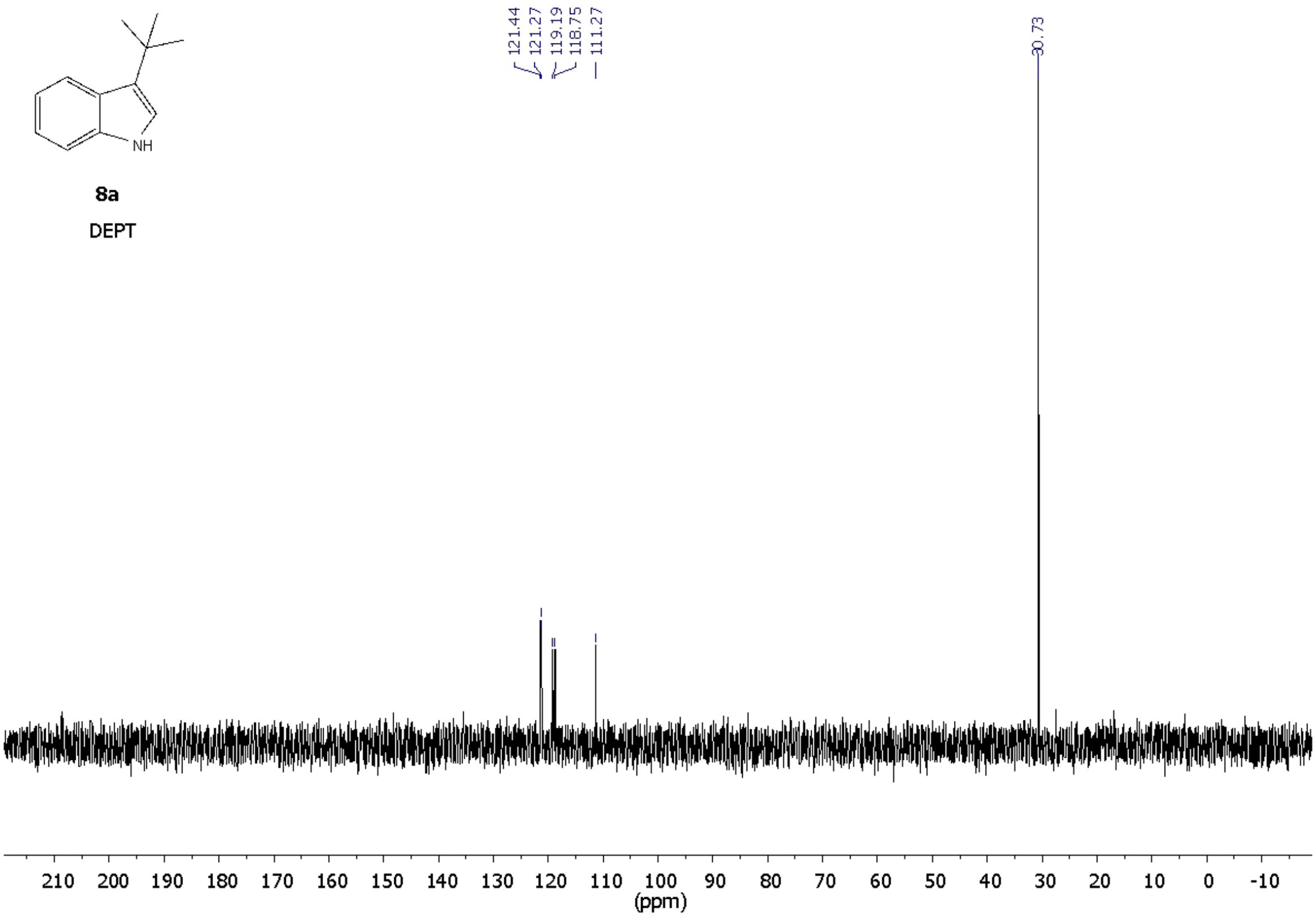

**8a**

DEPT

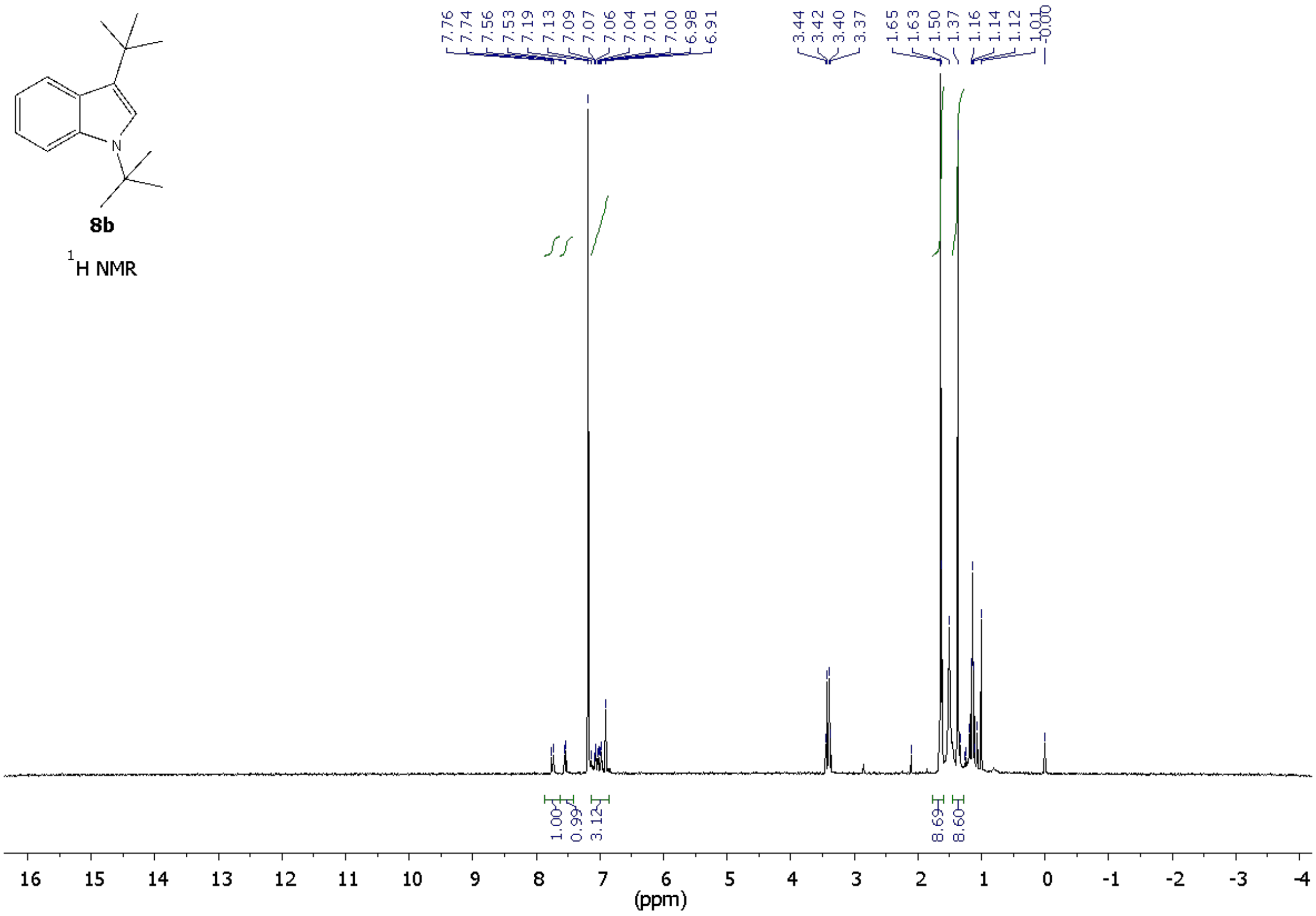

**8b**

$^{1}$H NMR

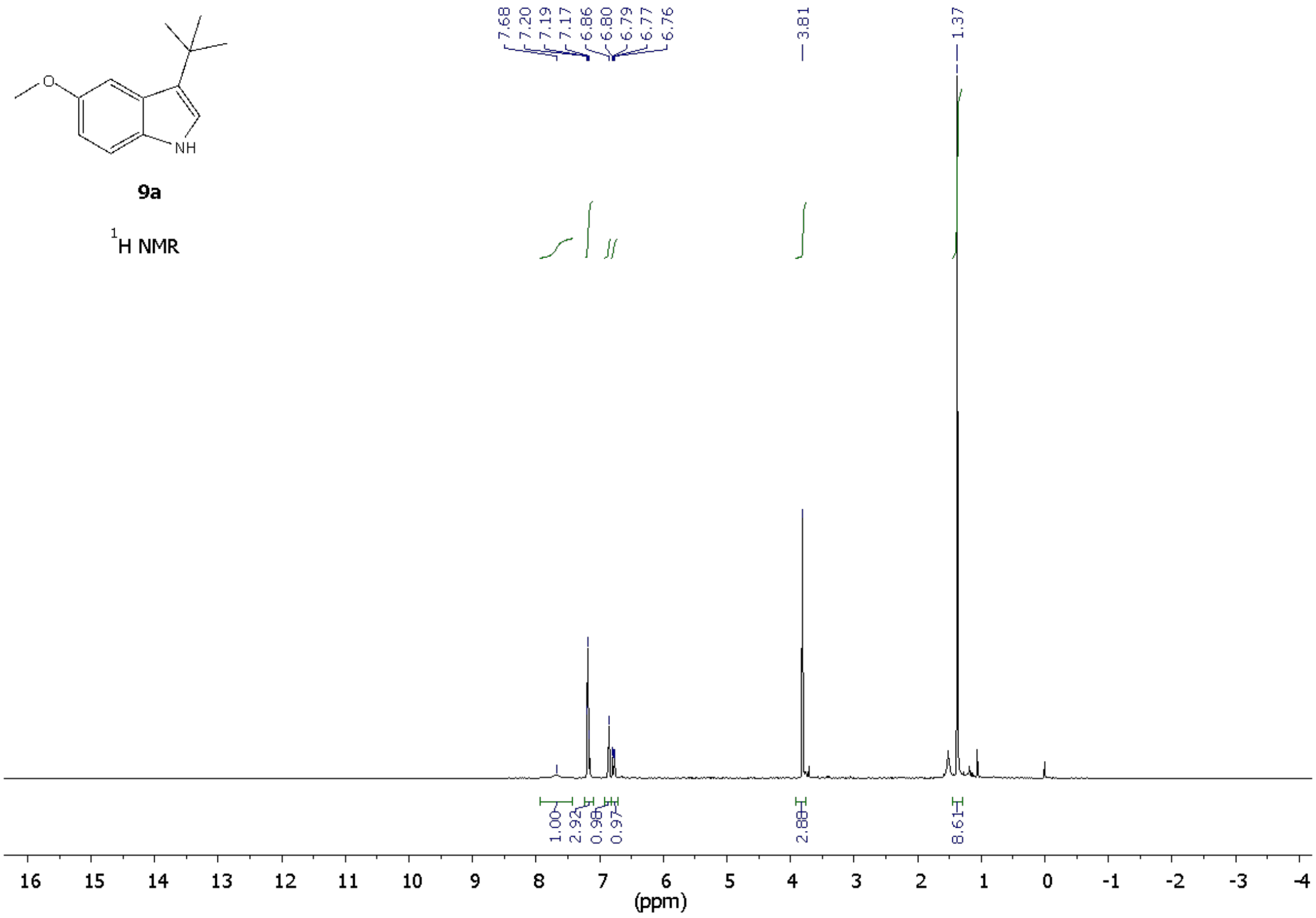

**9a**

$^{1}$H NMR

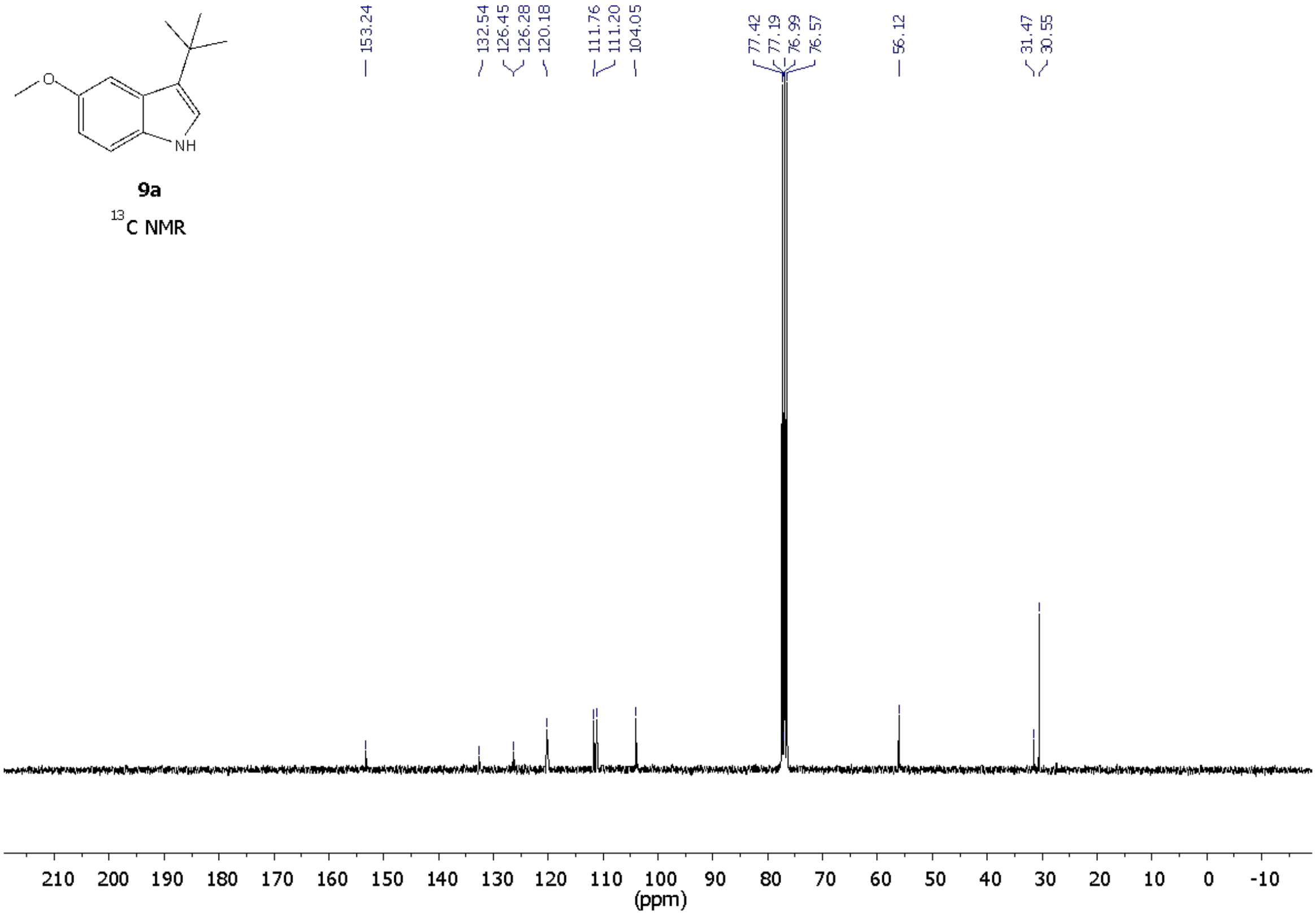

**9a**

$^{13}$C NMR

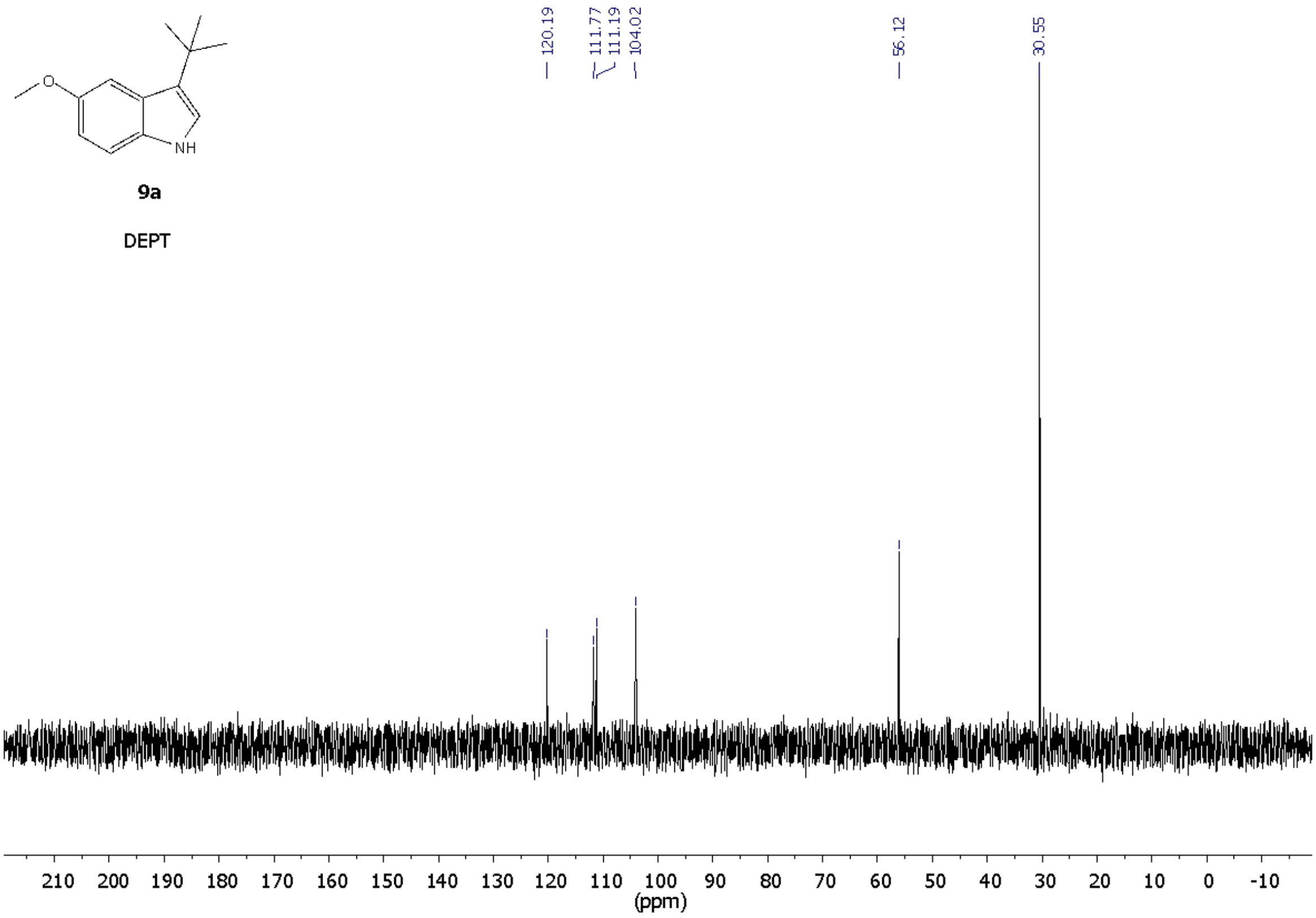

**9a**

DEPT

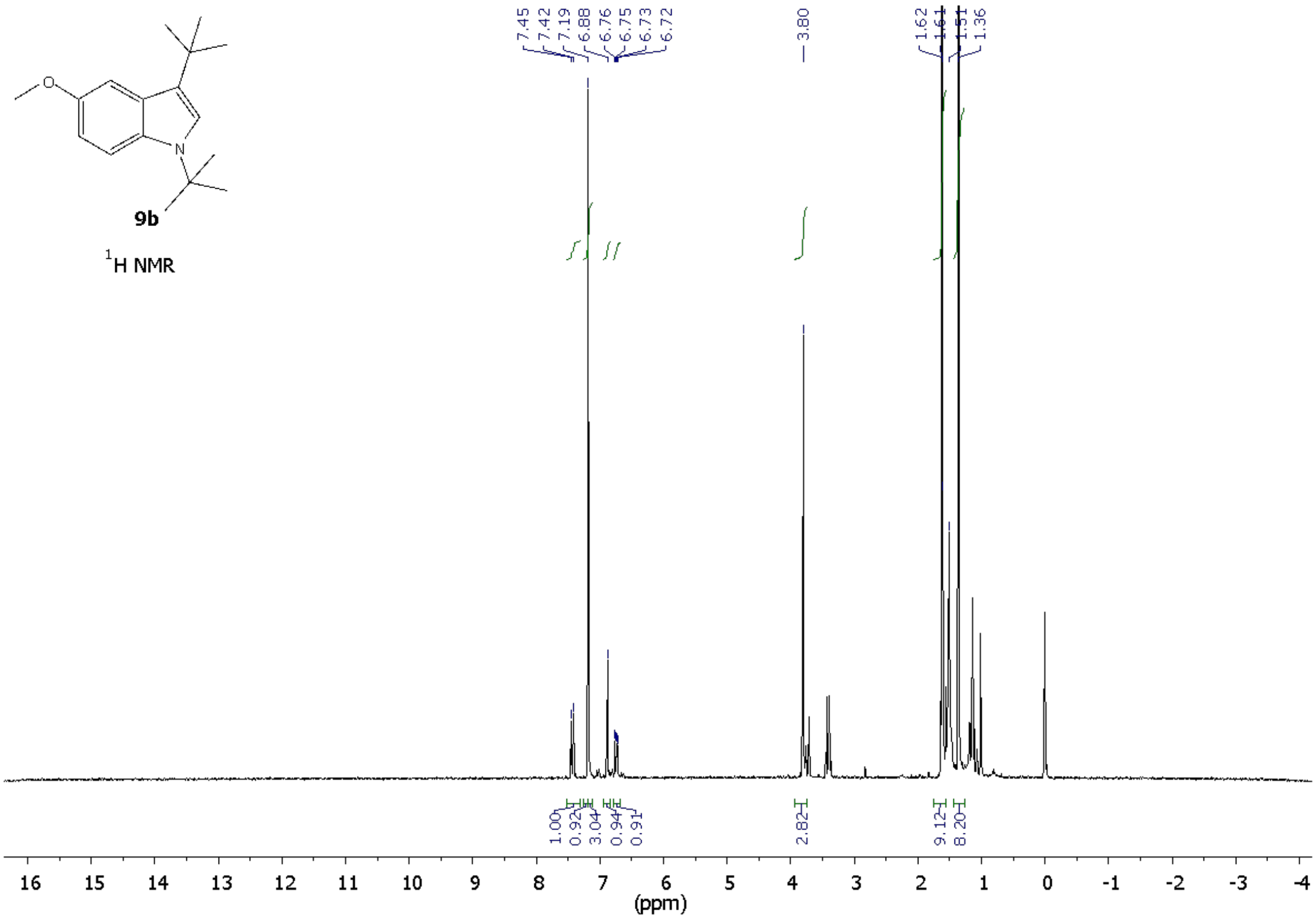

**9b**

[1]H NMR

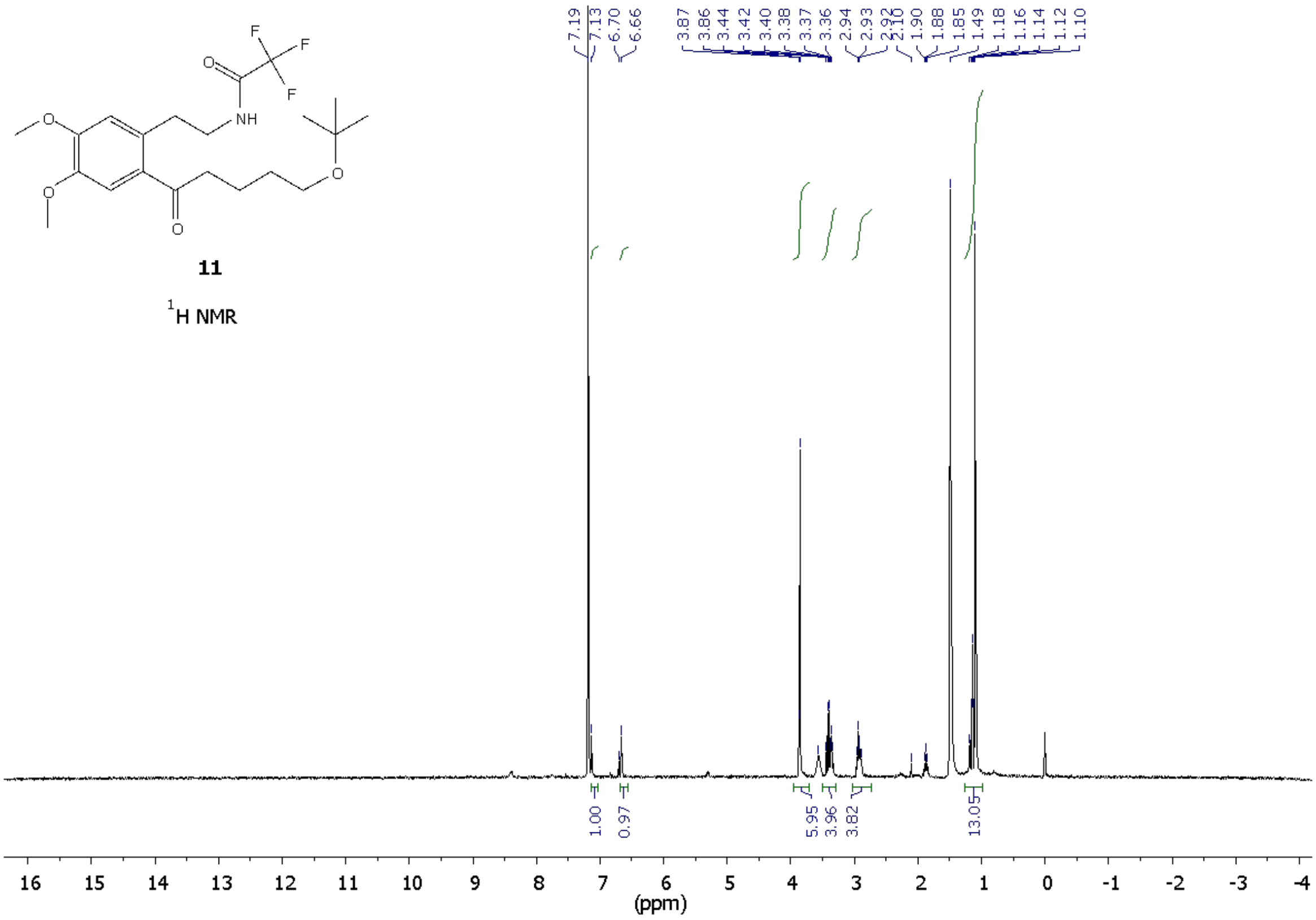

**11**

[1]H NMR

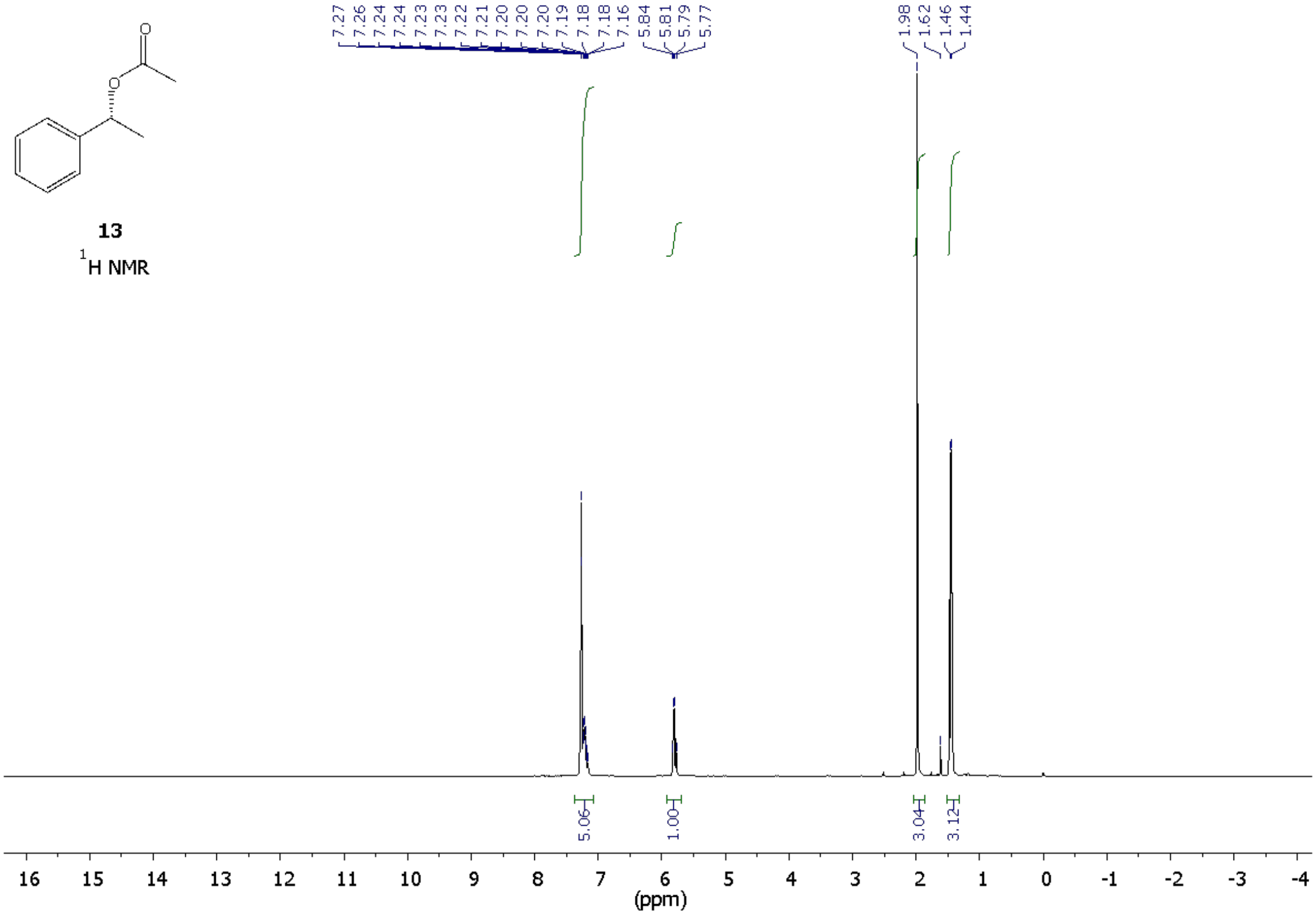

**13**

[1]H NMR

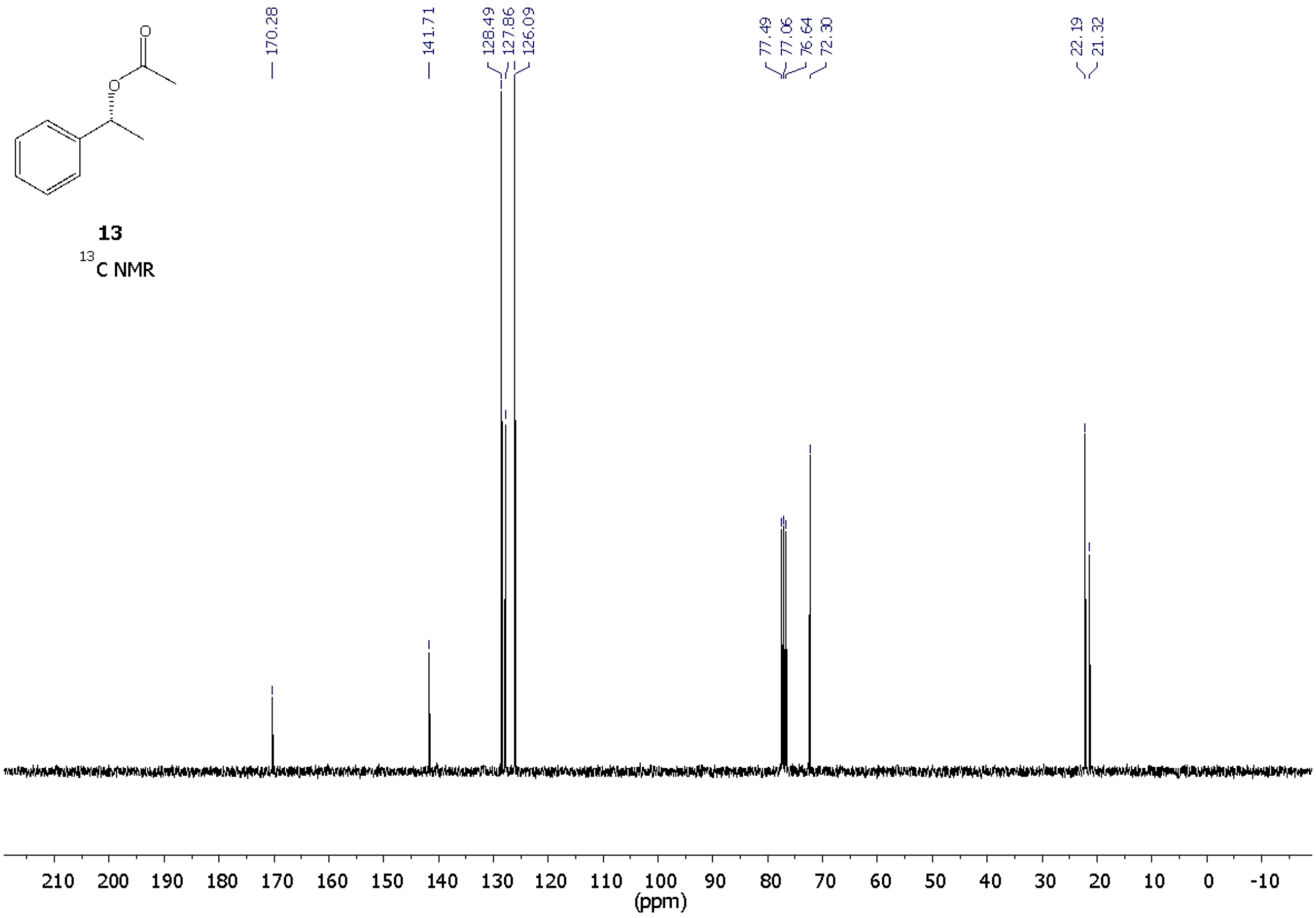
170.28
141.71
128.49
127.86
126.09
77.49
77.06
76.64
72.30
22.19
21.32
13
$^{13}$C NMR
210 200 190 180 170 160 150 140 130 120 110 100 90 80 70 60 50 40 30 20 10 0 -10
(ppm)

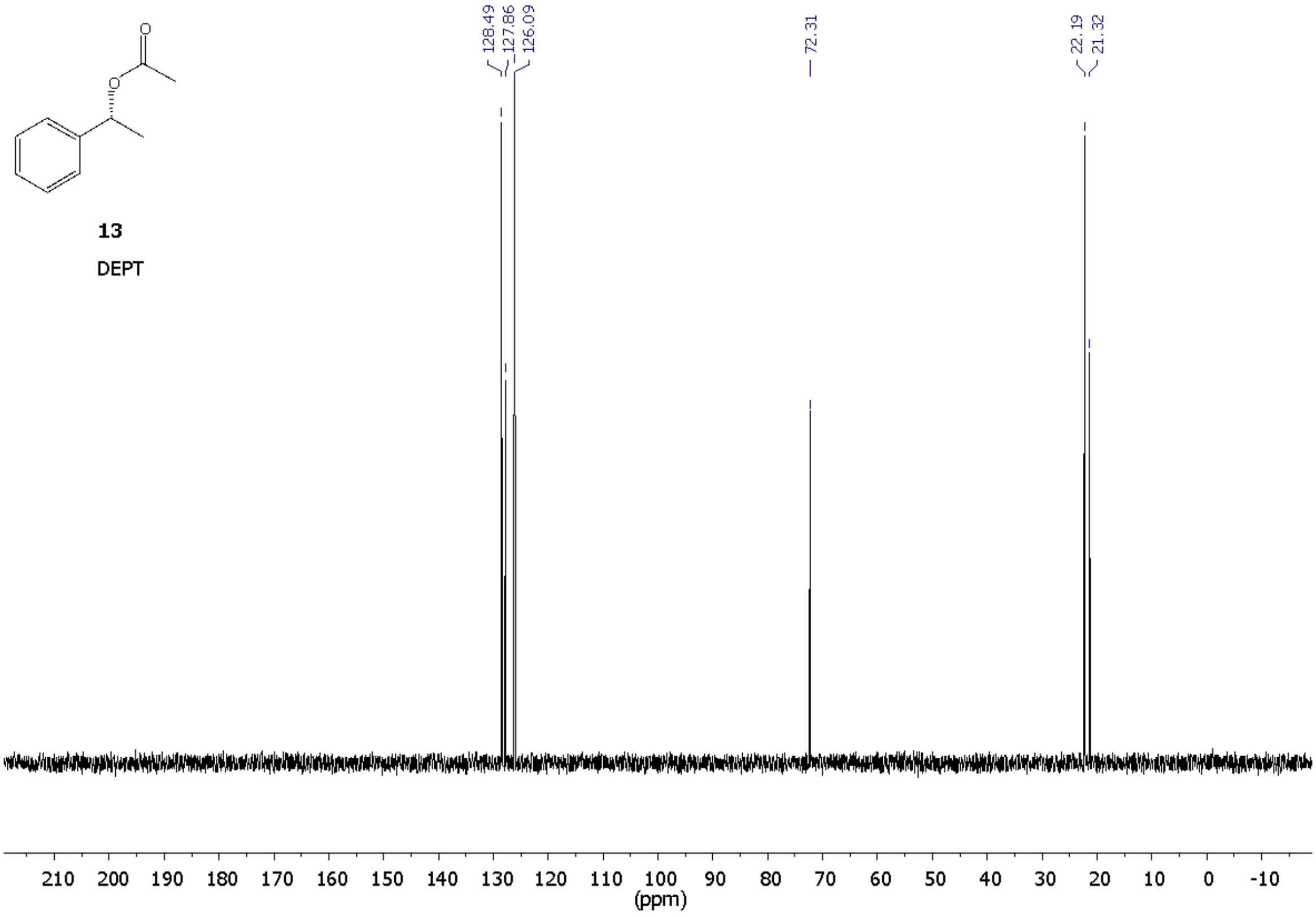
128.49
127.86
126.09
72.31
22.19
21.32
13
DEPT
210 200 190 180 170 160 150 140 130 120 110 100 90 80 70 60 50 40 30 20 10 0 -10
(ppm)

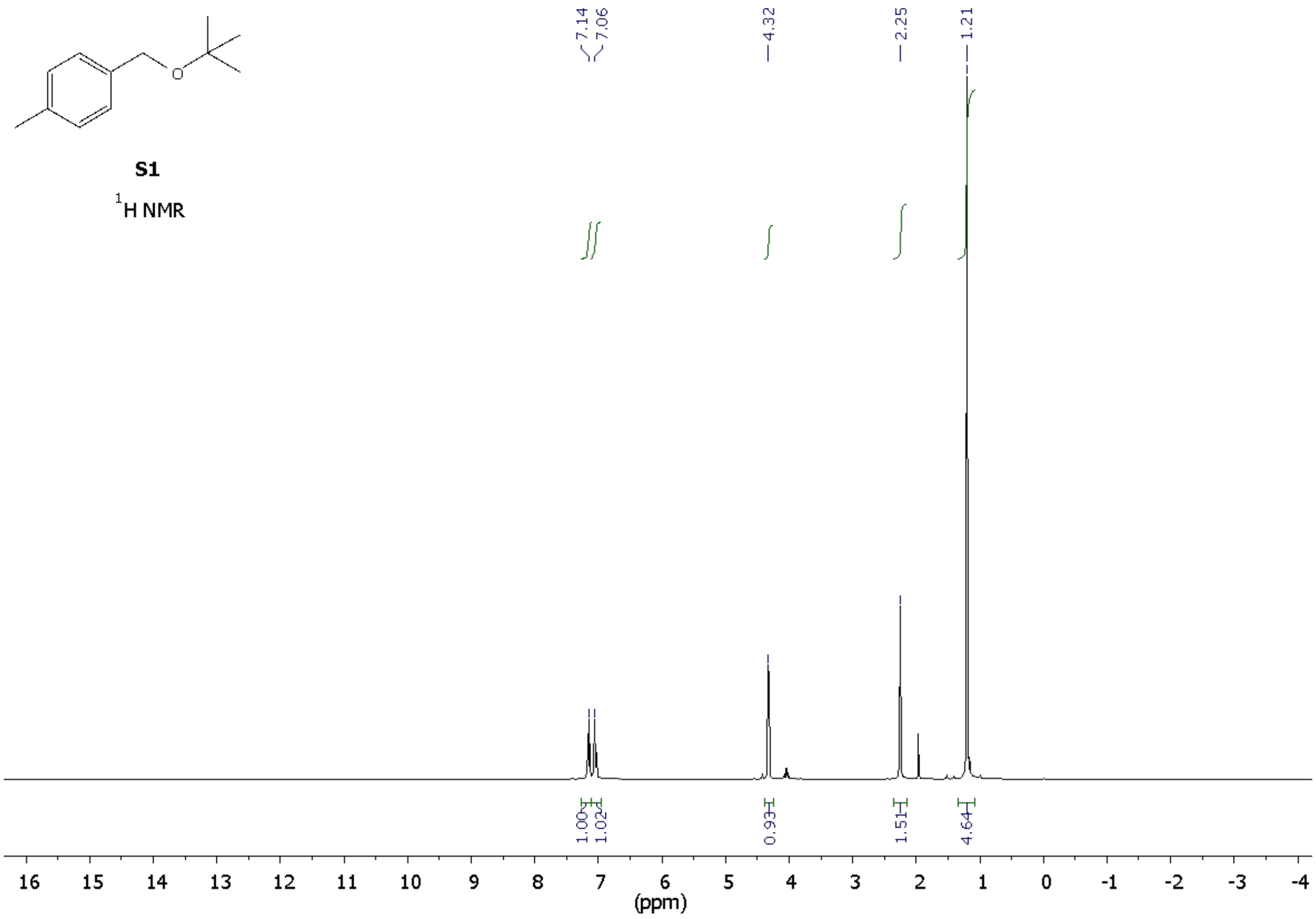
S1
1H NMR
7.14
7.06
4.32
2.25
1.21
1.00
1.02
0.93
1.51
4.64
16 15 14 13 12 11 10 9 8 7 6 5 4 3 2 1 0 -1 -2 -3 -4
(ppm)

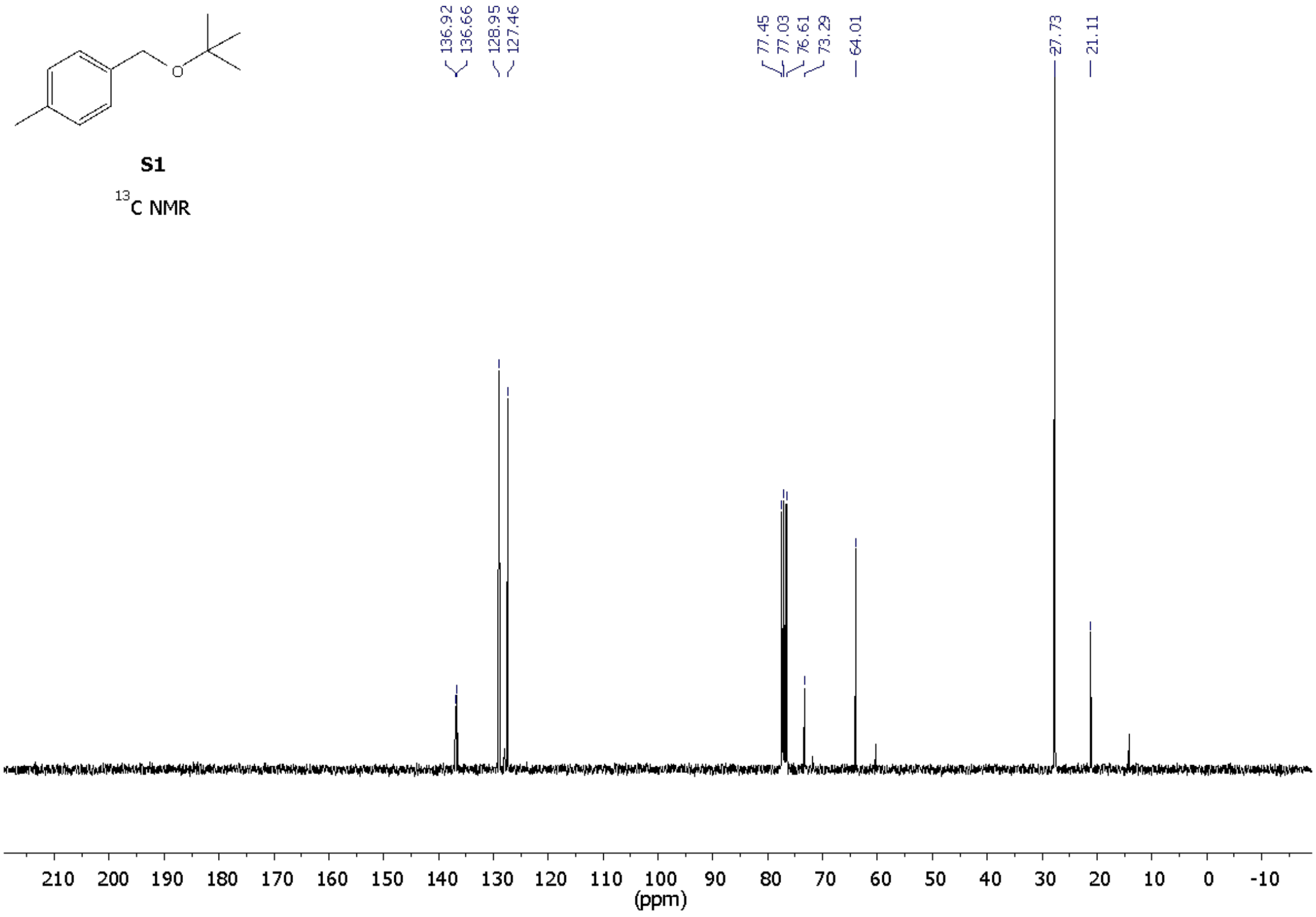
S1
13C NMR
136.92
136.66
128.95
127.46
77.45
77.03
76.61
73.29
64.01
27.73
21.11
210 200 190 180 170 160 150 140 130 120 110 100 90 80 70 60 50 40 30 20 10 0 -10
(ppm)

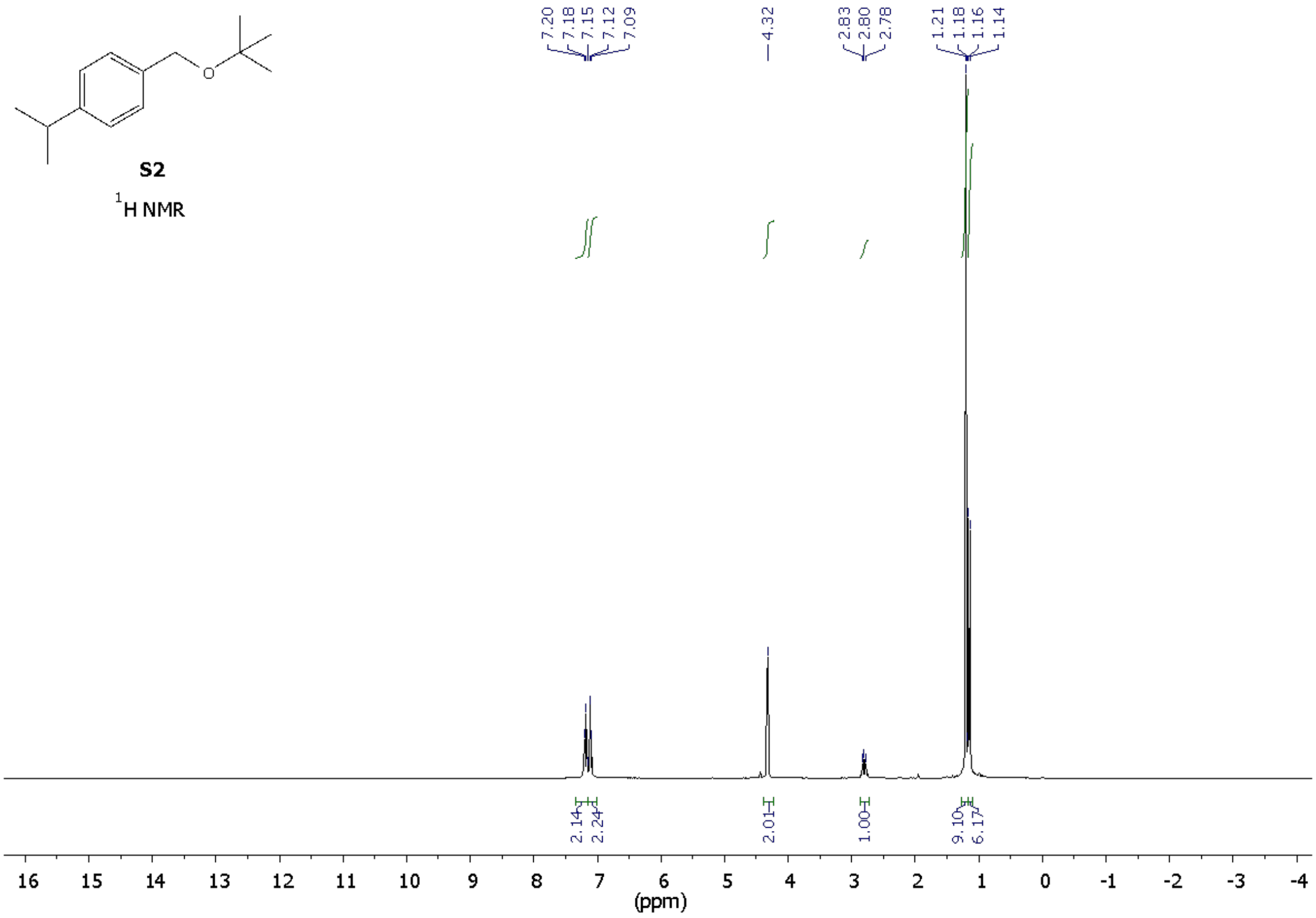
S2
1H NMR
7.20
7.18
7.15
7.12
7.09
4.32
2.83
2.80
2.78
1.21
1.18
1.16
1.14
2.14
2.24
2.01
1.00
9.10
6.17
16 15 14 13 12 11 10 9 8 7 6 5 4 3 2 1 0 -1 -2 -3 -4
(ppm)

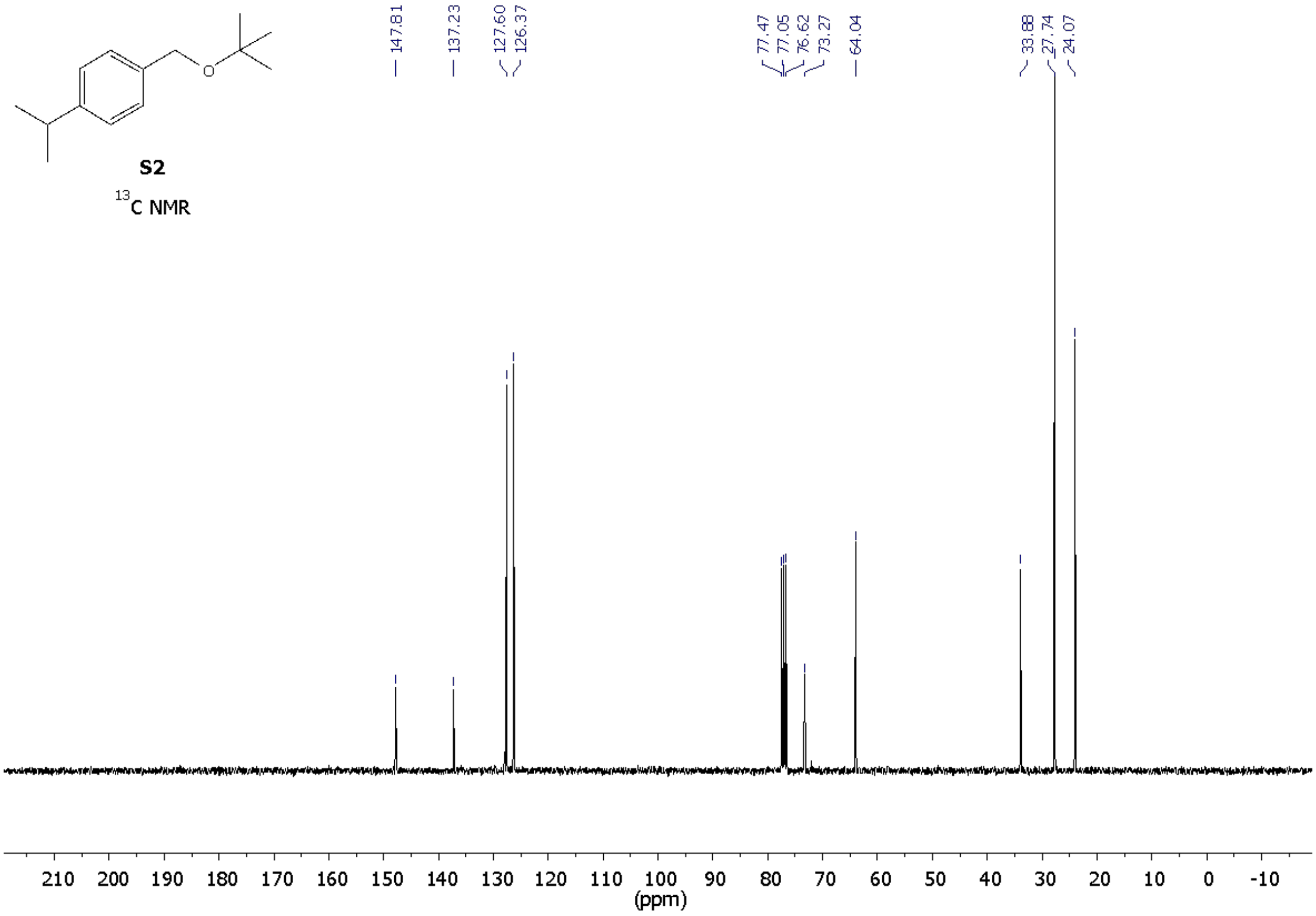
S2
13C NMR
147.81
137.23
127.60
126.37
77.47
77.05
76.62
73.27
64.04
33.88
27.74
24.07
210 200 190 180 170 160 150 140 130 120 110 100 90 80 70 60 50 40 30 20 10 0 -10
(ppm)

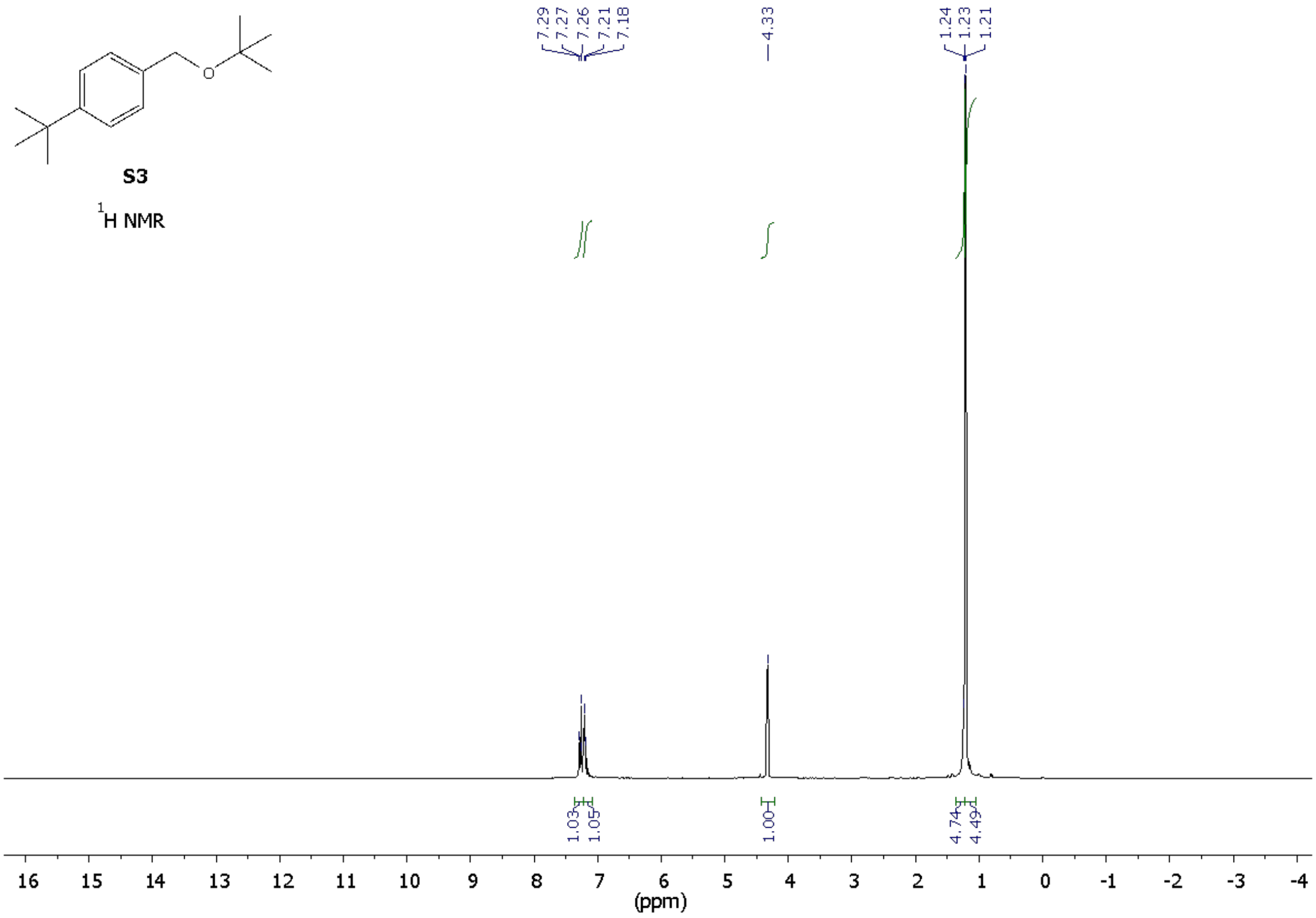
S3
1H NMR
7.29
7.27
7.26
7.21
7.18
4.33
1.24
1.23
1.21
1.03
1.05
1.00
4.74
4.49
16 15 14 13 12 11 10 9 8 7 6 5 4 3 2 1 0 -1 -2 -3 -4
(ppm)

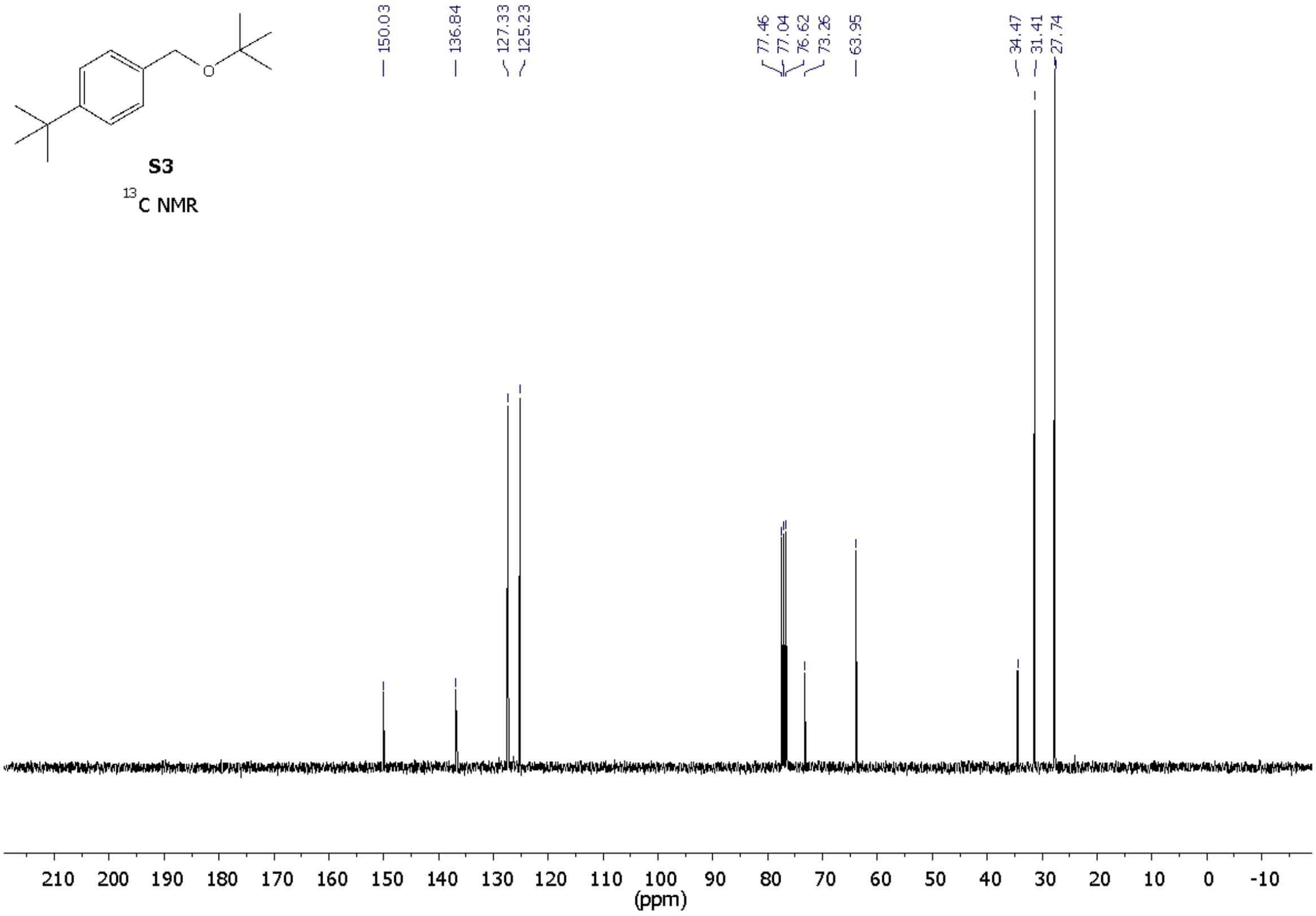
S3
13C NMR
150.03
136.84
127.33
125.23
77.46
77.04
76.62
73.26
63.95
34.47
31.41
27.74
210 200 190 180 170 160 150 140 130 120 110 100 90 80 70 60 50 40 30 20 10 0 -10
(ppm)

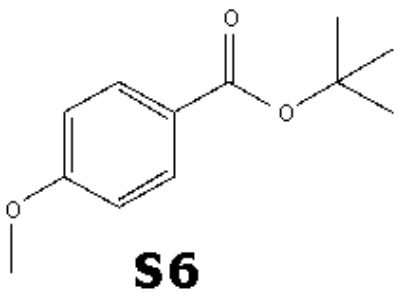

**S6**
1H
NMR

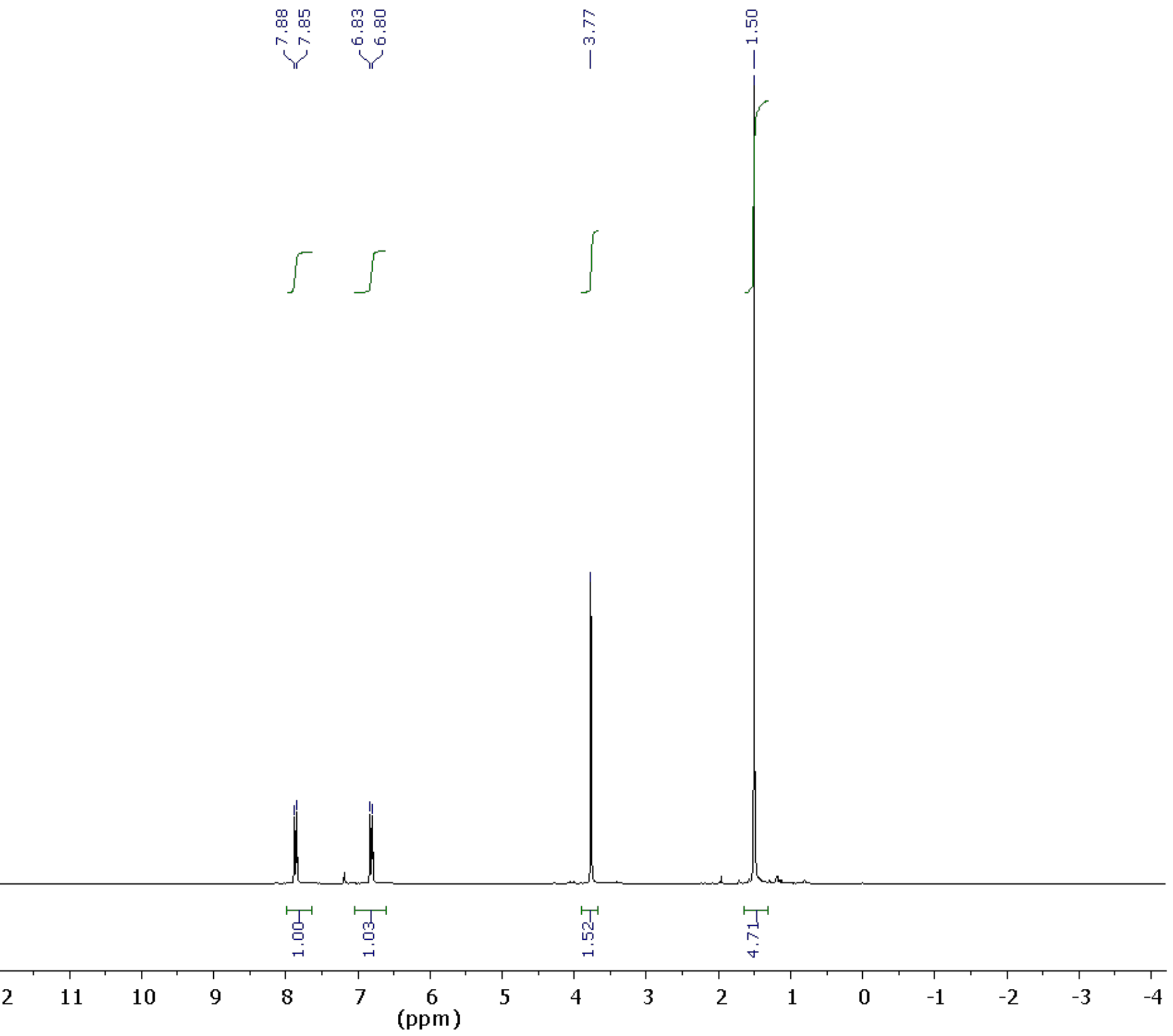

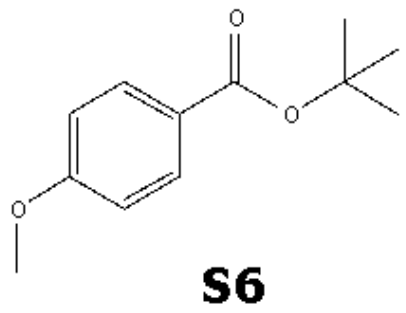

**S6**
13C
NMR

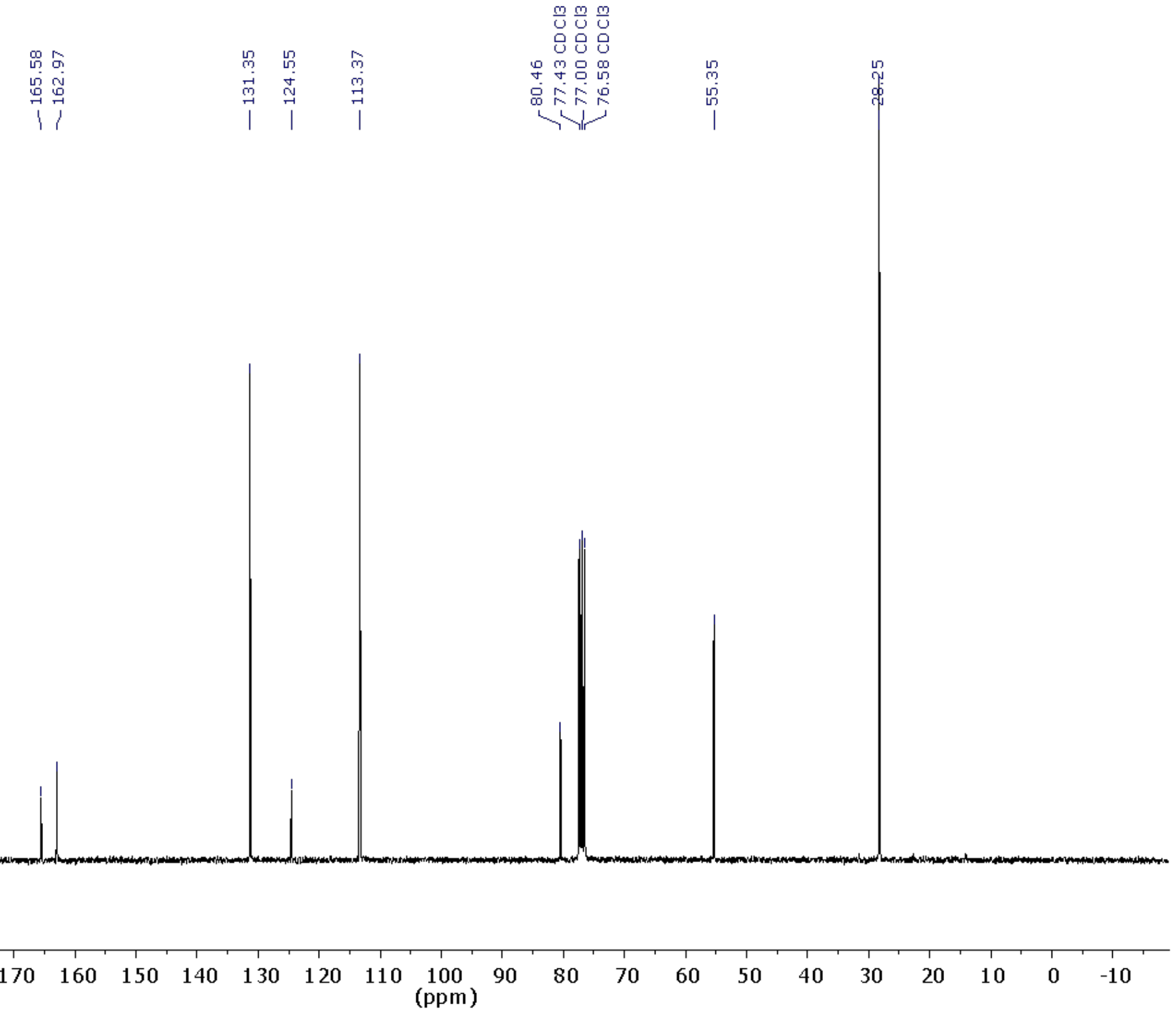

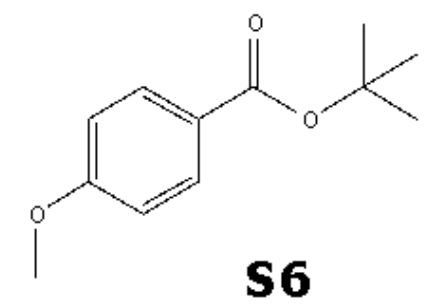

**S6**

DEPT

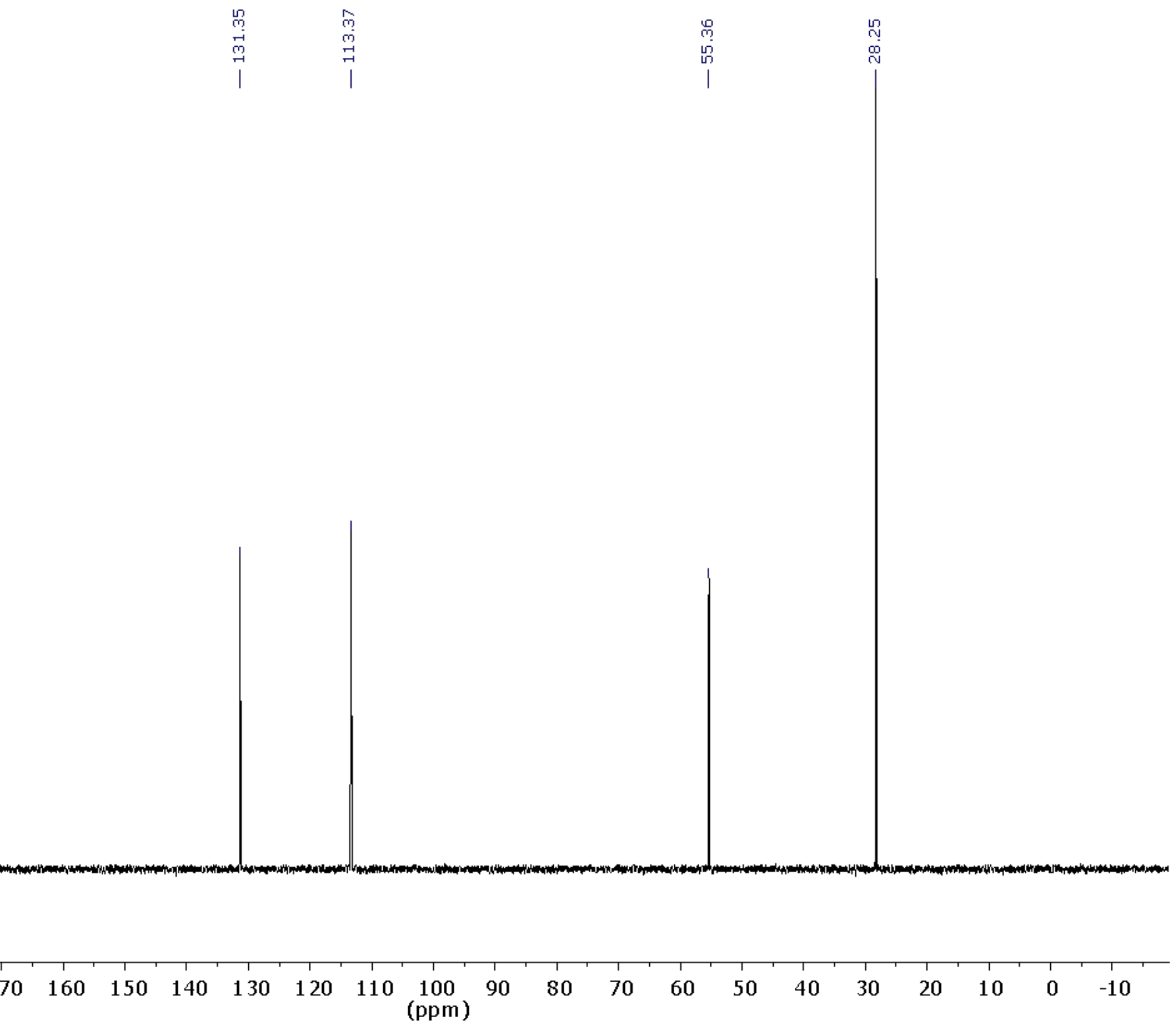

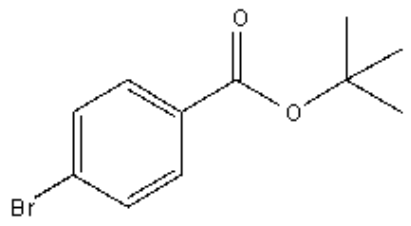

**S7**

1H NMR

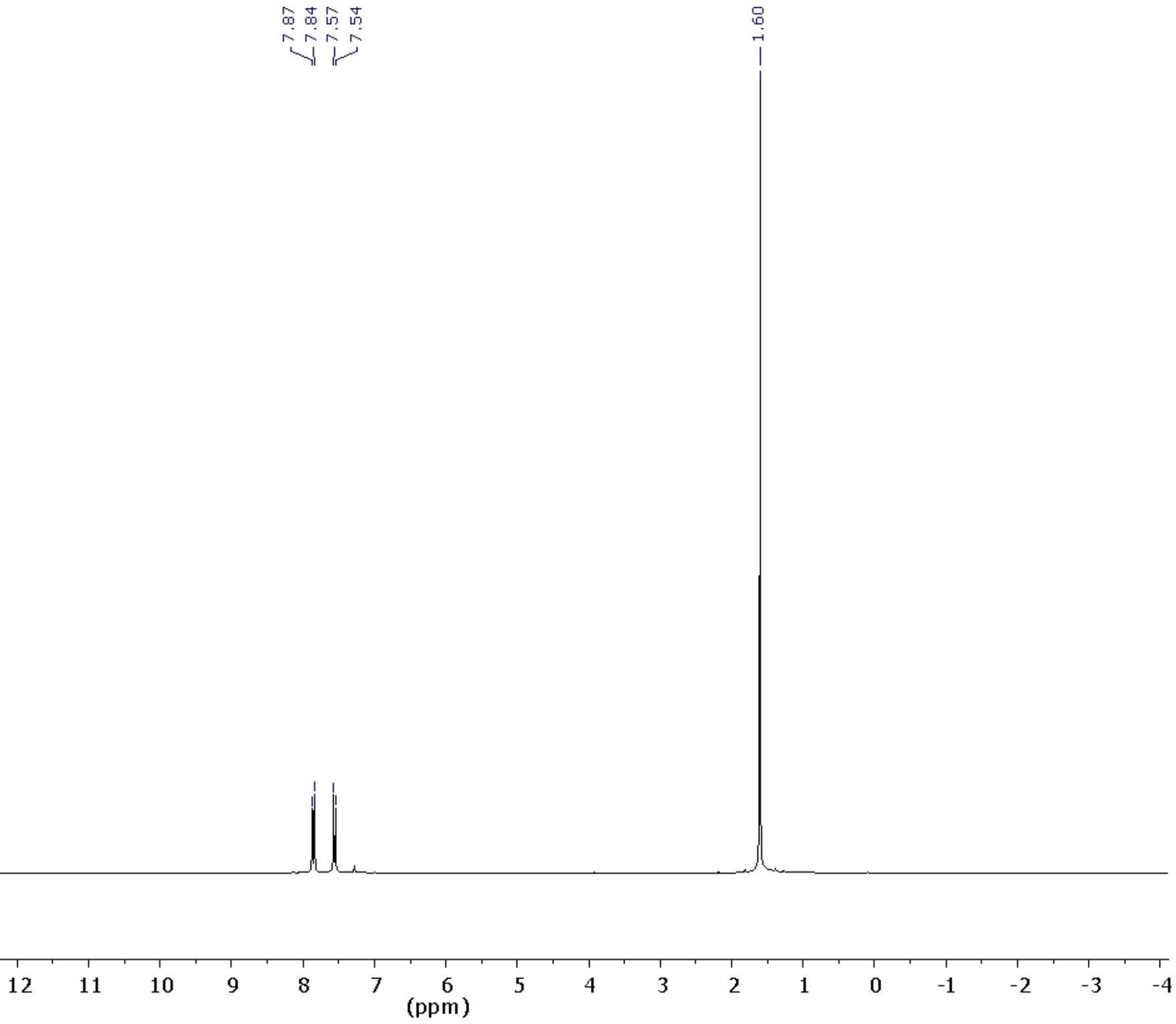

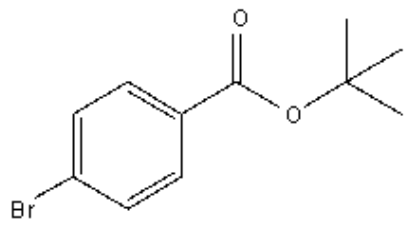

**S7**

13C NMR

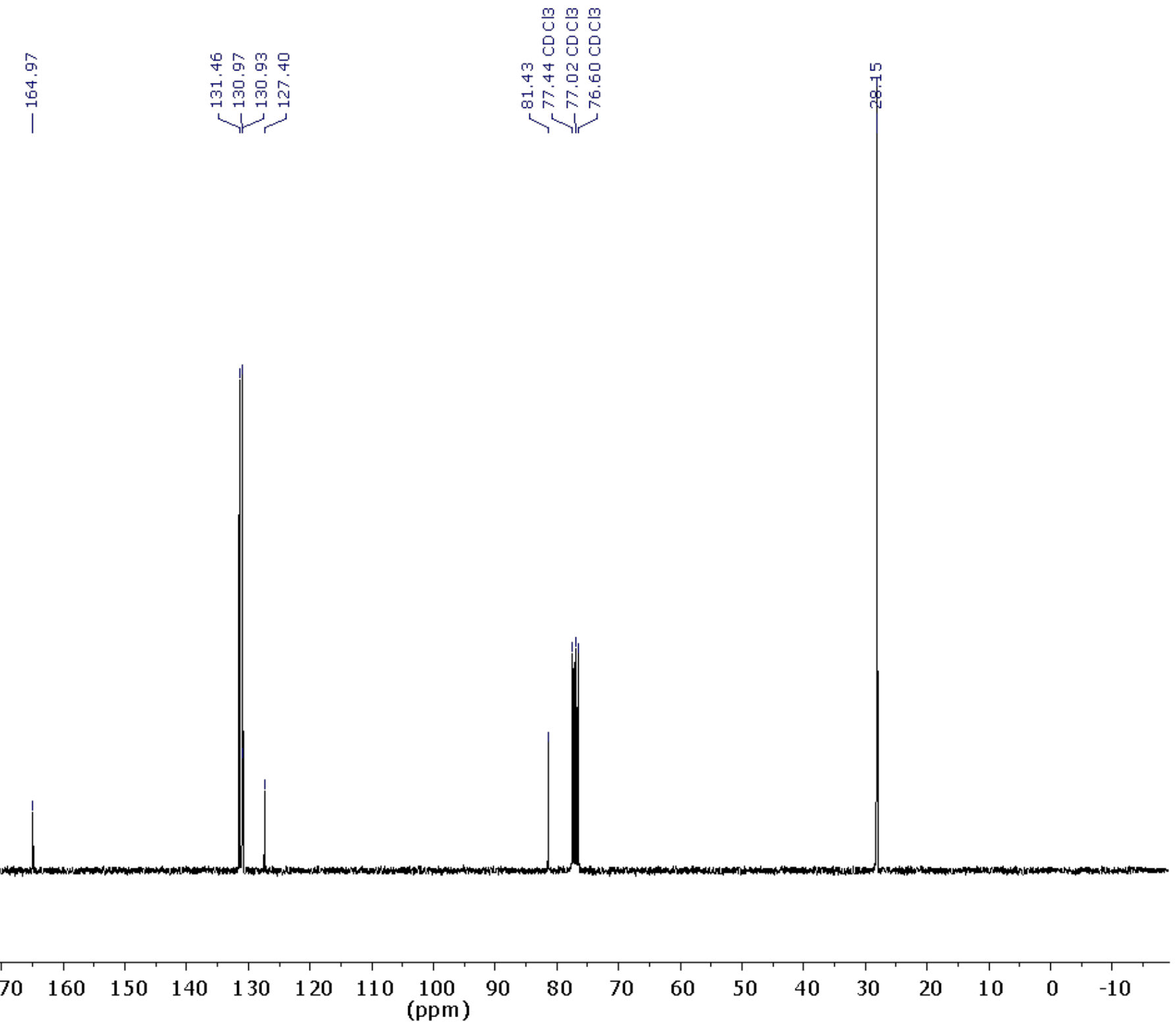

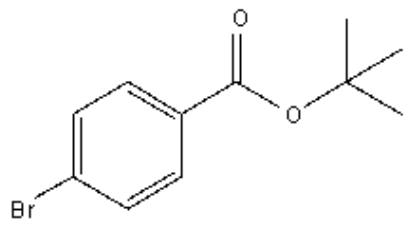

**S7**

DEPT

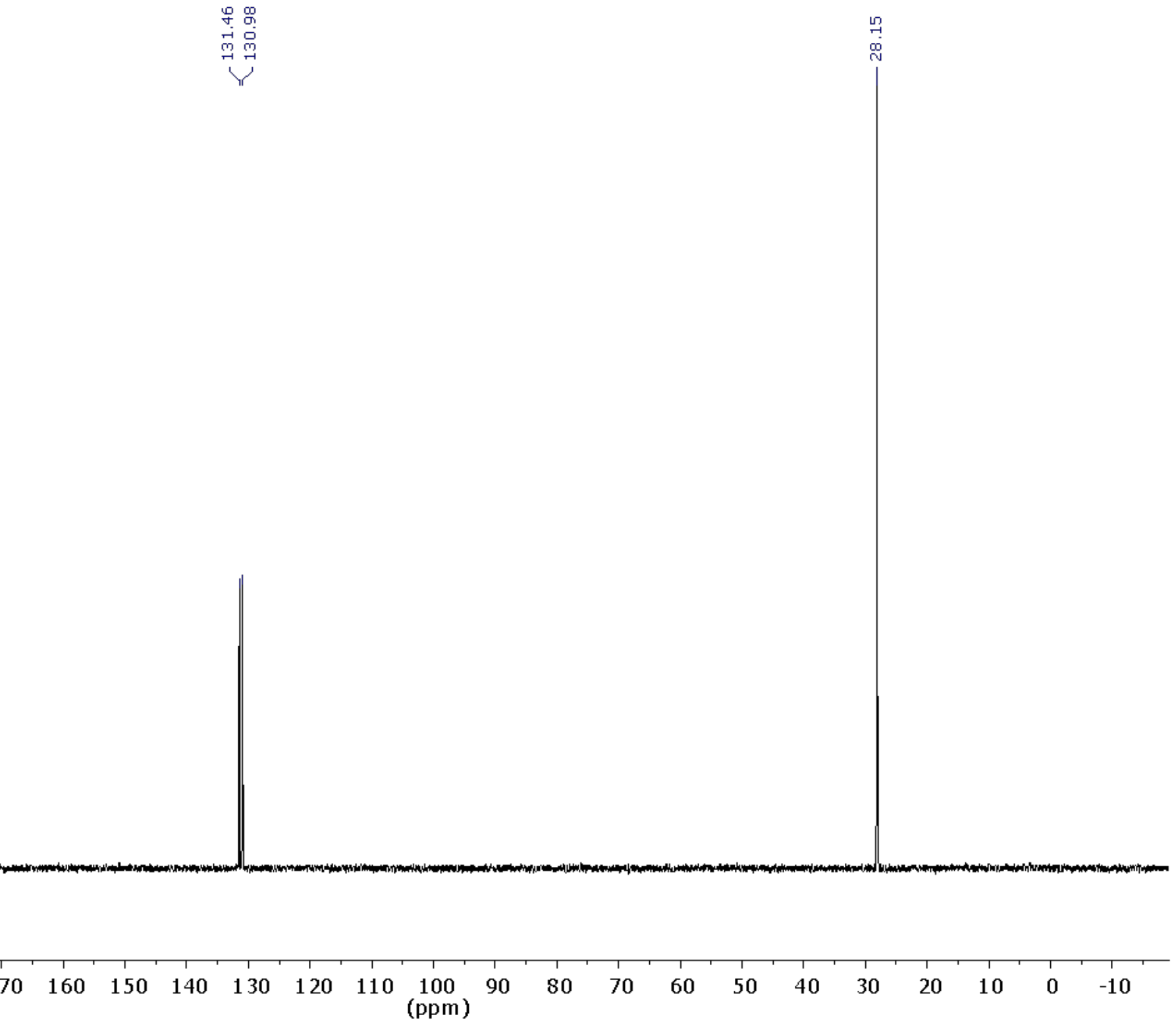

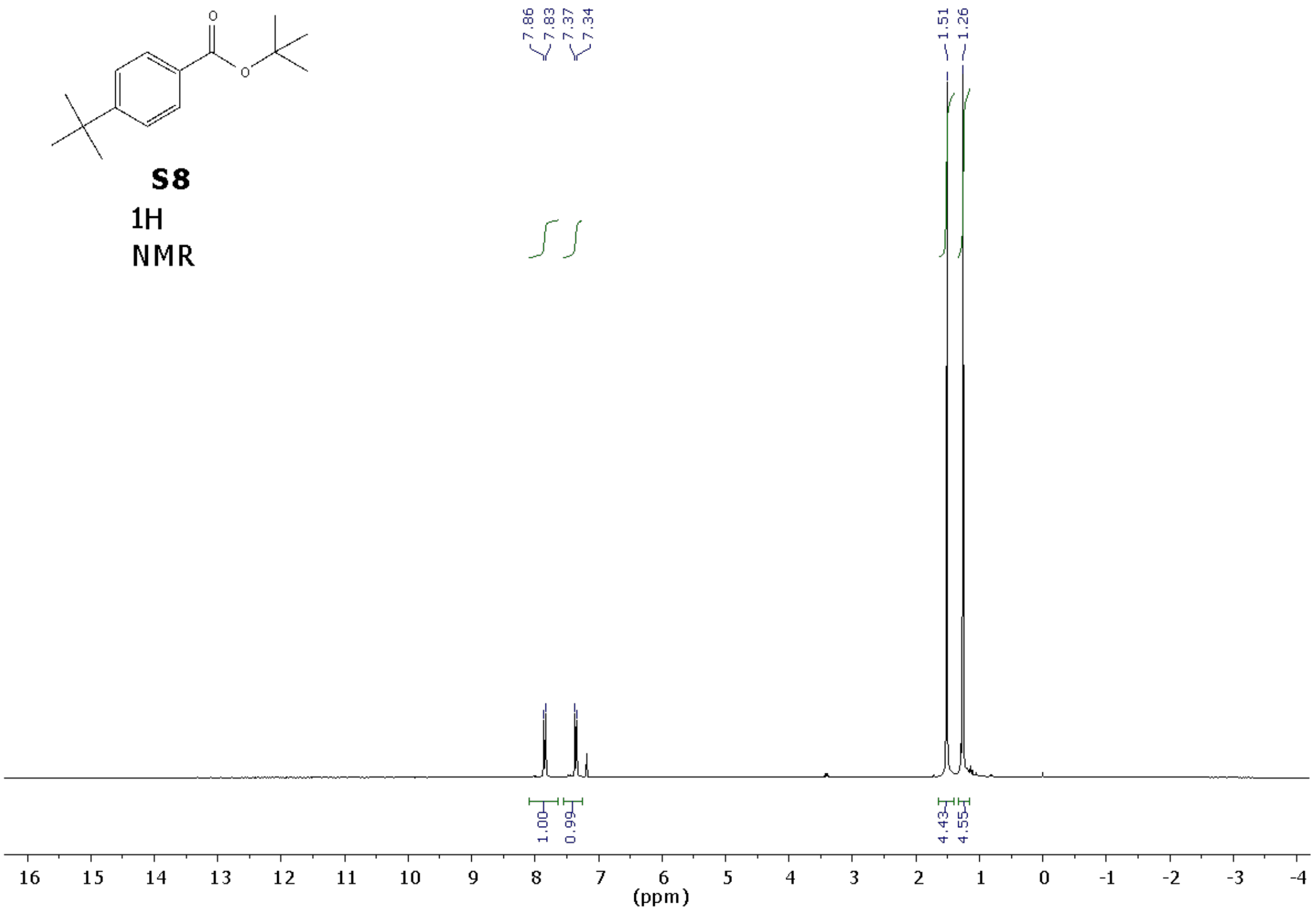
O
O
S8
1H
NMR
7.86
7.83
7.37
7.34
1.51
1.26
1.00
0.99
4.43
4.55
16 15 14 13 12 11 10 9 8 7 6 5 4 3 2 1 0 -1 -2 -3 -4
(ppm)

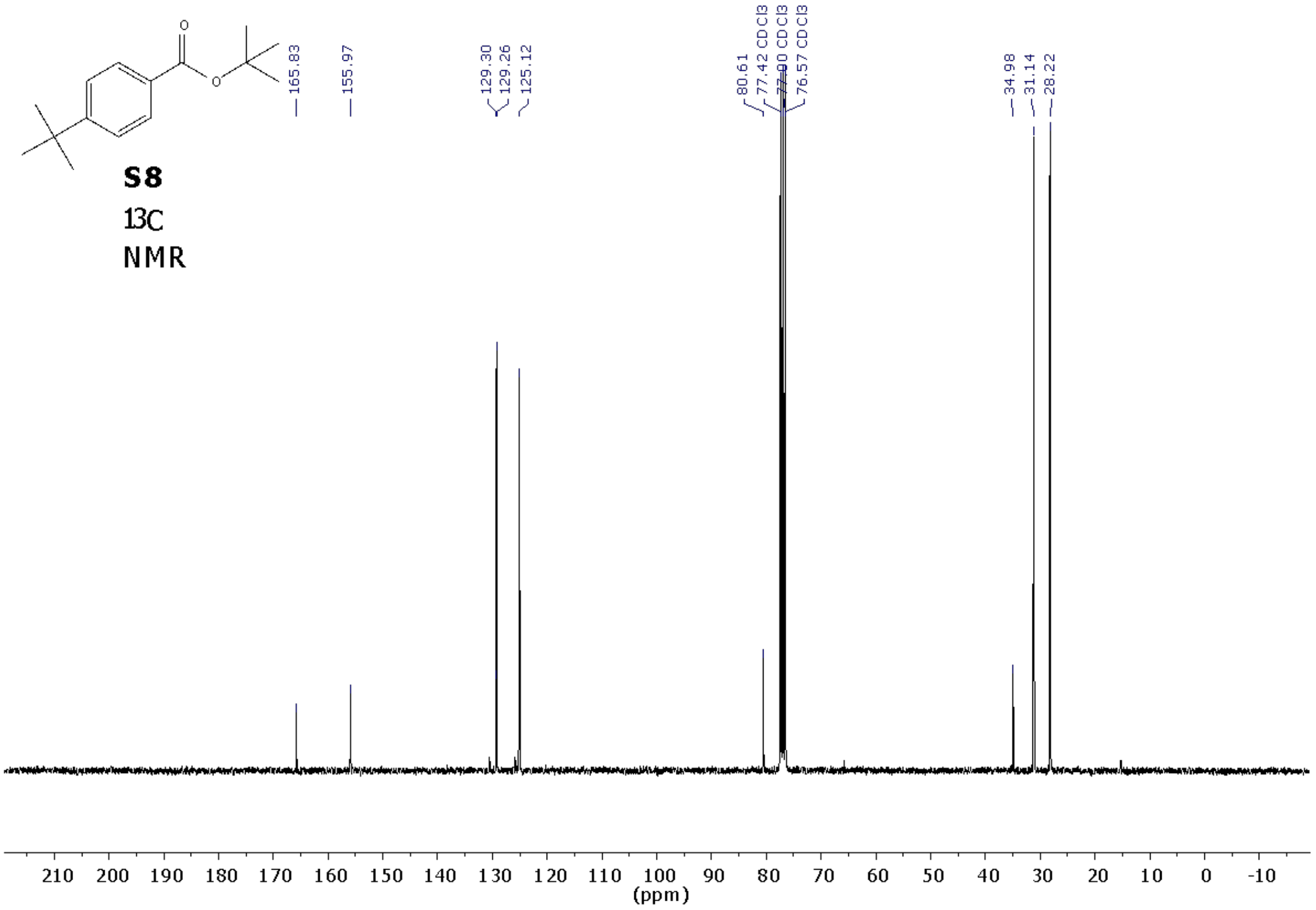
O
O
S8
13C
NMR
165.83
155.97
129.30
129.26
125.12
80.61
77.42 CDCl3
77.00 CDCl3
76.57 CDCl3
34.98
31.14
28.22
210 200 190 180 170 160 150 140 130 120 110 100 90 80 70 60 50 40 30 20 10 0 -10
(ppm)

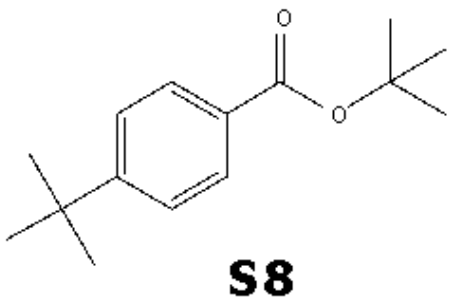

**S8**

DEPT

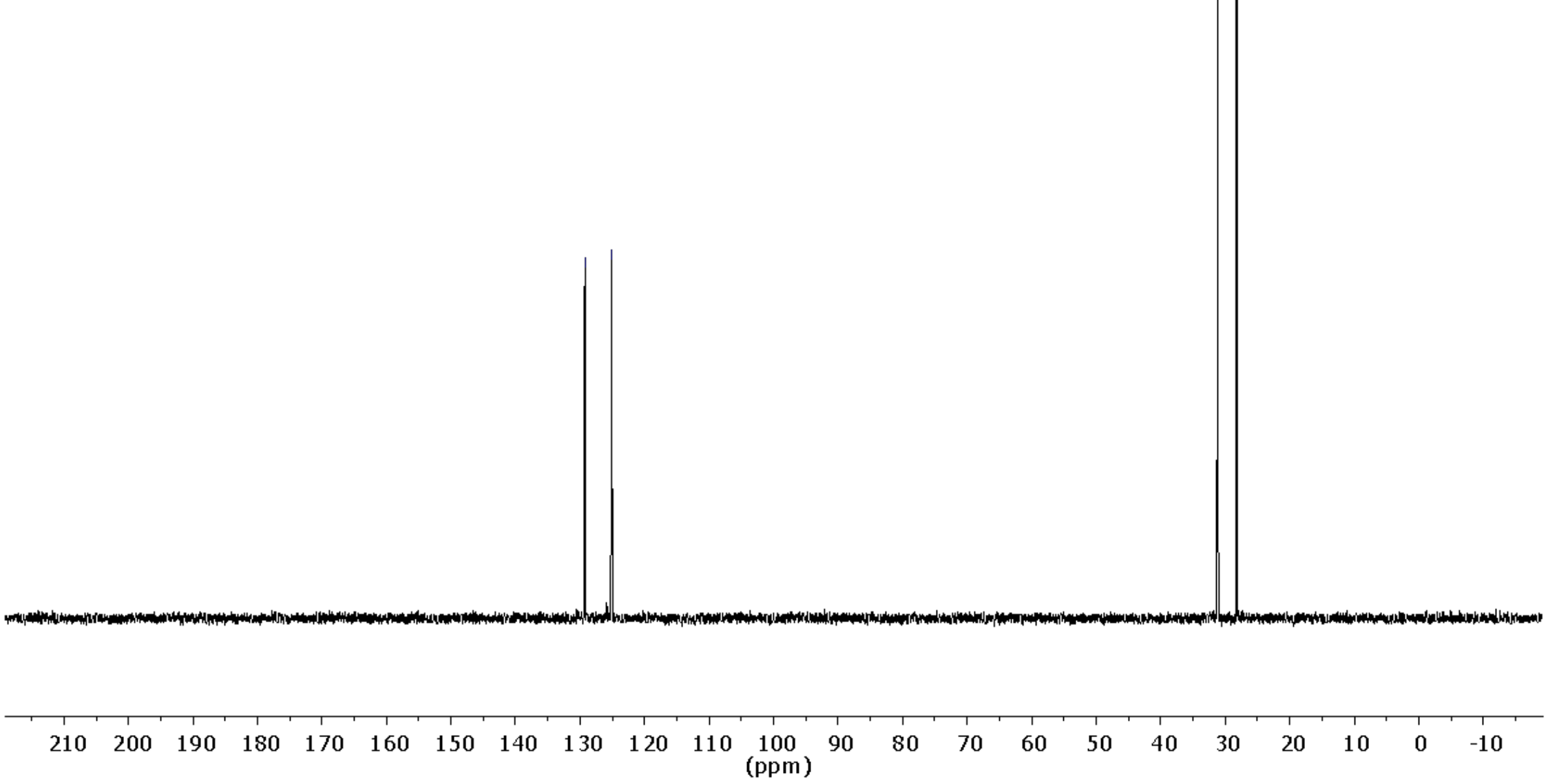

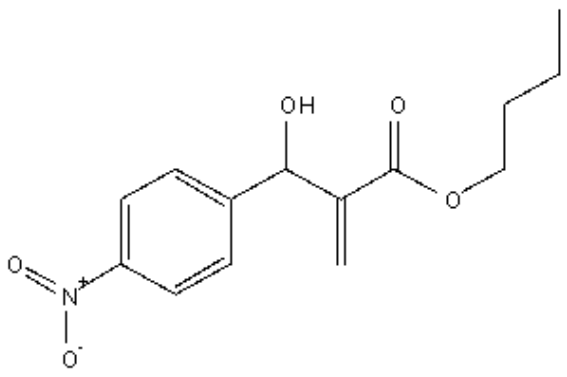

**S9**

1H
NMR

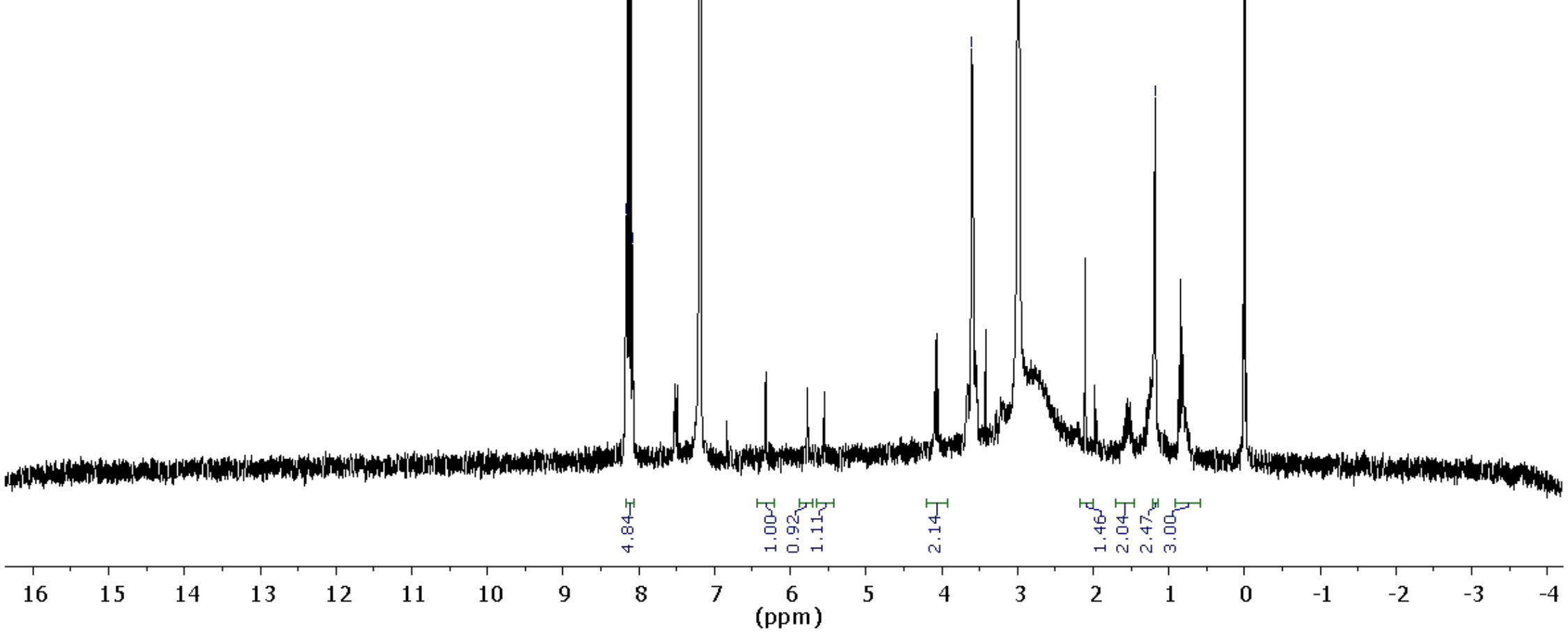

**Supplementary references.**